\documentclass[article]{aa}
\usepackage{graphicx} 
\usepackage{lscape} 
\usepackage{multirow}
\usepackage[caption=false]{subfig} 
\usepackage{txfonts}
\begin{document} 

   \title{Period-luminosity and period-luminosity-metallicity relations for Galactic 
   RR~Lyrae stars in the Sloan bands\thanks{Based on data from the Las Cumbres Observatory. 
   The light curves are only available in electronic form at the Araucaria Project webpage: 
   \url{https://araucaria.camk.edu.pl/} and the CDS via anonymous ftp to \url{cdsarc.u-strasbg.fr} 
   (130.79.128.5) or via \url{http://cdsweb.u-strasbg.fr/cgi-bin/qcat?J/A+A/}.}}


   \titlerunning{PL and PLZ relations for RR~Lyr stars in the Sloan bands}
   \authorrunning{W. Narloch et al.}

   \author{W. Narloch\inst{1}, 
          G. Hajdu\inst{1}, 
          G. Pietrzyński\inst{1}, 
          W. Gieren\inst{2}, 
          B. Zgirski\inst{2}, 
          P. Wielgórski\inst{1},
          P. Karczmarek\inst{2},
          M. Górski\inst{1}, \\
          \and
          D. Graczyk\inst{3}
          }

   \institute{Nicolaus Copernicus Astronomical Center, Polish Academy of Sciences, 
             Bartycka 18, 00-716 Warszawa, Poland\\
              \email{wnarloch@camk.edu.pl}
         \and
             Universidad de Concepci\'on, Departamento de Astronomia, 
             Casilla 160-C, Concepci\'on, Chile
         \and
             Nicolaus Copernicus Astronomical Center, Polish Academy of Sciences, 
             Rabia\'nska 8, 87-100 Toru\'n, Poland
             }

   \date{Received ..., 2024; accepted ..., 2024}

 
  \abstract
   {RR~Lyrae stars are excellent tracers of the old population~II 
   due to their period-luminosity 
   (PL) and period-luminosity-metallicity (PLZ) relations. While these relations have been investigated
   in detail in many photometric bands, there are few comprehensive studies about them in Sloan-like 
   systems.}
   {We present PL and PLZ relations 
   (as well as their counterparts in Wesenheit magnitudes) in the Sloan--Pan-STARSS 
   $g_{P1}r_{P1}i_{P1}$ bands obtained for Galactic RR~Lyrae stars in the vincinity of the Sun.}
   {The data used in this paper were collected with the network of $40$ cm telescopes of the 
   Las Cumbres Observatory, and geometric parallaxes were adopted from Gaia Data Release $3$.}
   {We derived PL and PLZ relations separately for RRab and RRc-type stars, as well as for the 
   mixed population of RRab+RRc stars.}
   {To our knowledge, these are the first PL and PLZ relations in the Sloan bands determined 
   using RR Lyrae stars in the Galactic field.}
   

   \maketitle
%

\section{Introduction}

   RR~Lyrae-type stars (hereafter RR~Lyr stars) are low-mass radially pulsating giants 
   of the horizontal branch on the Hertzsprung-Russell (HR) diagram. Initially, since 
   they were mostly discovered in globular clusters, they also used to be referred to 
   as cluster-type variables \citep[e.g.,][]{Bailey1902,BP1913}, but they were also 
   found in the Galactic field \citep[e.g.,][]{Kapteyn1890}, including the star RR~Lyr 
   \citep{Pickering1901}, which eventually became the prototype of this class of variables. 
   Originally, they were divided by \citet{Bailey1902} into three subclasses (a, b and c), 
   today simplified into just two: ab and c, where RRab stars are known to be fundamental 
   mode pulsators, while RRc stars are pulsating in the radial first-overtone. Much later 
   RRd stars were introduced into the nomenclature, \citep[e.g.,][]{JW1977,SKS1981,Nemec1985} 
   being double-mode pulsators, pulsating simultaneously in both fundamental and first-overtone 
   modes, as well as RRe stars \citep[e.g.,][]{DW1977} pulsating in the second-overtone only.

   Because of the similarities between RR~Lyr stars and Cepheids, \citet{Shapley1918} 
   incorporated them into his calibration of the Cepheid period-luminosity (PL) relation 
   and used it as a~tool for measuring distances to Galactic globular clusters which, in 
   turn, he used to estimate the distance to the center of the Milky Way (MW). This has 
   proven the usefulness of the RR~Lyr stars as distance indicators, which over time 
   became one of the most important standard candles, besides Cepheids. Despite the fact 
   that RR~Lyr stars are fainter than Cepheid variables, they serve as distance indicators 
   to old population~II stars, where young Cepheids are not observed, such as: globular 
   clusters \citep[e.g., $\omega$ Centauri from][]{Navarrete2017,Braga2018} and dwarf 
   galaxies \citep[e.g., Sculptor dSph, Carina, and Fornax galaxies from][]{Pietrzynski2008,Karczmarek2015,Karczmarek2017}. 
   The near-infrared (NIR) PL relations of RR~Lyr stars are of particular interest, as 
   they are very well defined and characterized with a~small scatter \citep[e.g.,][]{Zgirski2023,Bhardwaj2023,Bhardwaj2024}. 
   The amplitudes of the NIR light curves of the RR~Lyr stars are smaller than in the 
   optical domain, and the shapes are much more sinusoidal, so accurate mean magnitudes 
   can be estimated from only a~few photometric points. Moreover, the reddening in the 
   NIR is almost negligible. In the optical regime, the PL relations for RR~Lyr stars are 
   characterized by almost flat relations, dominated by the dependence on the metallicity 
   \citep[see, e.g.,][]{CaceresCatelan2008}, so their usefulness for distance determinations 
   is of less importance, compared to NIR. Nevertheless, the existing large-scale sky surveys, 
   such as the Sloan Digital Sky Survey \citep[SDSS;][]{Abazajian2003} or the upcoming 10~yr 
   Vera C. Rubin Observatory Legacy Survey of Space and Time \citep[Rubin-LSST;][]{Ivezic2019}, 
   use wide-band Sloan photometric filters \citep[$ugriz$;][]{Fukugita1996}, so there appeared 
   to be an urgent need to provide a~precise calibration of the PL relations in those filters 
   to effectively take advantage of these data sets.

   This need was addressed by studies from both the theoretical and observational sides. 
   On the one side, in their noteworthy works, \citet{Marconi2006} and \citet{CaceresCatelan2008}, 
   presented their theoretical period-magnitude-color and PL relations, respectively, for 
   RR~Lyr stars, calculated in the $ugriz$ SDSS photometric system. More recently, 
   \citet{Marconi2022} published theoretical period-luminosity-metallicity (PLZ) relations 
   for RR~Lyr stars in the Rubin-LSST filter system. 
   The empirical calibrations in the Sloan bands were done mostly using RR Lyr in globular 
   clusters. 
   \citet{Sesar2017} published PLZ relations in the $griz$ bands, using the first data release 
   from the Panoramic Survey Telescope And Rapid Response System 
   \citep[Pan-STARSS1,][]{Kaiser2010} based on the RR~Lyr stars from five globular clusters. 
   \citet{Vivas2017} published empirical PL relations for the $ugriz$ filter set of the Dark 
   Energy Camera \citep[DECam,][]{Flaugher2015} for RR~Lyr variables from the globular cluster 
   M5. 
   \citet{Bhardwaj2021} determined PL relations in the $gi$ bands for RR~Lyr stars from the 
   globular cluster M15, using data gathered with the 3.6~m Canada–France–Hawaii Telescope 
   (CFHT) and calibrated to the SDSS photometric system. 
   \citet{Ngeow2022RRL}, on the other hand, used data from the Zwicky Transient Facility 
   \citep[ZFT;][]{Bellm2018,Bellm2019,Dekany2020} in the Sloan $gri$ bands to derive the PLZ 
   and period-Wesenheit-metallicity (PWZ) relations based on hundreds of RR~Lyr stars from 
   $46$ globular clusters. 
   To the best of our knowledge, the empirical PL relations in the Sloan passbands for RR~Lyr 
   stars from the vincinity of the Sun have not been determined so far. Deriving them by 
   taking advantage of very high-quality geometric parallaxes provided by the Gaia mission 
   \citep{GaiaCol2023} would be an important contribution to this field.  
   
   Our aim in this work is to calibrate the PL and PLZ relations \citep[as well as their 
   counterparts in the Wesenheit magnitudes which are reddening-free by construction;][]{Madore1982} 
   for the MW RR~Lyr stars in the three Sloan $gri$ bands, and specifically as they are 
   implemented in the Pan-STARSS photometric system \citep{Tonry2018}, similarly as we did 
   in the case of the Galactic classical Cepheids in \citet{Narloch2023}. By this, we expect 
   to provide a~useful tool to determine distances in the universe, given the upcoming 
   era of the ambitious programs such as the Rubin-LSST. 
   
   The paper is organized as follows. In Section~\ref{sec:data} we provide a~description 
   of the data and reduction process, as well as methods of calculation of absolute magnitudes, 
   and source of metallicities of our stars. In Section~\ref{sec:relations} we describe the 
   derivation of the PL and PW, and PLZ and PWZ relations, which we further discuss in 
   Section~\ref{sec:discussion}. A~short Summary in Section~\ref{sec:summary} concludes the 
   paper.


\section{Data} \label{sec:data}


\subsection{Sample of stars} \label{ssec:sample}


   \begin{figure*}
   \centering
   \includegraphics{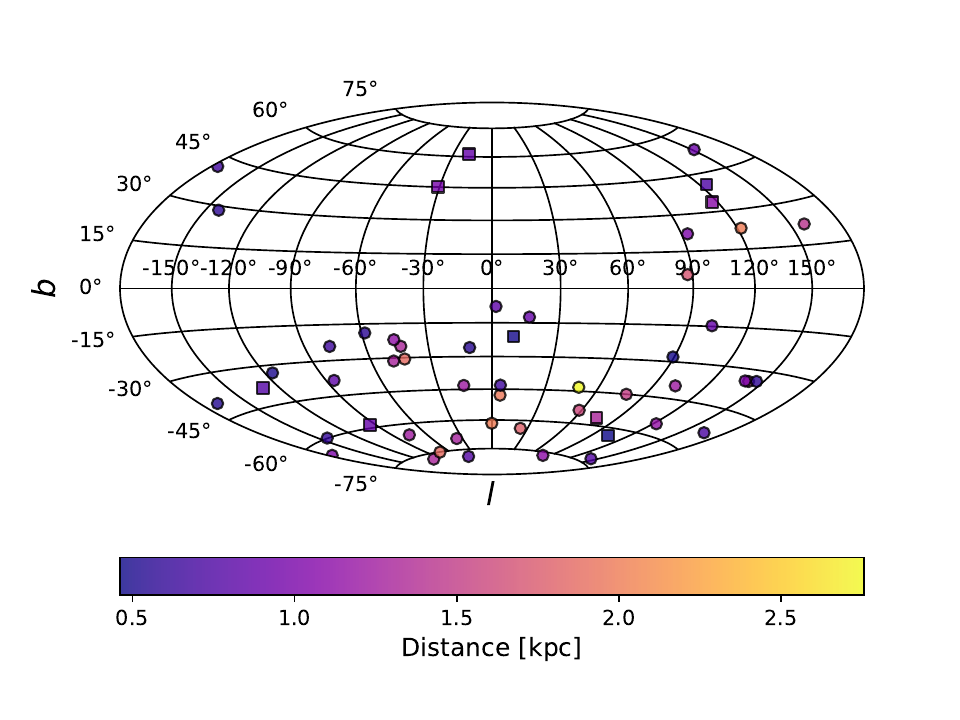}
   \caption{Location of the RR~Lyr stars (RRab, circles; RRc, squares) on the 
             sky used for establishing the PL relations in this paper given in 
             Galactic coordinates.}
              \label{fig:galmap}%
    \end{figure*}
   We have observed $53$~RR~Lyr type stars in total, among them $44$~fundamental 
   mode stars (hereafter RRab) and nine first overtone stars (hereafter RRc). The 
   stars are bright with Gaia G-magnitudes in the range from $8.97$ to $12.87$~mag 
   and they are distributed all over the sky, however, with the majority located 
   in the southern hemisphere. 
   The range of distances calculated as the inverse of the Gaia Data Release $3$ 
   (DR3) parallaxes \citep{GaiaCol2023} is between $\approx 0.5$ and $\approx 2.9$~kpc, 
   with a~median distance of about $1$~kpc. The sky distribution of the analyzed 
   RR~Lyr stars is shown in Figure~\ref{fig:galmap}, where circles denote RRab and 
   squares RRc stars. The pulsational periods of our sample stars were adopted from 
   the International Variable Star Index\footnote{\url{https://www.aavso.org/vsx/}} 
   (AAVSO) and range between about $0.36$ and $0.88$~days. 
   The physical parameters of stars are summarized in Table~\ref{tab:rrlyr}. 
   

\subsection{Data and reduction} \label{ssec:data_reduction}

   
   Data for the project were collected between 2021 August and 2022 July with $16$ 
   robotic 40~cm telescopes of the Las Cumbres Observatory (LCO) Global Telescope 
   Network\footnote{\url{https://lco.global/}} within observing programs CLN2021B-008 
   and CLN2022A-008. 
   Images were obtained in the Sloan $g'r'i'$ filters using $3\rm{K} \times 2\rm{K}$ 
   SBIG STL-6303 cameras with field of view $29.2 \times 19.5$~arcmin$^2$, and a~pixel 
   size of $0.571$~arcsec pixel$^{-1}$ without binning. The air mass range was 
   $1.0 - 1.6$, and average seeing was about $2.35$, $2.28$ and $2.30$~arcsec in the 
   Sloan $g$, $r$ and $i$ bands, respectively. 
   
   We downloaded images already reduced and processed with the LCO 
   BANZAI\footnote{\url{https://lco.global/documentation/data/BANZAIpipeline/}} pipeline 
   from the LCO Archive\footnote{\url{https://archive.lco.global/}}. The aperture 
   photometry and calibration of data was done the same way as in \citet{Narloch2023}, 
   which we recommend to see for details of the procedure. 
   The mean DAOPHOT photometric uncertainties were $\approx 0.03$~mag for the Sloan 
   $g$ filter and $\approx 0.02$~mag for Sloan $ri$ filters for all stars with Sloan 
   $g<14.0$~mag, which is the range of magnitudes from which most of the reference 
   stars are from. The average photomertic error for the range of magnitudes of the 
   RR~Lyr stars from our sample ($9.0<g<13.0$~mag) was about $0.01$~mag in all three 
   filters. 
   
   The influence of the nonlinearity of the LCO SBIG cameras was not as severe in 
   case of the RR~Lyr stars as it was in case of classical Cepheids in 
   \citet{Narloch2023}, as we were using photometric calibration reference stars with 
   brightnesses similar to the target stars. Nevertheless, we applied exactly the same 
   calibration equations as in \citet{Narloch2023} to account for that effect. 
   
   The final intensity-averaged mean apparent magnitudes obtained by fitting the 
   Fourier series to the RR~Lyr stars light curves are expressed in the photometric 
   system of the ATLAS All-Sky Stellar Reference Catalog version $2$ 
   \citep[ATLAS-REFCAT2;][]{Tonry2018}, which is in the Pan-STARSS system 
   \citep{Tonry2012} referred to as $g_{P1}r_{P1}i_{P1}$ later in the text. The applied 
   Fourier order ranged between $2$ and $13$, where the higher Fourier orders were 
   used to better fit the maximum brightness of specific cases. 
   Figures~\ref{fig:fig1} and \ref{fig:fig2} present the final light curves of our 
   RRab and RRc stars, respectively\footnote{The light curves are available at the 
   webpage of Araucaria Project: \url{https://araucaria.camk.edu.pl/} and the CDS.}. 


\subsection{Reddening} \label{ssec:redd}


   To deredden our data we used the E(B-V) color excess values from \citet{SF2011}, 
   available for all RR~Lyr stars from our sample, and integrated 
   up to the distance of our targets assuming the three-dimensional MW model of 
   \citet{DrimmelSpergel2001} 
   \citep[for details of the adopted model parameters see, e.g., ][]{Suchomska2015}.
   The extinction vectors 
   ($R_{\lambda}$) for the Sloan--Pan-STARSS $g_{P1}r_{P1}i_{P1}$ bands were 
   adopted from \citet[][see their Table~1]{Green2019}, and they are equal to 
   $R_{g} = 3.518$, $R_{r} = 2.617$ and $R_{i} = 1.971$. We used these values to 
   calculate three Wesenheit indices, reddening-free by the construction 
   \citep[for a~specific reddening law,][]{Madore1982}, and we defined 
   $W^{ri}_r = r - 4.051 (r-i) - \mu$, $W^{gr}_r = r - 2.905 (g-r) - \mu$, and 
   $W^{gi}_g = g - 2.274 (g-i) - \mu$, where $\mu$ is a~distance modulus (DM). 


\subsection{Distances} \label{ssec:dist}


   We applied four different techniques to determine the absolute 
   magnitudes ($M_{\lambda}$) calculated from the mean apparent magnitudes 
   ($m_{\lambda}$) of RR~Lyr stars from our sample:
   
   \begin{itemize} 
       \item Classical method, where we adopted the geometric parallaxes 
       from the Gaia~DR3 catalog and corrected for the zero point (ZP) offset 
       proposed by \citet{LindegrenBastian2021}\footnote{We calculated the 
       parallax ZP offset corrections using the dedicated Python code provided 
       for that purpose: \url{https://gitlab.com/icc-ub/public/gaiadr3_zeropoint}.} 
       and inserted them directly into the following formula: 
       
       \begin{equation} \label{eq:absmag}
        M_{\lambda} = m_{\lambda} + 5\log_{10} \varpi + 5,
        \end{equation}
        
       where $\varpi$ is a~parallax expressed in arcsec. 
       
       \item Astrometry-Based Luminosity \citep[ABL;][]{FC1997,AL1999} to avoid 
       the bias in the absolute magnitude caused by its nonlinear relation with 
       the parallax: 
       
       \begin{equation} \label{eq:abl}
        ABL_{\lambda} = 10^{0.2M_{\lambda}} = \varpi_{(arcsec)} 10^{\frac{m_{\lambda + 5}}{5}}.
       \end{equation} 
       
       \item Geometric distances from \citet{BJ2021} derived from Gaia parallaxes 
       using the parallax with a~direction-dependent prior on distance: 
       
       \begin{equation} \label{eq:absmagBJ}
        M_{\lambda} = m_{\lambda} - 5\log_{10}r + 5,
        \end{equation}
        
       where $r$ is a~distance in pc. 
       
       \item Photogeometric distances from \citet{BJ2021}, which were calculated 
       using additional priors on the color and apparent magnitude of a~star, 
       where the absolute magnitude can be calculated based on Equation~(\ref{eq:absmagBJ}).

\end{itemize}

   The Gaia parallaxes for our RR~Lyr stars corrected for the ZP offset derived 
   by \citet{LindegrenBastian2021} range between $0.36$ and $2.17$~mas (see column~$3$ 
   in Table~\ref{tab:rrlyr}), with a~median of $0.99$~mas, while the corrections range 
   from about $-5$ to $-57\,\mu$as (with a~mean of about $-30\,\mu$as). 
   \citet{LindegrenBastian2021} recommended to include an uncertainty of a~few $\mu$as 
   in the ZP, so we adopted $5\,\mu$as as a~systematic error.
   
   We rejected eight stars in total which did not meet our criteria for the parallax 
   quality, that is, having $\rm{RUWE}>1.4$ or $\rm{GOF}>12.5$, which we adopted after 
   \citet{Breuval2021} and \citet{Wielgorski2022}. 
   Because of the sensitivity of the RUWE parameter to photocentric motion of unresolved 
   objects, it might be an indicator of astrometric binaries. The GOF parameter is used 
   as an indicator of the level of asymmetry of a~source by, for example, \citet{Riess2021}. 
   Six RRab stars from our sample have $\rm{RUWE}>1.4$ (BB~Eri, BT~Aqr, DH~Hya, RV~Phe, 
   SZ~Gem, VW~Scl; marked with a-symbol in Table~\ref{tab:rrlyr}), among which only four 
   characterize with $\rm{GOF}>12.5$. From RRc stars two were rejected based on the RUWE 
   (MT~Tel, RU~Psc) and none based on the GOF parameter, meaning that the RUWE indicator 
   is more selective in our case. 
   
   Eight of our RRab and two of RRc stars fall into the problematic magnitude range 
   of $G=11.0 \pm 0.2$~mag, where a~transition of Gaia window classes occurs 
   \citep[see Figure~1 in][]{LindegrenBastian2021}, where the parallax ZP could be 
   affected. Another five RRab stars fall into the range of $G=12.0 \pm 0.2$~mag. 
   To account for possible additonal errors caused by this, we quadratically added 
   $10\,\mu$as to the parallax uncertainties of those stars, following 
   \citet{Breuval2021}. 
   Finally, all Gaia~DR3 parallax errors were increased by $10\%$, as suggested by 
   \citet{Riess2021} to account for a possible excess uncertainty. 
   The mean parallax uncertainty for our RR~Lyr stars given in column~$3$ of 
   Table~\ref{tab:rrlyr} is about $20\,\mu$as.


\subsection{Metallicities} \label{ssec:metal}


   For the purpose of derivation of the PLZ and PWZ relations we adopted the metallicity 
   values calculated from high-resolution spectra by \citet{Crestani2021a,Crestani2021b}. \citet{Crestani2021a} provide the metallicity for $32$ RRab and four RRc stars 
   from our sample, while \citet{Crestani2021b} complete this list with another 
   five RRab and one RRc stars (see column~$10$ of Table~\ref{tab:rrlyr}). 
   We lack the metallicity values from high-resolution spectroscopy for the remaining 
   seven RRab and four RRc stars (see column~$11$ of Table~\ref{tab:rrlyr}). 
   Nevertheless, we have decided not to adopt them from other sources, to assure 
   a~similar quality of the used values, but rather exclude those stars from further 
   analysis of the PLZ and PWZ relations. 
   Metallicities for remaining RRab stars have a~wide range from $-0.03$ to $-2.59$~dex, 
   with the mean of about $-1.40$~dex, while the metallicity range for RRc stars is 
   from $-1.49$ to $-2.60$~dex, with the mean of $-1.93$~dex.


\section{Derived relations} \label{sec:relations}


\subsection{Period-luminosity relations} \label{ssec:plr} 


   We derived the PL relations for the MW RR~Lyr stars using absolute magnitudes 
   in the three Sloan--Pan-STARSS $g_{P1}r_{P1}i_{P1}$ bands with all four methods 
   listed in Section~\ref{ssec:dist}. 
   For the absolute magnitudes calculated from Equation~(\ref{eq:absmag}) using 
   the Gaia~DR3 parallaxes and Equation~(\ref{eq:absmagBJ}) using distances from 
   \citet{BJ2021} we fit the linear relation:
   
   \begin{equation} \label{eq:plr} 
     M_{\lambda} = a_{\lambda} (\log P - \log P_0) + b_{\lambda},
   \end{equation}
   
   \noindent where $a_{\lambda}$ and $b_{\lambda}$ are the searched slope and intercept. 
   The logarithm of a pivot period ($\log P_0$) is subtracted in order to minimize 
   correlation between the two parameters. Following \citet{Zgirski2023}, we set it 
   to be $\log P_0 = -0.25$ in case of the RRab and mixed population of RRab+RRc 
   stars, and $\log P_0 = -0.45$ for RRc stars, where RRc stars were fundamentalized 
   according to the recipe from \citet{Iben1974}, $\log P_{RR_{ab}} = \log P_{RR_c} + 0.127$. 
   
   For the absolute magnitudes calculated with the ABL method given by 
   Equation~(\ref{eq:abl}) we fit the following relation: 
   
   \begin{equation} \label{eq:abl_plr} 
     ABL_{\lambda} = 10^{0.2[a_{\lambda}(\log P - \log P_0) + b_{\lambda}]}. 
   \end{equation} 
   
   We performed the fitting of above relations using the {\tt curve\_fit} function 
   from the {\tt scipy} Python library with $3\sigma$ clipping. The fitted coefficients 
   for all four methods with the corresponding statistical uncertainties taken as errors 
   returned by the fitting procedure are given in Table~\ref{tab:plr}. They all agree 
   well within their $1\sigma$ uncertainty. Figure~\ref{fig:plr} presents the resulting 
   PL relations with the absolute magnitudes calculated using Gaia~DR3 parallaxes, 
   where the error bars are the uncertainties derived from the error propagation based 
   on the statistical errors on the mean magnitudes and parallaxes. 
   
   The PL relations derived using four different methods result in almost identical rms, 
   and the obtained coefficients agree well within their $1\sigma$ uncertainty. Therefore, 
   on this basis we cannot determine which method is the most reliable one for distance 
   calculation. For the presentation of the results, we decided on the conceptually 
   simplest and most direct method of inverted parallaxes, which we also recommend to 
   use for the distance calculation purposes. Nevertheless, we have included the 
   coefficients for each of the method so that the reader could use them depending on 
   their preferences.
   
   In case of the PL relations for RRab stars, $38$ stars were used for the fit. Their 
   distances ranged from about $0.48$ to $2.76$~kpc, with a~median of about $0.97$~kpc. 
   The final rms of the fitted relations is $0.18$ for all four methods in the 
   Sloan--Pan-STARSS $g_{P1}$~band, and $0.14-0.15$ in the $r_{P1}$ and $0.13$ in the 
   $i_{P1}$~band. 
   In case of the PL relations for RRc variables, $7$ stars were used for the fitting 
   only, from the range of distances between $0.46$ to $1.31$~kpc, with a~median 
   $\approx 0.88$~kpc. The final rms is $0.11$ for the $g_{P1}$, $0.10$ for the $r_{P1}$ 
   and $0.09-0.10$ for the $i_{P1}$~band. 
   In case of the mixed population, $45$ stars were used from the full distance range 
   from $0.46$ to $2.76$~kpc, with a~median of $0.92$~kpc. The obtained rms is larger 
   than for the RRab and RRc populations separately, and are $0.21$, $0.17$ and $0.15$ 
   for the Sloan--Pan-STARSS $g_{P1}r_{P1}i_{P1}$~bands, respectively.  
   

\subsection{Period-Wesenheit relations} \label{ssec:pwr} 


   We also performed the fitting using the Wesenheit indices ($W^{ri}_{r}$, $W^{gr}_{r}$, 
   $W^{gi}_{g}$) as defined in Section~\ref{ssec:redd} and applying equations analogous 
   to Equations~(\ref{eq:plr}) and (\ref{eq:abl_plr}): 
   
   \begin{equation} 
     W = a (\log P - \log P_0) + b \label{eq:pwr}, 
   \end{equation} 
   
   \begin{equation} 
     ABL_{W} = 10^{0.2[a(\log P - \log P_0) + b]} \label{eq:abl_pwr},
   \end{equation} 
   
   \noindent where $a$ and $b$ are the slope and intercept we are looking for and 
   $\log P_0$ is the logarithm of the pivot period as adopted in Section~\ref{ssec:plr}. 
   
   For the fit of the PW relations we used the same number of stars as in the case 
   of the PL relations ($38$ RRab, seven RRc and $45$ RRab+RRc stars). The only exceptions 
   were $W^{gr}_{r}$ and $W^{gi}_{g}$ calculated based on photo-geometric distances 
   from \citet{BJ2021} where $44$ stars were used for the mixed population, as CP~Aqr 
   was rejected from the fit after $3\sigma$-clipping. 
   The rms of fitted relations are between $0.10 - 0.12$ for RRab, $0.07 - 0.10$ for 
   RRc and $0.09 - 0.12$ for RRab+RRc stars. 
   The final coefficients are given in Table~\ref{tab:plr} and the resulting relations 
   for distances calculated based on Gaia parallaxes are presented in Figure~\ref{fig:pwr}.
   

\subsection{Period-luminosity-metallicity relations} \label{ssec:plz}

   
   We performed the fitting of the PLZ relations using the metallicities from 
   \citet{Crestani2021a,Crestani2021b} for $42$ of our stars ($37$ RRab and four 
   RRc stars, see Section~\ref{ssec:metal}). After application of selection criteria 
   for RUWE and GOF we were left with $31$ RRab and only four RRc. Because of the 
   small number of the latter we decided not to perform the PLZ relation fitting for 
   RRc stars, however, we used them to fit the relations for the mixed population 
   after prior recalculation of their $\log P$ to the fundamental ones (see 
   Section~\ref{ssec:plr}). 
   We used equations similar to Equations~(\ref{eq:plr})--(\ref{eq:abl_pwr}) but 
   extended by the metallicity term in the form: 
   
   \begin{equation} \label{eq:plz} 
     M_{\lambda} = a_{\lambda} (\log P - \log P_0) + b_{\lambda} + c_{\lambda}(\rm{[Fe/H]} - \rm{[Fe/H]}_0),
   \end{equation}
   
   \begin{equation} \label{eq:abl_plz} 
     ABL_{\lambda} = 10^{0.2[a_{\lambda}(\log P - \log P_0) + b_{\lambda} + c_{\lambda}(\rm{[Fe/H]} - \rm{[Fe/H]}_0)]},
   \end{equation}
   
   \noindent where the new coefficient $c_{\lambda}$ is the~metallicity slope and 
   $\rm{[Fe/H]}_0$ is the~pivot metallicity, chosen to be $-1.5$~dex, a~value close 
   to our median metallicity. 
   We performed the fitting for all four methods of distance calculation. The resulting 
   coefficients for the Sloan~$gri$~bands and Wesenheit magnitudes with their corresponding 
   errors calculated as in Section~\ref{ssec:plr} are given in Table~\ref{tab:plzr2}. 
   Figure~(\ref{fig:plzr2}a) presents the residuals of the fit for the parallax method 
   for RRab stars and Fig.~(\ref{fig:plzr2}b) the same for mixed population. 
   
   The $c_{\lambda}$ coefficient values obtained for all four methods of the determination 
   of the absolute magnitudes are within the $1\sigma$ uncertainties of each other. The 
   metallicity slopes in the Sloan-Pan-STARRS $g_{P1}r_{P1}i_{P1}$ bands are quite 
   significant (about $0.2$ to $0.3$~mag/dex) and they decrease toward longer wavelengths. 
   Our values show up to $\approx 0.1$~mag/dex larger metallicity dependence than the 
   metallicity slopes obtained by \citet[][see their Table~$4$]{Ngeow2022RRL}, but the 
   trend is similar. 
   The resulting $c_{\lambda}$ values for Wesenheit indices, on the other hand, are 
   smaller than $0.1$~mag/dex, being largest for $W_{r}^{ri}$ and smallest for $W_{r}^{gr}$. 
   In comparison, the coefficients from \citet{Ngeow2022RRL} are up to $\approx 0.1$~mag/dex 
   larger in this case, although still show a~similar trend.  
   

\section{Discussion} \label{sec:discussion} 

   

\subsection{Fundamentalization} \label{ssec:fund} 


   As indicated in Section~\ref{ssec:plr}, we fundamentalized the sample of RRc 
   stars using the recipe from \citet{Iben1974} which is simply shifting the 
   $\log P$ of RRc stars by a~constant value ($0.127$ in this case). This method 
   is supposed to align RRc with RRab stars and is commonly used in the literature 
   \citep[e.g.,][] {Ngeow2022RRL,Zgirski2023}. The main reasons behind derivation of 
   PL and PLZ relations (and their Wesenheit versions) for mixed population of RRab+RRc 
   stars is to increase the number of stars used for the fitting, but also in order 
   to compare the results with theoretical predictions \citep[such as][]{CaceresCatelan2008}. 
   The latter does not differentiate between RRab and RRc stars as it is difficult 
   to separate them on the HR diagram where their instability strips overlap with 
   each other \citep[see for example Fig.~4 from][]{Kollth2002}.  
   
   However, a~simple shift in $\log P$ might not be enough to correctly align those 
   two types of variables. As already noticed by, for instance, \citet{Zgirski2023} in NIR 
   data, and what is also visible in our results based on the Sloan-Pan-STARRS 
   $g_{P1}r_{P1}i_{P1}$ bands (see Tables~\ref{tab:plr} and \ref{tab:plzr2}), the slopes 
   of the relations for both types of RR~Lyr stars are different, with slopes of RRc 
   stars being flatter than RRab. 
   After fundamentalization of RRc stars, the PL and PLZ relations for mixed population 
   resulted in even flatter slopes than for RRab and RRc stars, separately. This is 
   mostly a~consequence of existing misalignment. 
   \citet{Zgirski2023} determined new $\log P$ shifts for their sample of stars in 
   $JHK_{s}$~bands. We decided to follow their method of simultaneous $\Delta \log P$ 
   fitting, despite small number of RRc stars (seven and four for PL and PW, and PLZ and 
   PWZ relations, respectively). We obtained the shifts presented in Table~\ref{tab:dlogP}. 
   The shift values are higher for bluer passbands, which is most probably an effect 
   of flatter slopes in shorter wavelengths, as the shift in $\log P$-axis is constant. 
   The applied method minimizes the systematic error coming from the classical 
   fundamentalization formula, but does not solve the problem. 
   
   The constant value of $\Delta \log P$ from $\log P_{RRab} = \log P_{RRc} + \Delta \log P$ 
   fundamentalization formula was originally derived from pulsational models, from the 
   ratio of the period of the fundamental to the first-overtone modes at the intersection 
   of blue edges of the instability strips for those two modes \citep{IbenHuchra1971}. 
   After discovery of the classical RRd type stars \citep{JW1977}, the ratio of the 
   first-overtone to the fundamental period in those stars became the source of the shift 
   value \citep[e.g.,][]{Catelan2009}, which agreed remarkably well with the original one. 
   However, as can be seen clearly in \citet{Nemec2024} this ratio is not constant, 
   but change nonlinearly with period (see their Fig.~$1$ and Equation~$4$). 
   So it seems that more correct approach to the fundametalization would be to apply this 
   nonlinear relation \citep[see Figure~$4$ in][]{Nemec2024} and shift RRc stars 
   according to it. The obtained $\Delta \log P$ values for stars from our sample ranged 
   between $0.128 - 0.153$ (with a~mean of $\approx 0.133$). After recalculation of the 
   PL and PLZ relations, the slopes indeed became slightly steeper, however, the change was 
   well within $1 \sigma$ uncertainties. The intercept, on the other hand, was not affected. 
   So it turns out that a~simple shift by $\Delta \log P = 0.127$ is a~sufficient approximation 
   in our case, after all. 
   
   Nevertheless, the procedure of fundamentalization of the RRc stars introduces 
   a~systematic error into the derived PL and PLZ relations. The luminosities of RRab stars 
   are generally overestimated, which influences the distances determined using such 
   equations. As we showed in Table~\ref{tab:dlogP}, this bias will dependent on the used 
   filter. Although the shift in $\log P$-axis is constant, the slopes of PL and PLZ relations 
   are different in different bands. We provided the coefficients for PL and PLZ relations 
   for the mixed population in Tables~\ref{tab:plr} and \ref{tab:plzr2}, but for the distance 
   determination purposes we recommend to use the relations for RRab and RRc populations 
   separately. 


\subsection{The influence of the parallax ZP on the PL and PLZ relations} \label{ssec:ZPinfluence} 


   For the purpose of derivation of our PL and PLZ relations (and their Wesenheit 
   versions) we first corrected Gaia parallaxes for the ZP offset based on 
   \citet[][see Section~\ref{ssec:dist}]{LindegrenBastian2021}, particularly for 
   parallax and ABL methods, as the geometric and photogeometric distances from 
   \citet{BJ2021} have these corrections already incorporated. 
   To check the influence of that value on our results we performed two control tests. 
   First, we have not introduced the corrections at all, and secondly we calculated 
   the corrections according to the prescription given by \citet{Groenewegen2021}, 
   where the new mean value was again about $-30\mu as$ as for the 
   \citet[][see Section~\ref{ssec:dist}]{LindegrenBastian2021}. 
   In both cases the slopes for derived PL and PLZ (and Wesenheit) relations, as well 
   as metallicity slope for the PLZ relation agreed well within $1\sigma$ uncertainties. 
   In case of the ZPs of the relations, they again agreed very well with parallax 
   corrections determined based on \citet{Groenewegen2021}, but the difference was 
   significant when no corrections were applied. The shift of the intercept values 
   of the PL and PW relations was about $0.07$~mag for RRab and $0.05$~mag for RRc stars, 
   with intercepts being smaller when no parallax corrections are applied. The difference 
   in those two numbers might come from the fact that RRc stars from our sample are 
   located on average closer to us than RRab stars, so the uncertainty coming from 
   parallax ZP has a~smaller influence on them. 
   The ZP shift of the PLZ relations was even slightly larger (about $0.08$~mag) 
   when no ZP offset corrections were applied.  


\subsection{Comparison of the PL and PLZ relations with the literature} \label{ssec:comp_with_lit} 


   The general trend of the slopes of our PL and PLZ relations follows the trends 
   also presented in the results of other works such as \citet{Vivas2017,Ngeow2022RRL} 
   in the Sloan bands or \citet{Zgirski2023} for NIR data. 
   The slopes of the relations become steeper toward longer wavelengths while 
   the metallicity slopes flatten at the same time. Also, the PL and PLZ relations 
   are better constrained for longer wavelengths, as their dispersion decreases 
   in redder filters. 
   The slopes of PLZ relations for RRab stars are steeper than for RRab+RRc stars, 
   which again was observed in the Sloan passbands by, for example, \citet{Ngeow2022RRL}. 
   The direct comparison of our results with different authors is difficult to perform 
   for several reasons. Nevertheless, in the following Sections we present the 
   comparison with the existing PLZ relations, both theoretical and empirical.


\subsubsection{Comparison with theoretical PLZ relation} \label{sssec:theoretical} 


   \citet{CaceresCatelan2008} presented a~theoretical study of the RR~Lyr PLZ 
   relation in the $ugriz$ SDSS filter system. They provided the PLZ relation in 
   the SDSS $i$ band as: 
   
   \begin{equation} \label{eq:CaceresCatelan2008_1} 
     M_i = 0.908 - 1.035\log P + 0.220\log Z 
   \end{equation} 
   
   \noindent where $\log Z$ can be calculated as: 
   
   \begin{equation} \label{eq:CaceresCatelan2008_2} 
     \log Z = \rm{[Fe/H]} + \log (0.638 \times 10^{\rm{[\alpha/Fe]}} + 0.362) - 1.765 
   \end{equation} 
   
   As the theoretical models do not differentiate between RRab and RRc stars, in 
   Fig.~\ref{fig:comp_CaceresCatelan2008} we show the $i$ band PLZ relation derived 
   for the mixed population in this work with the parallax method and recalculated 
   onto the SDSS photometric system according to the linear equations given in 
   Table~6 of \citet{Tonry2012} for two constant metallicities $\rm{[Fe/H]}=-0.5$ 
   and $-2.0$~dex. With red, green and yellow lines we overplotted the relations 
   given by Equation~(\ref{eq:CaceresCatelan2008_1}) for the same metallicity 
   values and three different values of $\rm[\alpha / Fe]=-0.1, +0.2$ and $+0.5$~dex, 
   respectively. 
   The disagreement between our result and relation from \citet{CaceresCatelan2008} is 
   significant. Firstly, the slope of $-1.035$ from \citet{CaceresCatelan2008} is flatter 
   than our $-1.469 \pm 0.328$.  The relations are also shifted in the ZP relative to ours, 
   and that shift is larger for lower metallicities. Finally, 
   Fig.~\ref{fig:comp_CaceresCatelan2008} shows that the change of the ZP strongly 
   depends on the $\rm{[\alpha / Fe]}$ value. For metallicity $\rm{[Fe/H]}=-0.5$~dex at 
   $\log P = -0.25$~days (equal to the pivot $\log P_0$), the ZP shift between our and 
   the theoretical relation is about $0.18, 0.13$ and $0.08$~mag for 
   $\rm{[\alpha / Fe]} = -0.1, +0.2$ and $+0.5$~dex, respectively. The same values for 
   $\rm{[Fe/H]}=-2.0$~dex are $0.21, 0.16$ and $0.11$. 
   The ZP shift for all four methods are given in Table~\ref{tab:ZPshift_teoretical}.
   The significant change of the ZP depending on the $\rm{[\alpha / Fe]}$ value might 
   partially explain the differences between our relations for RR~Lyr stars from the 
   vincinity of the Sun and RR~Lyr stars from the globular clusters, presented in the next 
   Section~\ref{sssec:empirical}, as the $\rm{[\alpha / Fe]}$ ratio is different for RR~Lyr 
   stars in different environments \citep[see for example Figure~4 in][]{Pritzl2005}. Also, 
   our sample of stars is far more heterogeneous in terms of $\rm{[\alpha / Fe]}$ ratio than 
   a~typical globular cluster. 

   \citet{Marconi2022} published theoretical PLZ relations for RR~Lyr stars in the Rubin-LSST 
   filter system, using theoretical bolometric corrections based on the expected performance 
   throughput curves for the Rubin-LSST photometric system. In absence of transformation 
   equations between the (observed) Rubin-LSST bands and the PS1 system, comparing their 
   relations to our result would be a nontrivial task, so we decided not to perform it.


\subsubsection{Comparison with empirical PLZ relations} \label{sssec:empirical} 


   \citet{Sesar2017} derived empirical PLZ relations based on $55$ RRab stars 
   located in five globular clusters with magnitudes in the $griz$ Pan-STARSS1 
   bands. They used a~probabilistic approach to constrain their relations. 
   In Figure~\ref{fig:comp_Sesar2017} we show the comparison of our original PLZ 
   relations for RRab stars derived with parallax method, with overplotted relations 
   from \citet{Sesar2017} given with their Equation~4, where the coefficient values 
   were taken from their Table~1. The comparison is made for two constant metallicities 
   ($\rm{[Fe/H]}=-0.5$ and $-2.0$~dex). 
   The first evident disagreement between our result and \citet{Sesar2017} are the 
   slopes of the relations. Both the dependences of the period and metallicity in 
   \citet{Sesar2017} are roughly constant in all three Sloan $g_{P1}r_{P1}i_{P1}$ bands, 
   while in our case they change depending on the filter. The slopes of the period 
   dependence in \citet{Sesar2017} are steep (of about $-1.7$), while our $g_{P1}$~band 
   PLZ relation is almost flat ($-0.527$), then becoming steeper at about $-1.47$ in the 
   $i_{P1}$~band. 
   Also, the metallicity slopes in \citet{Sesar2017} are close to zero in all three 
   bands while our slopes are between about $0.2-0.3$~mag/dex and flatten toward 
   longer wavelengths. 
   \citet{Sesar2017} used the stars from a~much narrower metallicity range than we 
   (from $-1.02$ to $-2.37$~dex versus from $-0.03$ to $-2.59$~dex) which might partially 
   explain their very small dependence on the metallicity. 
   The shift in the ZP between compared relations is quite large for higher metallicities. 
   For $\rm{[Fe/H]}=-0.5$~dex and $\log P = -0.25$~days it is about $0.2$ in all three 
   filters while for $\rm{[Fe/H]}=-2.0$~dex it is about $-0.04, 0.04$ and $0.08$ for 
   $g_{P1}r_{P1}i_{P1}$~bands, respectively. 
   The values of the ZP shifts calculated for all four methods are presented in 
   Table~\ref{tab:ZPshift_empirical}.
   
   \citet{Vivas2017} provided empirical PLZ relations for the RR~Lyr stars from the 
   globular cluster M5. Their data for $47$ RRab and $14$ RRc stars were gathered with 
   DECam and $ugriz$ filters. In Figure~\ref{fig:comp_Vivas2017} we present the comparison 
   of our original $g_{P1}r_{P1}i_{P1}$~band PLZ relations for RRab stars derived with 
   the parallax method for a~constant metallicity $\rm{[Fe/H]}=-1.25$~dex, as adopted by 
   \citet{Vivas2017} for M5, with overplotted relations recalculated onto the Pan-STARSS 
   photometric system using the calibration equations for a~Dark Energy Survey (DES) 
   photometric system from Appendix~B of \citet{Abbott2021}. Red lines show the derredened 
   and scaled relations given by Equation~(3) in \citet{Vivas2017} for the DM of $14.44$~mag 
   ($\pm 0.02$~mag) adopted by them, while the magenta line show the relations from Table~8 
   shifted by $\rm{DM} = 14.37$~mag ($\pm 0.02$~mag) taken from Table~1 of 
   \citet{BaumgardtVasiliev2021}. 
   The slopes of our relations and \citet{Vivas2017} are in a~good agreement with each 
   other, and agree within $1 \sigma$ uncertainties. The ZP, however, is quite different. 
   For a~$\log P = -0.25$~days and $\rm{DM} = 14.44$~mag the shift in the ZP is about 
   $0.33, 0.25$ and $0.23$~mag for $g_{P1}r_{P1}i_{P1}$ filters, respectively. 
   Using the alternative value of DM would change those shifts to about $0.26, 0.18$ and 
   $0.17$~mag. This comparison shows that the choice of the DM value used for putting the 
   ZP of a~globular cluster to an absolute scale is critical. In this work we used Gaia DR3 
   parallaxes for the calculation of the distance to our stars, which is the most direct 
   and precise method known today.
   
   \citet{Bhardwaj2021} presented a~study of the PL relations of RR~Lyr stars in the 
   globular cluster M15 in SDSS $gi$ bands. To compare their results with ours, we 
   recalculated our PLZ relations for RRab stars and mixed population obtained with 
   parallax method onto the SDSS system \citep{Tonry2012}. The slope values for both 
   relations agree well within $1 \sigma$ uncertainties. However, the difference of the 
   ZP is, again, significant. The shift of the ZP calculated for a~constant metallicity 
   $\rm{[Fe/H]}=-2.33$~dex and $\rm{DM} = 15.15$~mag ($\pm 0.02$~mag) adopted by 
   \citet{Bhardwaj2021} for M15 at $\log P = -0.25$~days is about $0.17$ and $0.21$~mag 
   for RRab stars, and $0.18$ and $0.20$~mag for RRab+RRc stars for SDSS $gi$~bands, 
   respectively.

   \citet{Ngeow2022RRL} derived their PL relations for the $gri$ Pan-STARSS magnitudes 
   as well as Wesenheit indices defined in an identical way as the ones used in this 
   work (see Section~\ref{ssec:redd}) for RRab, RRc and RRab+RRc stars located in $46$ 
   globular clusters. In Figure~\ref{fig:comp_Ngeow2022RRL} we show the comparison of 
   our original PLZ relations for $g_{P1}r_{P1}i_{P1}$ magnitudes and Wesenheit indices 
   of RRab stars derived with the parallax method. Overplotted with red lines are relations 
   from \citet{Ngeow2022RRL} given with Equation~(1) and coefficients from their Table~4 
   for two constant metallicities $\rm{[Fe/H]}=-0.5$ and $-2.0$~dex. Once again, the slopes 
   of the period dependence agree well within $1 \sigma$ uncertainties between both results, 
   while our slopes are systematically steeper than those of \citet{Ngeow2022RRL}. The 
   agreement of the metallicity slope is slightly worse, as the slopes agree within 
   $2 \sigma$. 
   One of the possible reasons of this could be that \citet{Ngeow2022RRL} use a~narrower 
   range of metallicities (from $-0.43$ to $-2.36$~dex versus our from $-0.03$ to $-2.59$~dex).
   The difference of the ZP in case of the $g_{P1}r_{P1}i_{P1}$ magnitudes is significant. 
   The shift of the ZP for RRab stars at the $\log P = -0.25$~days for $\rm{[Fe/H]}=-0.5$~dex 
   is about $0.41$~mag for the $g_{P1}$~band and $0.32,0.28$~mag in the $r_{P1}i_{P1}$~bands, 
   respectively. The same shifts for $\rm{[Fe/H]}=-2.0$~dex are smaller ($0.26,0.22$ and 
   $0.24$~mag). A~better agreement we see, however, for the Wesenheit magnitudes, where 
   the shift for $\rm{[Fe/H]}=-0.5$~dex at the $\log P = -0.25$~days for RRab stars is 
   about $0.02$ and $0.04$~mag for $W^{ri}_{r}$ and $W^{gi}_{g}$, and slightly larger for 
   $W^{gr}_{g}$ ($\approx 0.11$~mag). 
   Those values grow for $\rm{[Fe/H]}=-2.0$~dex and are, respectively, about $0.19, 0.16$~mag 
   and $0.15$~mag. 
   
   The slopes of the period dependence of the PLZ relations derived in this work are 
   in a~quite good agreement with other studies in the globular clusters 
   \citep[especially][]{Vivas2017,Bhardwaj2021,Ngeow2022RRL}. 
   Larger differences occur for the metallicity slopes, which might be due to the use 
   of different metallicity ranges. In case of the globular clusters this is generally 
   much narrower \citep[e.g.,][]{Sesar2017,Ngeow2022RRL}. 
   The largest differences, however, we see in the ZPs of the relations. Our ZPs are 
   systematically fainter than all other studies. There might be several potential reasons 
   for this state of affairs. 
   For example, as we can see in Figure~\ref{fig:comp_CaceresCatelan2008}, the value of 
   $\rm{[\alpha / Fe]}$ ratio might influence noticebly the ZP of the PLZ relation, which 
   might be another factor differentiating the results in globular clusters and those 
   from the vicinity of the Sun. 
   The metallicity slope uncertainty also influences the ZP of the relation, so the more 
   stars from larger metallicity range are used, the better. Even though we can boast 
   a~wide range of metallicities, the number of stars used for our fitting is rather small.
   An additional difficulty might be the fact that metallicities are on different metallicity 
   scales. 
   The knowledge of the accurate distances of stars is crucial in determination of the 
   absolute PL and PLZ relations, which still is a~problem in case of globular clusters (as 
   we can see for example in Figure~\ref{fig:comp_Vivas2017}). The sample of RR~Lyr stars 
   used in the current study is located less than $3$~kpc around the Sun, so the good 
   quality Gaia~DR3 parallaxes were available for them. Also, calculation of the reddening 
   for the stars was performed differently in different studies 
   \citep[e.g.,][]{Vivas2017,Bhardwaj2021}.  
   Last but not least, the fact that our relations seem to be shifted systematically 
   relative to globular clusters suggests a~possible problem of blending of those stars 
   with much fainter background or forground stars in the globular clusters which, in 
   a~consequence, might slightly change their apparent magnitudes. 
   For the discussion of the influence of the blending see, for instance, the study of 
   \citet{Majaess2012}. This problem does not appear to exist in our data.


\subsection{Revision of the photometric correctness} \label{sssec:vosa}


   The significant discrepancy between the ZPs determined by us and those in the 
   literature motivated us to analyze the spectral energy distribution (SED) for 
   some of our stars, in order to check the correctness of the mean magnitudes of 
   RR~Lyr stars in the Sloan--Pan-STARSS bands. We used Virtual Observatory Spectral 
   Energy Distribution Analyzer \citep[VOSA,][]{Bayo2008} to compare the SED of $18$ 
   RR~Lyr stars common with the sample of stars from \citet{Monson2017}, where they 
   compiled in their Table~5 mean magnitudes of RR~Lyr stars in the 
   $UBVR_{C}I_{C}JHK_{\mathrm{s}}W_{1}W_{2}$ bands. Our mean apparent magnitudes 
   matched very well the ones from \citet{Monson2017} on the SED, except for V675~Sgr, 
   where our photometry in the Sloan--Pan-STARSS $g_{P1}r_{P1}i_{P1}$ bands is 
   systematically fainter (by about $0.1$~mag). The reason for that is unclear. 
   V675~Sgr is located in a~dense field toward the Galactic Bulge, which may affect 
   aperture photometry. Another possibilty is that the ATLAS-REFCAT2 itself might have 
   a~systematic bias in the magnitudes of stars in this area. 
   We marked this problematic star in Table~\ref{tab:rrlyr} with a~star-symbol, and 
   we recommend using the mean magnitudes given for it with caution. 
   Nevertheless, this one problematic star is not influencing the overall result much, 
   and consequently, do not explain the obtained discrepancy in the ZP between this 
   work and the literature.


\section{Summary} \label{sec:summary}


   In this work, we have derived PL and PLZ relations in the $g_{P1}r_{P1}i_{P1}$ 
   Sloan--Pan-STARSS bands, together with their Wesenheit index counterparts 
   ($W_{r}^{ri}$, $W_{r}^{gr}$, and $W_{g}^{gi}$) for Galactic RR~Lyr stars located 
   within a~radius of about $3$~kpc around the Sun. 
   Data for the project were acquired with the $40$ cm LCO Telescope Network in the 
   $g'r'i'$ Sloan filters, and calibrated using the ATLAS-REFCAT2 \citep{Tonry2018} 
   catalog, which is on the Pan-STARRS implementation of the Sloan photometric system. 
   Well covered light curves for $44$ RRab and nine RRc type stars are presented in 
   Figures~\ref{fig:fig1} and \ref{fig:fig2}, respectively, in the Appendix~\ref{app:lc}. 
   The mean magnitudes obtained based on the resulting light curves are given in 
   Table~\ref{tab:rrlyr}. 
   The adopted reddening values from \citet{SF2011} were available for all our stars, 
   and the extinction coefficients were taken from \citet{Green2019}. 
   The PL and PW relations were derived using $38$ RRab and seven RRc stars, separately, 
   as well as for a~mixed population of RRab+RRc stars ($45$ stars in total). The 
   PLZ and PWZ relations were derived based on $31$ RRab stars and $35$ stars in total 
   for the mixed population, as not all stars in the original sample had metallicity 
   values available in \citet{Crestani2021a,Crestani2021b}. 
   The relations were determined using absolute magnitudes calculated in four ways: 
   directly from geometric Gaia~DR3 parallaxes, the ABL method, and using geometric 
   and photogeometric distances from \citet{BJ2021}. 
   The resulting coefficients are given in Tables~\ref{tab:plr} and \ref{tab:plzr2}. 
   
\begin{figure*}
   \centering
   \includegraphics[width=0.9\hsize]{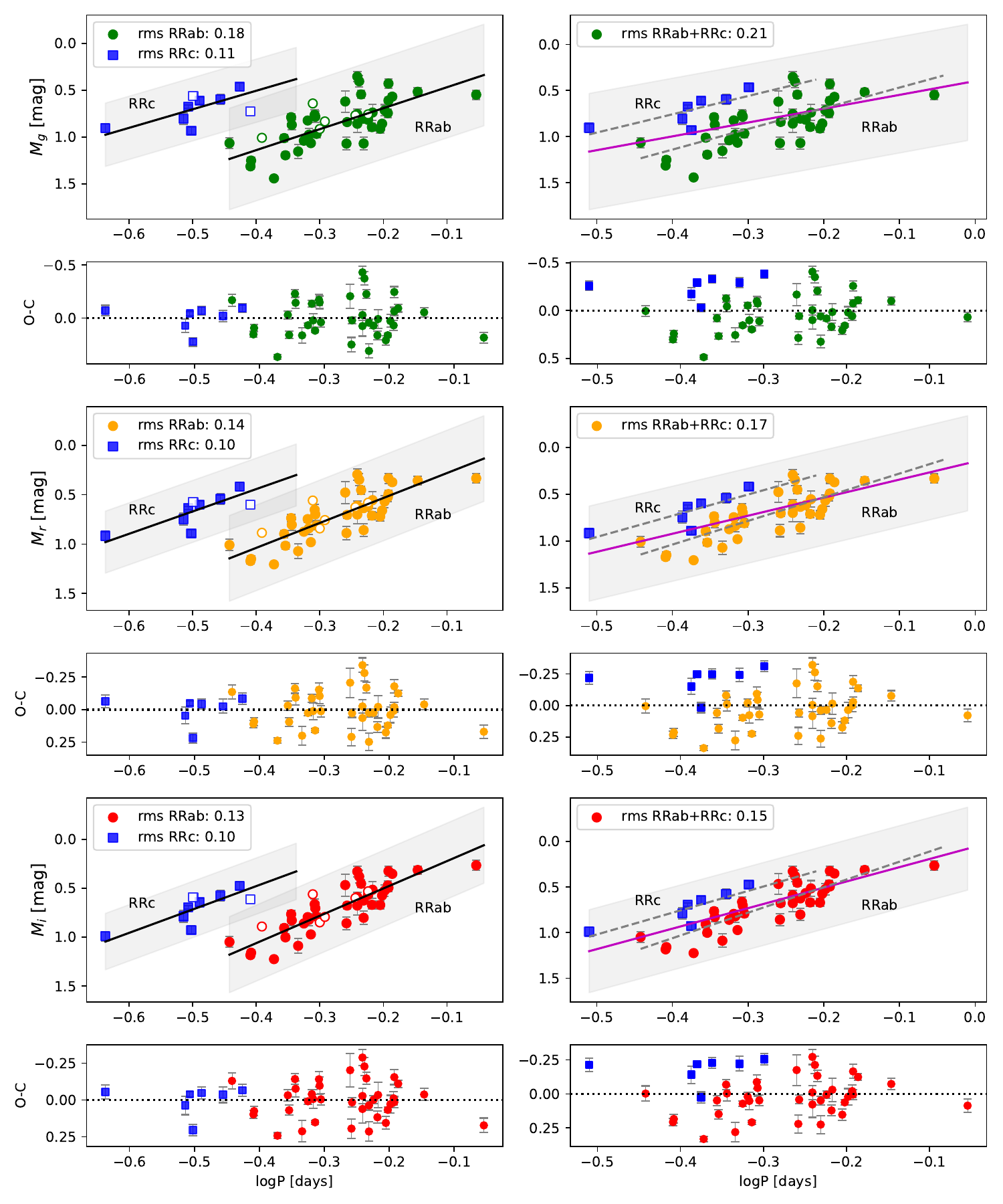}
   \caption{PL relations for RR~Lyr stars (RRab$+$RRc) based on mean reddenings 
            from \citet{SF2011}, the reddening vector ($R_{\lambda}$) from \citet{Green2019}, 
            and Gaia~DR3 parallaxes. 
            Filled circles and squares mark RRab and RRc stars adopted for derivation of the 
            PL relations, respectively; 
            open circles and squares: RRab and RRc stars with $RUWE>1.4$, respectively; 
            black solid and dashed gray lines: the fit to Equation~(\ref{eq:plr}) for RRab 
            and RRc, separately; 
            magenta solid line: the fit to Equation~(\ref{eq:plr}) for RRab+RRc stars; 
            shaded areas: $\pm 3$rms.
            }
      \label{fig:plr}%
    \end{figure*}

\begin{figure*}
   \centering
   \includegraphics[width=0.9\hsize]{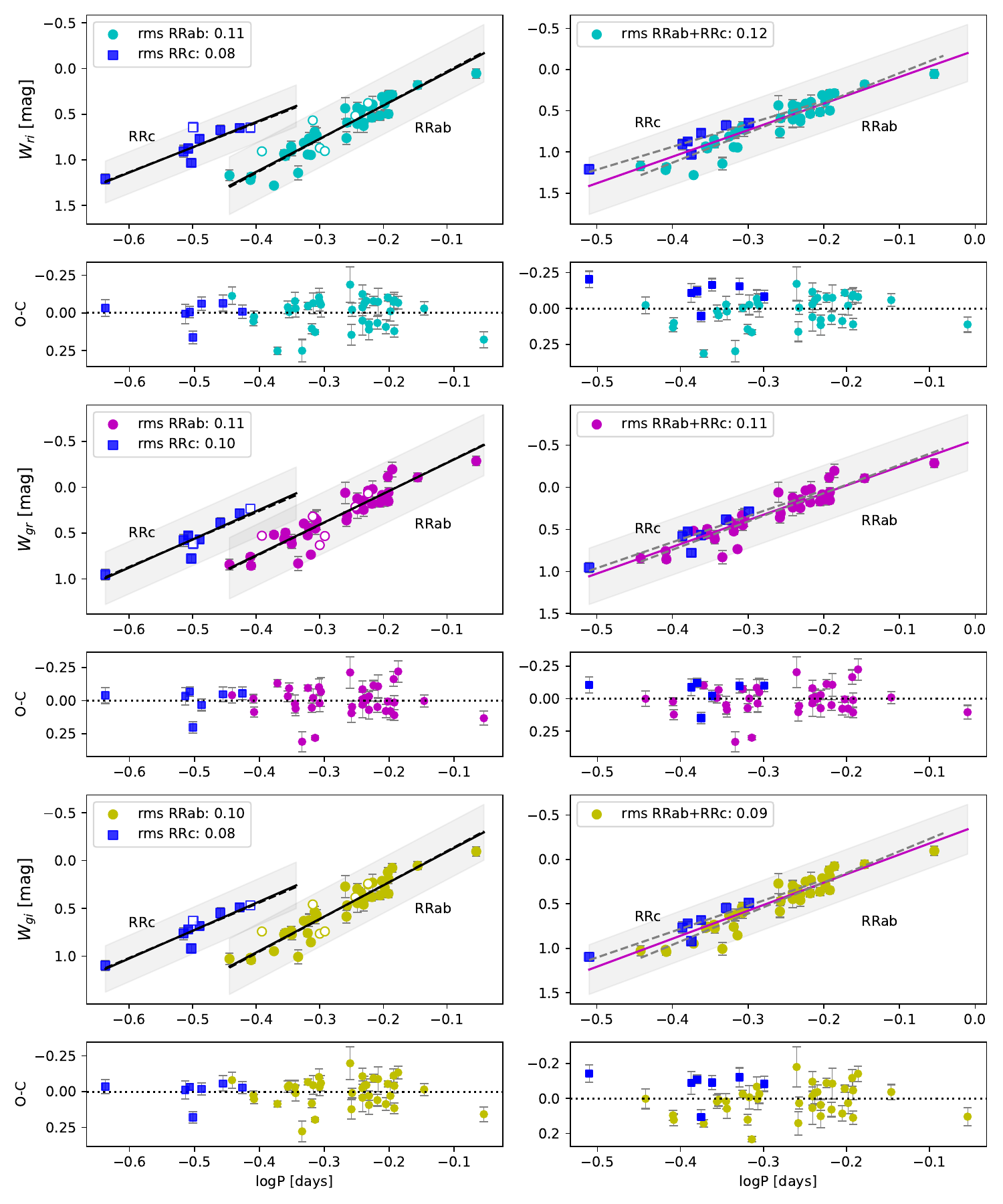}
   \caption{PW relations for RR~Lyr stars (RRab$+$RRc) based on mean reddenings 
            from \citet{SF2011}, the reddening vector ($R_{\lambda}$) from \citet{Green2019} 
            and Gaia~DR3 parallaxes. 
            Filled circles and squares mark RRab and RRc stars adopted for derivation of 
            the PW relations, respectively; 
            open circles and squares: RRab and RRc stars with $RUWE>1.4$, respectively; 
            black solid and dashed gray lines: the fit to Equation~(\ref{eq:pwr}) for RRab 
            and RRc, separately; 
            magenta solid line: the fit to Equation~(\ref{eq:plr}) for RRab+RRc stars; 
            shaded areas: $\pm 3$rms.
            }
      \label{fig:pwr}%
    \end{figure*}

\begin{figure*}
   \centering
   \includegraphics[width=0.9\hsize]{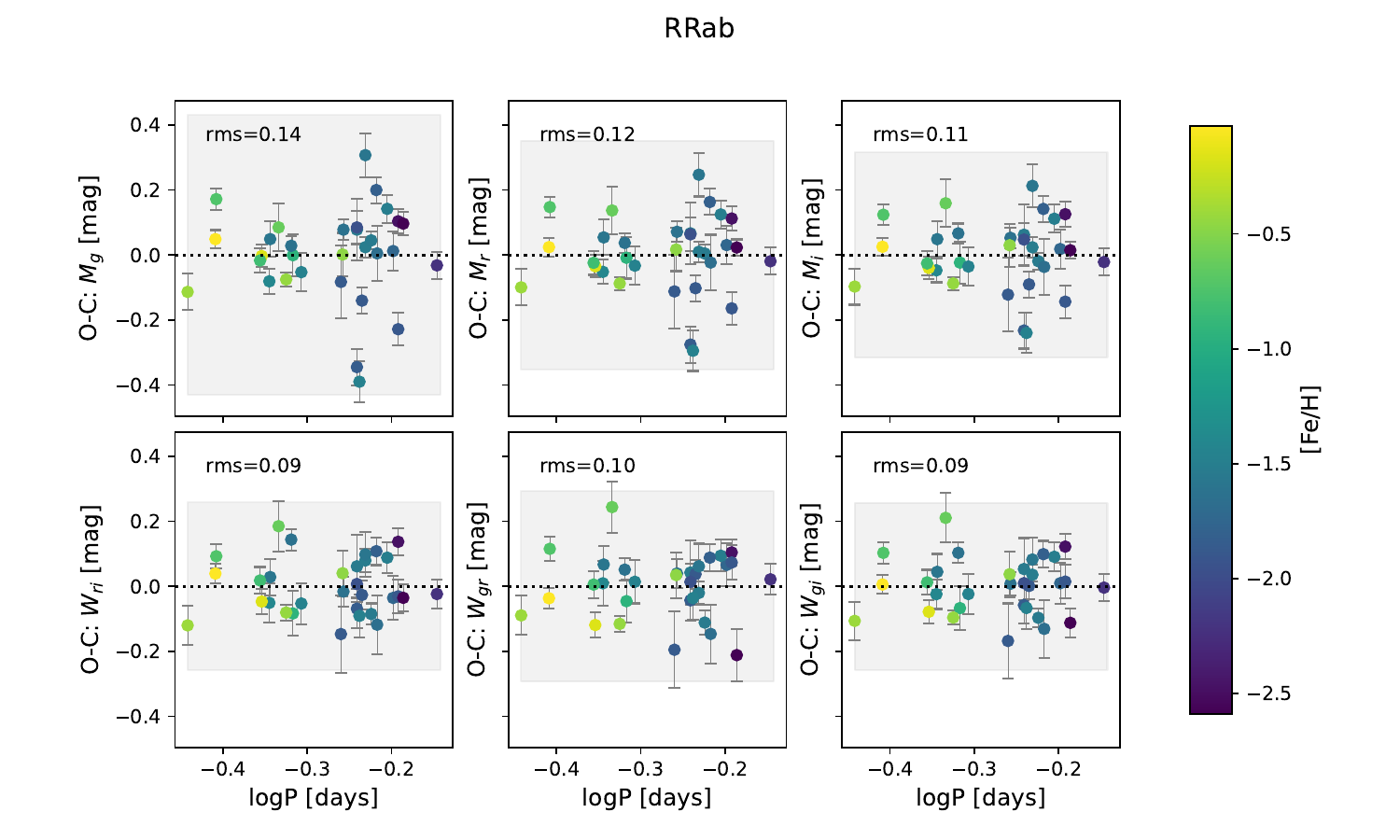}
   \includegraphics[width=0.9\hsize]{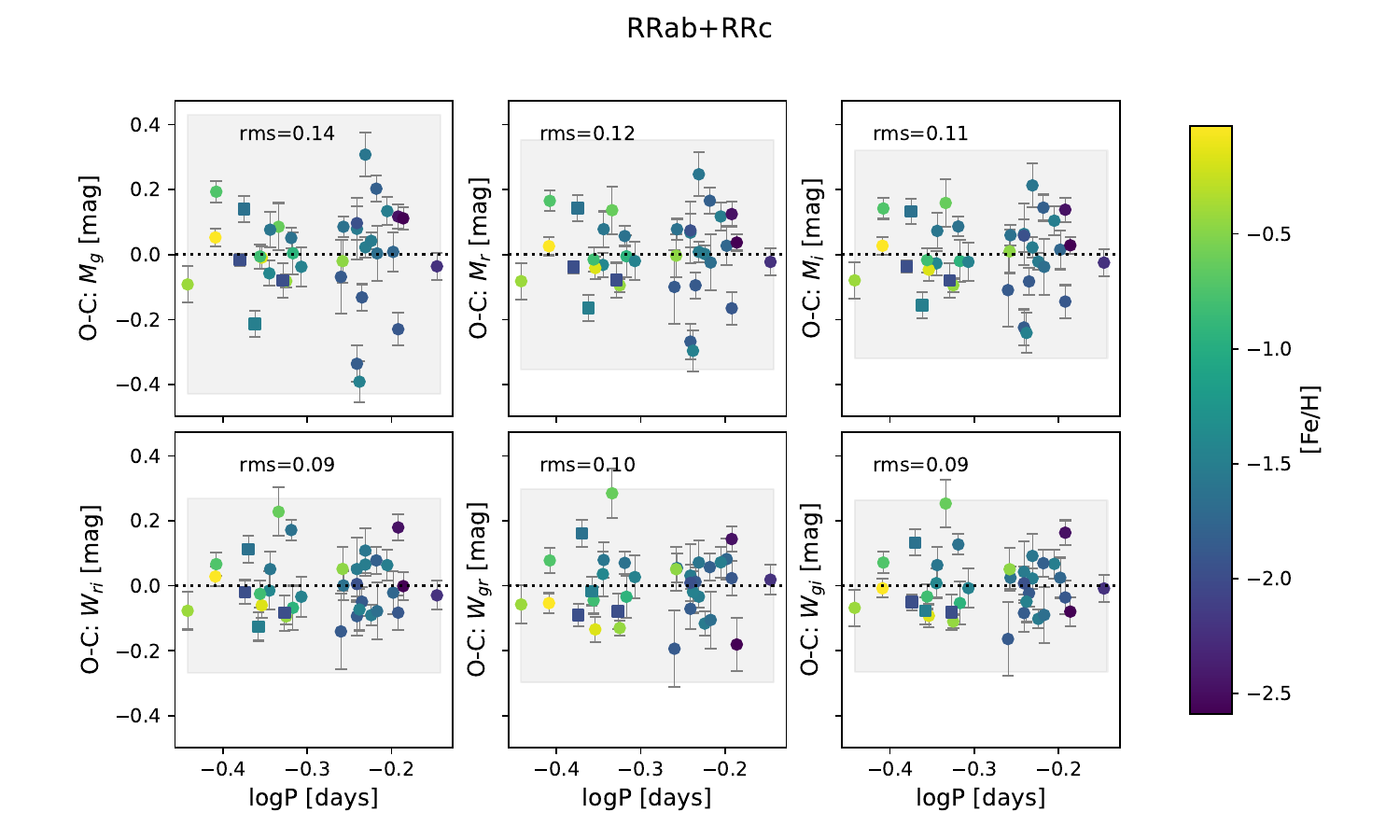}
   \caption{Residuals of the PLZ and PWZ relations for the Sloan~$gri$ bands and Wesenheit 
            indices for RR~Lyr stars (RRab and RRab$+$RRc) based on the parallax method. 
            Circles mark RRab stars; 
            squares: RRc stars;
            shaded areas: $\pm 3$rms.
            }
      \label{fig:plzr2}%
    \end{figure*}

   We noticed that the fundamentalization procedure of the RRc stars from \citet{Iben1974} 
   failed to align the RRc and RRab stars correctly, and instead introduced the systematic 
   shift seen in Figures~\ref{fig:plr} and \ref{fig:pwr}, which depends on the filter used. 
   We discussed different approaches to the fundamentalization procedure of the RRc stars, 
   first by following the method of \citet{Zgirski2023}, and then the nonlinear relation 
   showed in \citet{Nemec2024}. 
   None of the proposed approaches fully removes the existing offset, as it is a manifestation 
   of the temperature (and hence color) difference propagating into the PL and PLZ relations. 
   For this reason, for distance determination purposes we recommend to use the relations 
   derived separately for RRab and RRc stars.  
   
   We tested the behavior of our PL, PW, PLZ and PWZ relations in case of applying parallax 
   corrections from \citet{Groenewegen2021} or not applying them at all. In the first case, 
   the differences in slopes and ZPs of the original and new relations are statisticaly 
   insignificant. In the second case, the new intercepts were noticebly smaller. This 
   exercise shows that uncertainties related to the ZP of Gaia parallaxes are significant. 
   
\begin{figure*}
   \sidecaption
   \includegraphics[width=12cm]{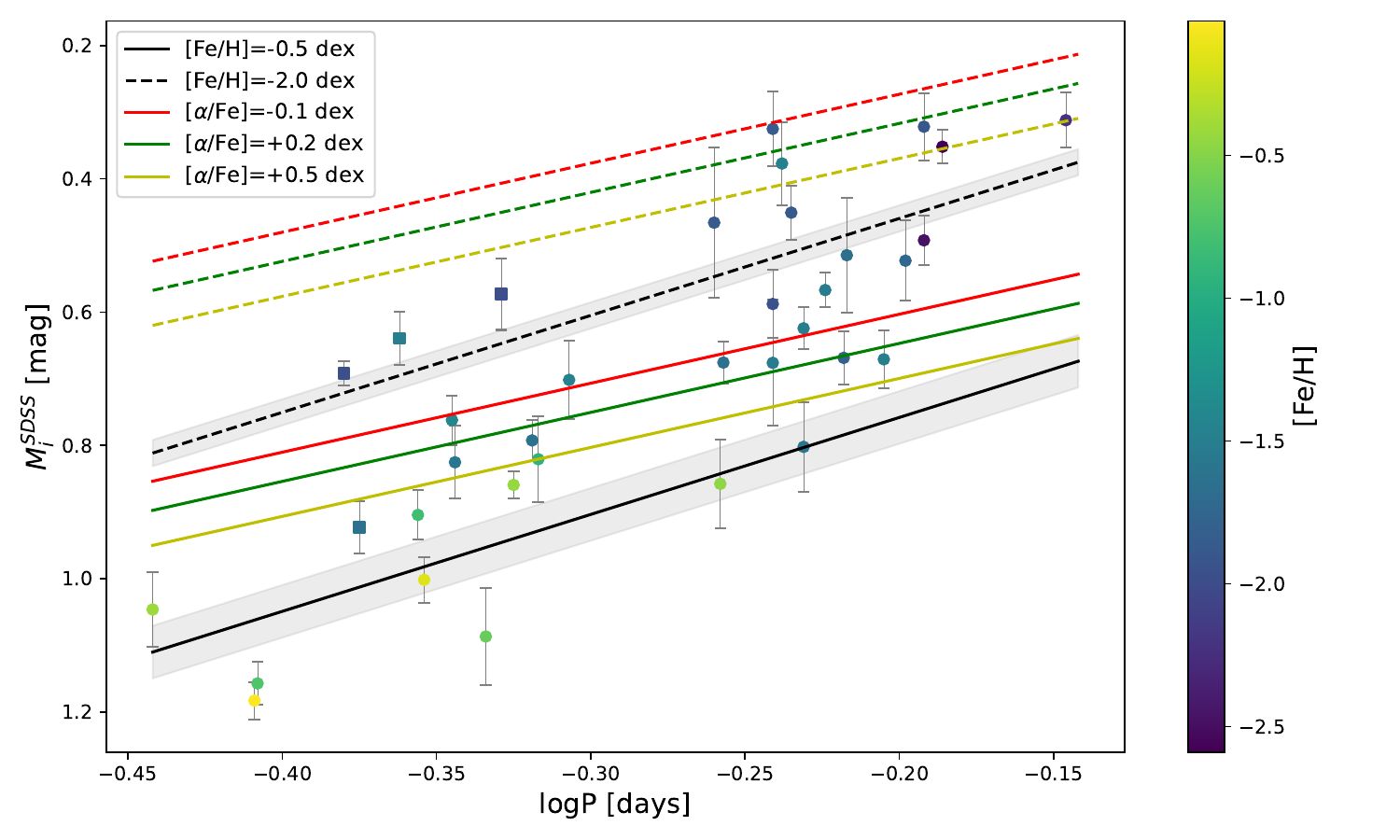}
   \caption{Comparison of the PLZ relation derived in this work with 
            the theoretical PLZ relation provided by \citet{CaceresCatelan2008}. 
            Black lines mark the PLZ relation derived in this work for a~constant metallicity 
            of $\rm{[Fe/H]} = -0.5;-2.0$~dex; 
            gray, shaded areas: metallicity slope $c_i \pm \delta c_i$ from Equation~\ref{eq:plz}; 
            red, green, yellow lines: theoretical PLZ relation from \citet{CaceresCatelan2008} 
            for $[\alpha / Fe] = -0.1;+0.2;+0.5$~dex at a~constant metallicity of 
            $\rm{[Fe/H]} = -0.5;-2.0$~dex.
            }
      \label{fig:comp_CaceresCatelan2008}%
    \end{figure*}

\begin{figure*}
   \sidecaption
   \includegraphics[width=12cm]{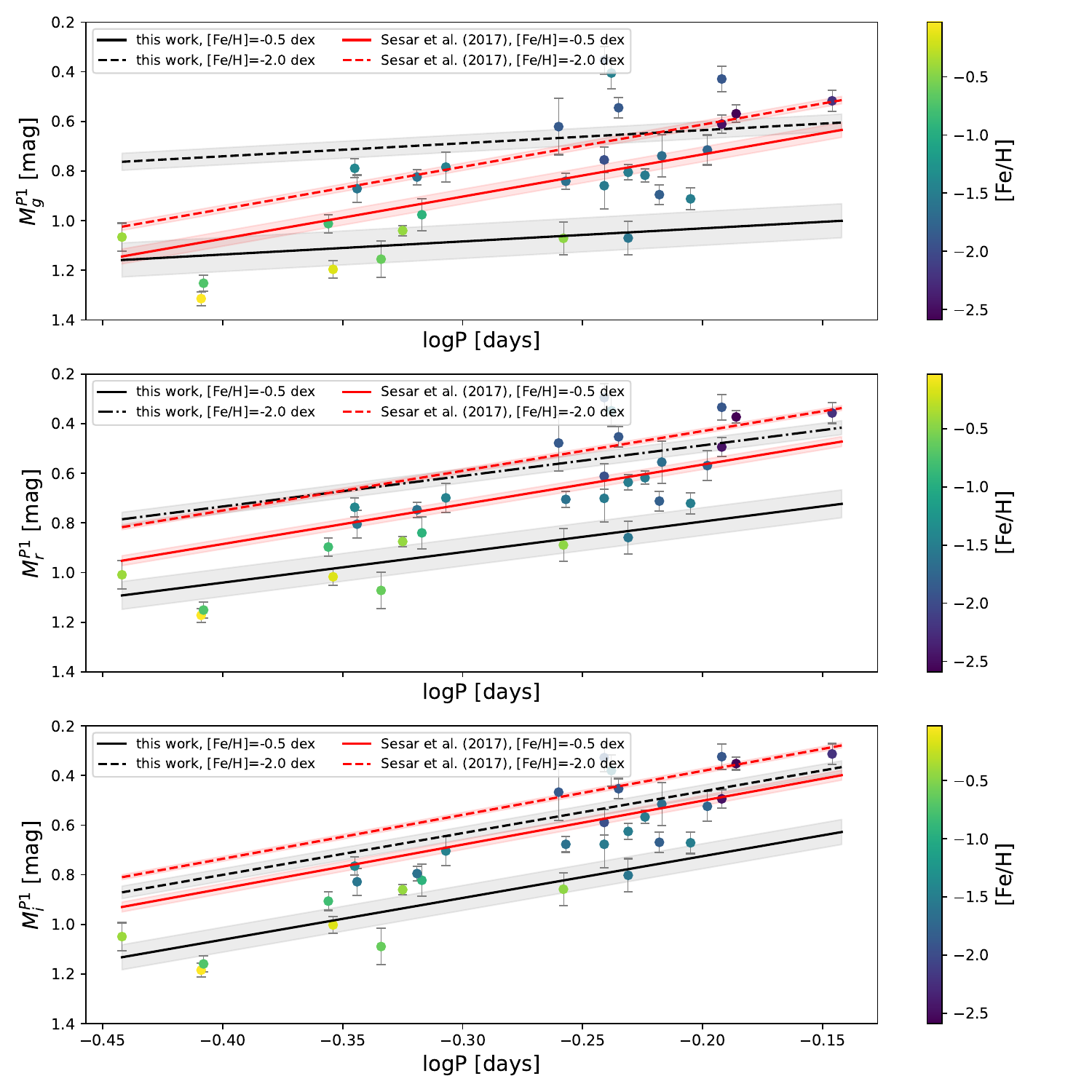}
   \caption{Comparison of the PLZ relations derived in this work with empirical 
            PLZ relations provided by \citet{Sesar2017}. 
            Black lines mark the PLZ relations derived in this work for a~constant metallicity 
            of $\rm{[Fe/H]} = -0.5;-2.0$~dex; 
            gray, shaded areas: metallicity slope $c_{\lambda} \pm \delta c_{\lambda}$ 
            from Equation~\ref{eq:plz}; 
            red lines: empirical PLZ relation from \citet{Sesar2017} for a~constant metallicity 
            of $\rm{[Fe/H]} = -0.5;-2.0$~dex; 
            red, shaded areas: metallicity slope $\beta \pm \delta \beta$ from Table~1 in 
            \citet{Sesar2017}.
            }
      \label{fig:comp_Sesar2017}%
    \end{figure*}

\begin{figure*}
   \sidecaption
   \includegraphics[width=12cm]{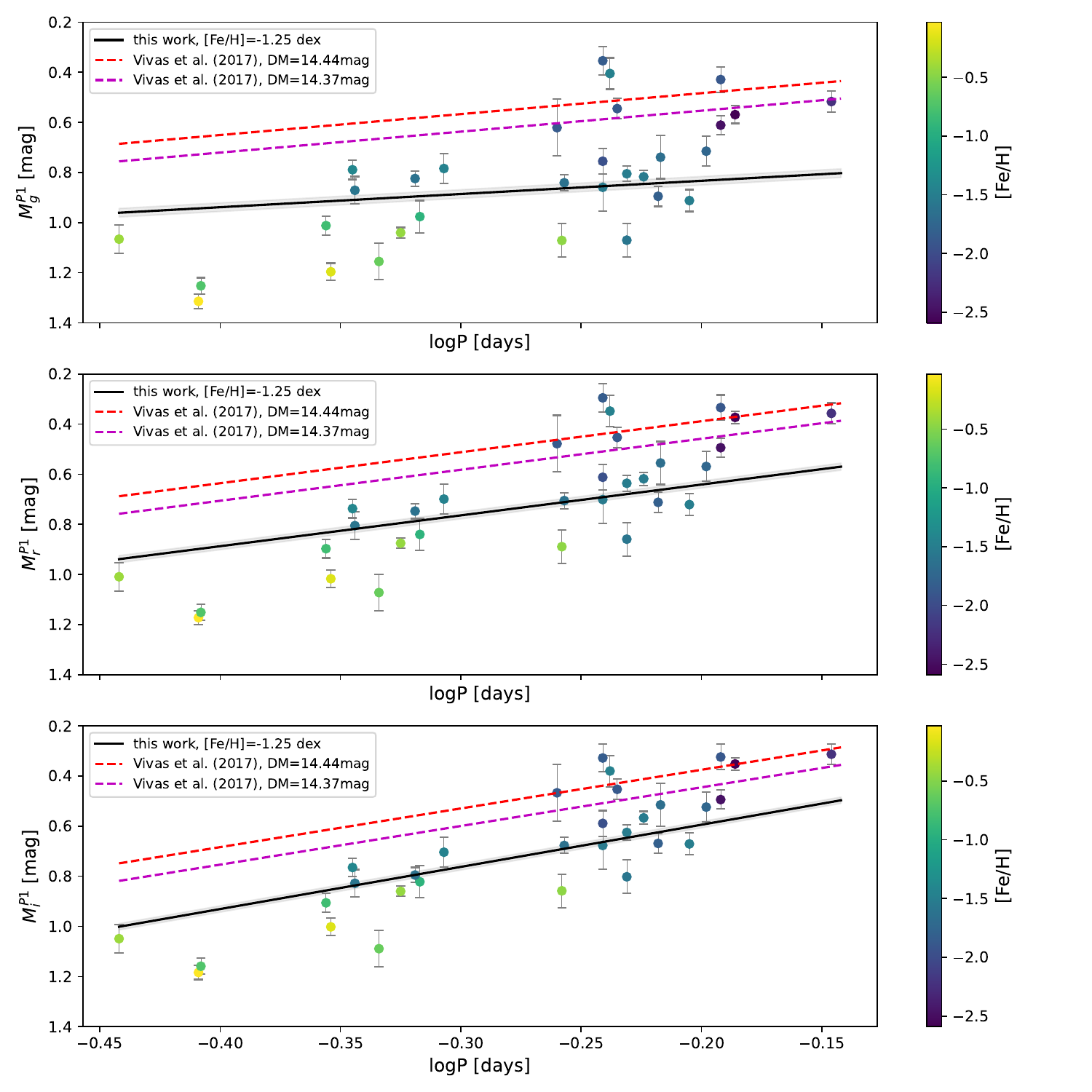}
   \caption{Comparison of the PLZ relations derived in this work with empirical 
            PLZ relations provided by \citet{Vivas2017} for the globular cluster M5. 
            Comparison is given for two distances to M5. 
            Black lines: the PLZ relations derived in this work for a~constant metallicity 
            of $\rm{[Fe/H]} = -1.25$~dex; 
            gray, shaded areas: metallicity slope $c_{\lambda} \pm \delta c_{\lambda}$ 
            from Equation~(\ref{eq:plz}); 
            red lines: empirical PLZ relation from \citet{Vivas2017} for a~constant 
            metallicity of $\rm{[Fe/H]} = -1.25$~dex and distance modulus equal to $14.44$~mag; 
            magenta lines: same as red lines but for distance modulus equal to $14.37$~mag 
            from Table~1 of \citet{BaumgardtVasiliev2021}.
            }
      \label{fig:comp_Vivas2017}%
    \end{figure*}

\begin{figure*}
   \centering
   \includegraphics[width=\hsize]{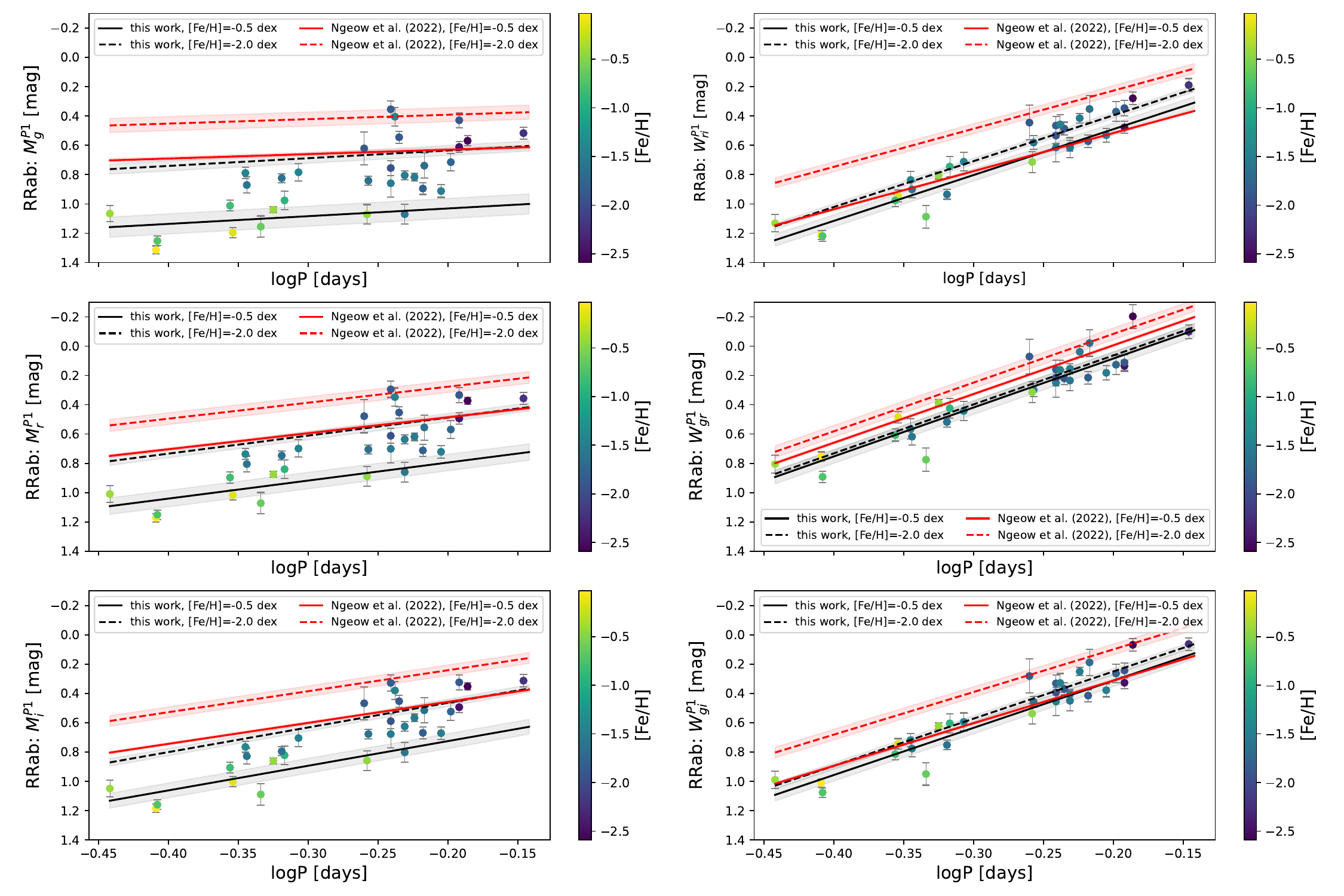}
   \caption{Comparison of the PLZ and PWZ relations derived in this work with 
            empirical PLZ and PWZ relations provided by \citet{Ngeow2022RRL} from 
            RR~Lyr in $46$ globular clusters. 
            Black lines mark the PLZ and PWZ relations derived in this work for a~constant 
            metallicity of $\rm{[Fe/H]} = -0.5;-2.0$~dex; 
            gray, shaded areas: metallicity slope $c_{\lambda} \pm \delta c_{\lambda}$ 
            from Equation~(\ref{eq:plz}); 
            red lines: empirical PLZ relation from \citet{Ngeow2022RRL} for a~constant 
            metallicity of $\rm{[Fe/H]} = -0.5;-2.0$~dex; 
            red, shaded areas: metallicity slope $c \pm \delta c$ from Table~4 in 
            \citet{Ngeow2022RRL}.
            }
      \label{fig:comp_Ngeow2022RRL}%
    \end{figure*}

   The comparison of the PLZ and PWZ relations presented in this work with available literature 
   studies showed that our slopes of the period dependence are in a~good agreement with 
   some of the works \citep[e.g.,][]{Vivas2017,Bhardwaj2021,Ngeow2022RRL}, becoming steeper 
   for longer wavelengths, but do they not agree with theoretical study of 
   \citet{CaceresCatelan2008} or the empirical PL relations from \citet{Sesar2017}. The 
   metallicity slopes obtained in this work are steeper than other literature values, and 
   the reason for that might be, that the metallicity range we use is larger than the metallicity 
   range of globular clusters studied by other authors. 
   The ZPs obtained in this work are systematically fainter than the literature ZPs, however, 
   this conclusion should be treated with caution because the comparison of our and literature 
   results is not straightforward due to the number of reasons. We are comparing different 
   populations of stars which might differ according to, for example, $\rm{[\alpha/Fe]}$ 
   content which is influencing the ZP of the PLZ relation (as shown in 
   Figure~\ref{fig:comp_CaceresCatelan2008}). Also, the metallicity slope have an influence 
   on the ZP, and that slope is steeper in our case. Our results are also not based on many 
   stars, although, as the study of the SED showed, our photometry is of very good quality, 
   and we also use good quality Gaia~DR3 parallaxes, which further assures us about the 
   correctness of procedures applied in this work. 
   
   Presented PL, PW, PLZ and PWZ relations in the Sloan--Pan-STARSS bands, to the best of 
   our knowledge, are the first of their kind for RR~Lyr stars from the vicinity of the 
   Sun. So we hope that as such they will be useful for distance determinations of RR Lyr 
   stars in the upcoming Rubin-LSST era.


%

\begin{sidewaystable*}
\scriptsize
\caption{\label{tab:rrlyr} MW RR Lyr star sample and their main parameters.}
\begin{tabular}{l@{\extracolsep{2.5em}}llcccrrrcc}
\hline\hline
Star & Period & $\varpi_{DR3}$ & RUWE & GOF & E(B-V) & $<g>$ & $<r>$ & $<i>$ & [Fe/H] & Source \\
     & (days) & (mas)  &  & (mag) &  & (mag) & (mag) & (mag) & (dex) & \\ 
\hline
\multicolumn{11}{l}{RRab} \\ 
\hline
AA~Aql    & 0.3617877 & 0.7615 $\pm$ 0.0179 & 1.14 &  3.96 & 0.0710 $\pm$ 0.0059 & 11.907 $\pm$ 0.004 & 11.786 $\pm$ 0.003 & 11.780 $\pm$ 0.003 & -0.42 $\pm$ 0.11 & 1 \\
AV~Peg    & 0.3903814 & 1.5075 $\pm$ 0.0176 & 1.25 &  5.44 & 0.0520 $\pm$ 0.0007 & 10.606 $\pm$ 0.003 & 10.417 $\pm$ 0.002 & 10.395 $\pm$ 0.002 & -0.03 $\pm$ 0.07 & 1 \\
BB~Eri$^{a}$    & 0.5699097 & 0.7216 $\pm$ 0.0238 & 1.74 & 22.82 & 0.0430 $\pm$ 0.0015 & 11.628 $\pm$ 0.003 & 11.447 $\pm$ 0.003 & 11.392 $\pm$ 0.003 & -1.66 $\pm$ 0.07 & 1 \\
BH~Peg    & 0.640993 & 1.1813 $\pm$ 0.0226 & 1.22 &  3.97 & 0.0700 $\pm$ 0.0008 & 10.629 $\pm$ 0.008 & 10.346 $\pm$ 0.005 & 10.247 $\pm$ 0.005 & - & - \\
BK~Tuc    & 0.5500676 & 0.3630 $\pm$ 0.0172 & 1.34 &  8.98 & 0.0240 $\pm$ 0.0004 & 12.906 $\pm$ 0.007 & 12.741 $\pm$ 0.006 & 12.715 $\pm$ 0.006 & -1.86 $\pm$ 0.01 & 1 \\ 
BR~Aqr    & 0.4818717 & 0.8075 $\pm$ 0.0217 & 0.90 & -2.09 & 0.0240 $\pm$ 0.0003 & 11.525 $\pm$ 0.002 & 11.367 $\pm$ 0.003 & 11.333 $\pm$ 0.005 & -0.94 $\pm$ 0.07 & 1 \\ 
BT~Aqr$^{a}$    & 0.4063595 & 0.5365 $\pm$ 0.0220 & 1.41 & 12.11 & 0.0400 $\pm$ 0.0002 & 12.502 $\pm$ 0.003 & 12.343 $\pm$ 0.002 & 12.322 $\pm$ 0.003 & -0.24 $\pm$ 0.04 & 1 \\ 
CD~Vel     & 0.5735076 & 0.5874 $\pm$ 0.0137 & 1.26 &  7.44 & 0.1920 $\pm$ 0.0055 & 12.185 $\pm$ 0.006 & 11.953 $\pm$ 0.005 & 11.862 $\pm$ 0.004 & -1.84 $\pm$ 0.09 & 1 \\ 
CP~Aqr     & 0.4634018 & 0.7754 $\pm$ 0.0236 & 1.04 &  0.78 & 0.0500 $\pm$ 0.0016 & 11.883 $\pm$ 0.007 & 11.755 $\pm$ 0.004 & 11.740 $\pm$ 0.003 & -0.62 $\pm$ 0.04 & 1 \\ 
DH~Hya$^{a}$    & 0.4890007 & 0.5033 $\pm$ 0.0266 & 1.55 &  16.42 & 0.0370 $\pm$ 0.0031 & 12.263 $\pm$ 0.003 & 12.147 $\pm$ 0.003 & 12.125 $\pm$ 0.003 & -1.78 $\pm$ 0.04 & 1 \\
DN~Aqr    & 0.6337558 & 0.7785 $\pm$ 0.0195 & 0.95 & -0.97 & 0.0220 $\pm$ 0.0007 & 11.336 $\pm$ 0.006 & 11.170 $\pm$ 0.004 & 11.111 $\pm$ 0.006 & -1.74 $\pm$ 0.03 & 1 \\
DX~Del    & 0.4726191 & 1.7584 $\pm$ 0.0150 & 0.99 & -0.43 & 0.0800 $\pm$ 0.0018 & 10.096 $\pm$ 0.002 & 9.859 $\pm$ 0.002 & 9.792 $\pm$ 0.003 & -0.43 $\pm$ 0.08 & 1 \\ 
EW~Cam    & 0.628408 & 1.7499 $\pm$ 0.0136 & 1.04 &  1.30 & 0.0180 $\pm$ 0.0007 & 9.706 $\pm$ 0.003 & 9.492 $\pm$ 0.002 & 9.394 $\pm$ 0.002 & - & - \\ 
HH~Pup    & 0.3907454 & 1.1377 $\pm$ 0.0151 & 1.07 &  1.82 & 0.1240 $\pm$ 0.0054 & 11.408 $\pm$ 0.005 & 11.195 $\pm$ 0.004 & 11.123 $\pm$ 0.003 & -0.73 $\pm$ 0.02 & 1 \\ 
RR~Cet    & 0.55302836 & 1.6217 $\pm$ 0.0213 & 1.01 &  0.43 & 0.0200 $\pm$ 0.0003 & 9.835 $\pm$ 0.006 & 9.681 $\pm$ 0.007 & 9.640 $\pm$ 0.006 & -1.63 $\pm$ 0.10 & 1 \\
RR~Gru    & 0.5524640 & 0.5105 $\pm$ 0.0142 & 1.05 &  1.45 & 0.0190 $\pm$ 0.0007 & 12.598 $\pm$ 0.003 & 12.399 $\pm$ 0.003 & 12.355 $\pm$ 0.003 & -0.47 $\pm$ 0.04 & 1 \\ 
RR~Leo    & 0.4524021 & 1.0836 $\pm$ 0.0248 & 1.12 &  2.75 & 0.0340 $\pm$ 0.0035 & 10.816 $\pm$ 0.003 & 10.720 $\pm$ 0.002 & 10.721 $\pm$ 0.001 & -1.58 $\pm$ 0.08 & 1 \\ 
RU~Scl    & 0.4933549 & 1.2800 $\pm$ 0.0317 & 1.24 &  5.20 & 0.0170 $\pm$ 0.0004 & 10.308 $\pm$ 0.008 & 10.207 $\pm$ 0.005 & 10.201 $\pm$ 0.004 & -1.45 $\pm$ 0.06 & 1 \\
RV~Cet    & 0.62341 & 0.9762 $\pm$ 0.0177 & 1.14 &  3.21 & 0.0270 $\pm$ 0.0006 & 11.059 $\pm$ 0.005 & 10.844 $\pm$ 0.004 & 10.776 $\pm$ 0.004 & -1.50 $\pm$ 0.03 & 1 \\
RV~Phe$^{a}$    & 0.5964071 & 0.5606 $\pm$ 0.0197 & 1.44 & 14.27 & 0.0070 $\pm$ 0.0005 & 12.039 $\pm$ 0.003 & 11.856 $\pm$ 0.003 & 11.801 $\pm$ 0.004 & -1.54 $\pm$ 0.04 & 1 \\
RX~Cet    & 0.57373 & 0.7647 $\pm$ 0.0303 & 1.33 &  6.21 & 0.0240 $\pm$ 0.0007 & 11.526 $\pm$ 0.006 & 11.346 $\pm$ 0.005 & 11.307 $\pm$ 0.004 & -1.53 $\pm$ 0.04 & 1 \\
RX~Eri    & 0.5872453 & 1.7229 $\pm$ 0.0225 & 1.28 &  8.44 & 0.0530 $\pm$ 0.0018 & 9.810 $\pm$ 0.003 & 9.593 $\pm$ 0.003 & 9.548 $\pm$ 0.003 & -1.51 $\pm$ 0.08 & 1 \\
S~Ara     & 0.451848 & 1.1271 $\pm$ 0.0169 & 1.04 &  0.94 & 0.0880 $\pm$ 0.0027 & 10.839 $\pm$ 0.015 & 10.707 $\pm$ 0.010 & 10.679 $\pm$ 0.009 & -1.40 $\pm$ 0.04 & 1 \\
SS~For    & 0.49543 & 1.2868 $\pm$ 0.0205 & 1.27 &  8.44 & 0.0130 $\pm$ 0.0002 & 10.465 $\pm$ 0.024 & 10.297 $\pm$ 0.019 & 10.269 $\pm$ 0.010 & - & - \\
ST~Pic    & 0.4857445 & 2.0822 $\pm$ 0.0123 & 1.00 & -0.03 & 0.0270 $\pm$ 0.0012 & 9.568 $\pm$ 0.002 & 9.459 $\pm$ 0.002 & 9.433 $\pm$ 0.002 & - & - \\
SV~Eri    & 0.713877 & 1.3609 $\pm$ 0.0236 & 0.98 & -0.40 & 0.0780 $\pm$ 0.0023 & 10.122 $\pm$ 0.005 & 9.892 $\pm$ 0.004 & 9.798 $\pm$ 0.002 & -2.22 $\pm$ 0.03 & 1 \\
SW~And    & 0.442262 & 1.9954 $\pm$ 0.0284 & 1.20 &  5.23 & 0.0330 $\pm$ 0.0011 & 9.812 $\pm$ 0.003 & 9.603 $\pm$ 0.004 & 9.567 $\pm$ 0.004 & -0.17 $\pm$ 0.05 & 2 \\
SX~For    & 0.6053423 & 0.8680 $\pm$ 0.0146 & 0.97 & -0.99 & 0.0120 $\pm$ 0.0009 & 11.245 $\pm$ 0.003 & 11.051 $\pm$ 0.003 & 11.000 $\pm$ 0.003 & -1.81 $\pm$ 0.01 & 1 \\
SZ~Gem$^{a}$    & 0.5011303 & 0.7004 $\pm$ 0.0248 & 1.44 & 10.70 & 0.0380 $\pm$ 0.0004 & 11.822 $\pm$ 0.002 & 11.715 $\pm$ 0.002 & 11.697 $\pm$ 0.002 & -1.87 $\pm$ 0.10 & 2 \\
TT~Lyn    & 0.597434355 & 1.4798 $\pm$ 0.0160 & 0.86 & -3.57 & 0.0150 $\pm$ 0.0005 & 10.019 $\pm$ 0.006 & 9.806 $\pm$ 0.005 & 9.746 $\pm$ 0.003 & -1.53 $\pm$ 0.01 & 2 \\
U~Lep     & 0.5814789 & 0.9889 $\pm$ 0.0167 & 1.33 &  9.72 & 0.0290 $\pm$ 0.0015 & 10.671 $\pm$ 0.005 & 10.553 $\pm$ 0.003 & 10.534 $\pm$ 0.004 & -1.88 $\pm$ 0.28 & 1 \\
U~Pic     & 0.4403750 & 0.8236 $\pm$ 0.0128 & 0.92 & -2.09 & 0.0090 $\pm$ 0.0002 & 11.465 $\pm$ 0.004 & 11.342 $\pm$ 0.004 & 11.345 $\pm$ 0.004 & -0.82 $\pm$ 0.03 & 1 \\
UU~Cet    & 0.6060736 & 0.5334 $\pm$ 0.0191 & 1.20 &  4.83 & 0.0200 $\pm$ 0.0006 & 12.174 $\pm$ 0.004 & 11.972 $\pm$ 0.004 & 11.919 $\pm$ 0.004 & -1.66 $\pm$ 0.04 & 2 \\
V341~Aql  & 0.5780230 & 0.8822 $\pm$ 0.0230 & 1.21 &  5.61 & 0.0790 $\pm$ 0.0021 & 10.955 $\pm$ 0.003 & 10.827 $\pm$ 0.003 & 10.808 $\pm$ 0.003 & -1.47 $\pm$ 0.04 & 1 \\
V4424~Sgr & 0.4245034 & 1.7868 $\pm$ 0.0145 & 1.05 &  1.22 & 0.0890 $\pm$ 0.0029 & 10.496 $\pm$ 0.006 & 10.179 $\pm$ 0.003 & 10.140 $\pm$ 0.003 & - & - \\
V675~Sgr$^{*}$  & 0.6422935 & 1.1993 $\pm$ 0.0188 & 0.75 & -5.51 & 0.0910 $\pm$ 0.0055 & 10.536 $\pm$ 0.002 & 10.337 $\pm$ 0.002 & 10.279 $\pm$ 0.003 & -2.47 $\pm$ 0.02 & 2 \\
V~Ind     & 0.4796017 & 1.5058 $\pm$ 0.0191 & 1.00 &  0.02 & 0.0400 $\pm$ 0.0009 & 10.076 $\pm$ 0.005 & 9.963 $\pm$ 0.002 & 9.985 $\pm$ 0.003 & -1.63 $\pm$ 0.03 & 1 \\
VW~Scl$^{a}$    & 0.5109117 & 0.9027 $\pm$ 0.0338 & 1.74 & 26.98 & 0.0140 $\pm$ 0.0011 & 11.106 $\pm$ 0.005 & 11.016 $\pm$ 0.004 & 11.043 $\pm$ 0.004 & -1.38 $\pm$ 0.23 & 1 \\
W~Tuc     & 0.6422382 & 0.6276 $\pm$ 0.0133 & 0.98 & -0.58 & 0.0180 $\pm$ 0.0002 & 11.504 $\pm$ 0.003 & 11.393 $\pm$ 0.003 & 11.371 $\pm$ 0.003 & -1.90 $\pm$ 0.08 & 1 \\
WY~Ant    & 0.5743427 & 0.9787 $\pm$ 0.0208 & 1.11 &  3.29 & 0.0550 $\pm$ 0.0009 & 10.995 $\pm$ 0.005 & 10.803 $\pm$ 0.004 & 10.744 $\pm$ 0.004 & -1.95 $\pm$ 0.06 & 1 \\
X~Ari     & 0.6511796 & 1.8690 $\pm$ 0.0188 & 1.22 &  4.48 & 0.1670 $\pm$ 0.0053 & 9.798 $\pm$ 0.026 & 9.452 $\pm$ 0.006 & 9.323 $\pm$ 0.008 & -2.59 $\pm$ 0.05 & 1 \\
X~Ret     & 0.49201 & 0.6590 $\pm$ 0.0144 & 1.38 &  9.78 & 0.0400 $\pm$ 0.0022 & 11.809 $\pm$ 0.009 & 11.663 $\pm$ 0.007 & 11.647 $\pm$ 0.006 & - & - \\
XZ~Gru    & 0.8831035 & 0.8700 $\pm$ 0.0185 & 1.12 &  2.76 & 0.0040 $\pm$ 0.0006 & 10.864 $\pm$ 0.004 & 10.647 $\pm$ 0.004 & 10.575 $\pm$ 0.004 & - & - \\
Z~Mic     & 0.5869258 & 0.8190 $\pm$ 0.0230 & 0.96 & -0.83 & 0.0820 $\pm$ 0.0028 & 11.792 $\pm$ 0.003 & 11.507 $\pm$ 0.003 & 11.397 $\pm$ 0.002 & -1.58 $\pm$ 0.05 & 1 \\
\hline
\multicolumn{11}{l}{RRc} \\ 
\hline
AE~Boo     & 0.3148921 & 1.1432 $\pm$ 0.0190 & 1.06 &  1.45 & 0.0230 $\pm$ 0.0012 & 10.722 $\pm$ 0.004 & 10.662 $\pm$ 0.003 & 10.681 $\pm$ 0.002 & -1.62 $\pm$ 0.09 & 1 \\
CS~Eri     & 0.311331 & 2.1665 $\pm$ 0.0158 & 1.05 &  1.38 & 0.0170 $\pm$ 0.0004 & 9.056 $\pm$ 0.006 & 9.002 $\pm$ 0.006 & 9.050 $\pm$ 0.007 & -1.97 $\pm$ 0.09 & 1 \\
EV~Psc     & 0.3062573 & 1.1328 $\pm$ 0.0300 & 1.12 &  3.21 & 0.0300 $\pm$ 0.0008 & 10.640 $\pm$ 0.003 & 10.555 $\pm$ 0.003 & 10.575 $\pm$ 0.003 & - & - \\
IY~Eri     & 0.375026 & 0.7626 $\pm$ 0.0133 & 0.96 & -1.10 & 0.0230 $\pm$ 0.0011 & 11.133 $\pm$ 0.005 & 11.067 $\pm$ 0.004 & 11.109 $\pm$ 0.003 & - & - \\
LS~Her     & 0.230808 & 1.0254 $\pm$ 0.0206 & 0.98 & -0.36 & 0.0360 $\pm$ 0.0009 & 10.977 $\pm$ 0.009 & 10.957 $\pm$ 0.007 & 11.006 $\pm$ 0.005 & - & - \\
MT~Tel$^{a}$     & 0.3169011 & 2.0704 $\pm$ 0.0301 & 1.65 & 11.93 & 0.0340 $\pm$ 0.0026 & 9.102 $\pm$ 0.006 & 9.085 $\pm$ 0.007 & 9.079 $\pm$ 0.004 & -2.60 $\pm$ 0.18 & 1 \\
RU~Psc$^{a}$     & 0.390385 & 1.2783 $\pm$ 0.0291 & 1.53 &  1.25 & 0.0390 $\pm$ 0.0002 & 10.333 $\pm$ 0.010 & 10.171 $\pm$ 0.006 & 10.157 $\pm$ 0.005 & - & - \\
RU~Sex     & 0.350232 & 0.9155 $\pm$ 0.0206 & 1.33 &  8.54 & 0.0270 $\pm$ 0.0006 & 10.885 $\pm$ 0.003 & 10.806 $\pm$ 0.002 & 10.821 $\pm$ 0.002 & -1.97 $\pm$ 0.10 & 2 \\
T~Sex     & 0.32469759 & 1.3400 $\pm$ 0.0225 & 1.16 &  3.16 & 0.0420 $\pm$ 0.0032 & 10.124 $\pm$ 0.004 & 10.075 $\pm$ 0.003 & 10.090 $\pm$ 0.003 & -1.49 $\pm$ 0.10 & 1 \\ 
\hline
\multicolumn{11}{@{}l@{}}{\parbox{0.96\textwidth}{
   Star: name of a~RR~Lyr star;
          Period: period of a~RR~Lyr star adopted from AAVSO database; 
          $\varpi_{DR3}$: parallax from the Gaia~DR3 catalog corrected with
          RUWE: renormalized unit weight error from the Gaia~DR3 catalog; 
          GOF: goodness-of-fit from the Gaia~DR3 catalog; 
          E(B-V): reddening value applied from \citet{SF2011} reddening map 
          corrected for the MW model by \citet{DrimmelSpergel2001}; 
          $<g>$, $<r>$, $<i>$: intensity-averaged mean magnitude from Fourier 
          series fitting for the Pan-STARRS $g_{P1}r_{P1}i_{P1}$ filters, respectively; 
          [Fe/H]: the adopted iron abundance;
          Source: source of iron abundance ($1$~- \citealp{Crestani2021a}; 
          $2$~- \citealp{Crestani2021b}). \\ 
          $^{a}$ Stars rejected based on the RUWE and GOF parallax quality parameters 
          given by the Gaia~DR3 catalog. \\ 
          $^{*}$ The SED showed that the Sloan $gri$ mean magnitudes are systematically 
          fainter than the photomertic compilation from \citet{Monson2017}. 
}}
\end{tabular}
\end{sidewaystable*}

%
\begin{table*}
\caption{Determined PL and PW relations for Galactic RR~Lyr stars of type RRab and RRc, 
         as well as for the combined population.}             
\label{tab:plr}      
\centering          
\begin{tabular}{ll|cccc|cccc}     
\hline\hline       
band & type & $a_{\lambda}$ & $b_{\lambda}$ & rms & N & 
$a_{\lambda}$ & $b_{\lambda}$ & rms & N  \\
\hline     
& & \multicolumn{4}{l}{(parallax):} & \multicolumn{4}{l}{(geometric distances):} \\ 
\hline
\multirow{3}{*}{$g$} & RRab     & -2.242 $\pm$ 0.376 & 0.805 $\pm$ 0.031 & 0.18 & 38 & -2.162 $\pm$ 0.382 & 0.802 $\pm$ 0.031 & 0.18 & 38 \\ 
                     & RRc      & -1.984 $\pm$ 0.760 & 0.605 $\pm$ 0.062 & 0.11 &  7 & 
-1.983 $\pm$ 0.746 & 0.604 $\pm$ 0.061 & 0.11 &  7 \\ 
                     & RRab+RRc & -1.500 $\pm$ 0.368 & 0.773 $\pm$ 0.034 & 0.21 & 45 & 
-1.448 $\pm$ 0.368 & 0.771 $\pm$ 0.034 & 0.21 & 45 \\ 
\hline
\multirow{3}{*}{$r$} & RRab     & -2.528 $\pm$ 0.300 & 0.660 $\pm$ 0.025 & 0.14 & 38 & 
-2.447 $\pm$ 0.308 & 0.657 $\pm$ 0.025 & 0.15 & 38 \\
                     & RRc      & -2.266 $\pm$ 0.705 & 0.557 $\pm$ 0.058 & 0.10 &  7 & 
-2.266 $\pm$ 0.694 & 0.556 $\pm$ 0.057 & 0.10 &  7 \\ 
                     & RRab+RRc & -1.929 $\pm$ 0.296 & 0.634 $\pm$ 0.028 & 0.17 & 45 & 
-1.878 $\pm$ 0.297 & 0.633 $\pm$ 0.028 & 0.17 & 45 \\ 
\hline
\multirow{3}{*}{$i$} & RRab     & -2.799 $\pm$ 0.296 & 0.642 $\pm$ 0.022 & 0.13 & 38 & 
-2.718 $\pm$ 0.278 & 0.639 $\pm$ 0.023 & 0.13 & 38 \\ 
                     & RRc      & -2.393 $\pm$ 0.647 & 0.598 $\pm$ 0.053 & 0.10 &  7 & 
-2.392 $\pm$ 0.637 & 0.597 $\pm$ 0.052 & 0.09 &  7 \\ 
                     & RRab+RRc & -2.249 $\pm$ 0.266 & 0.619 $\pm$ 0.025 & 0.15 & 45 & 
-2.197 $\pm$ 0.267 & 0.618 $\pm$ 0.025 & 0.15 & 45 \\ 
\hline      
\multirow{3}{*}{$W^{ri}_r$} & RRab     & -3.626 $\pm$ 0.218 & 0.587 $\pm$ 0.018 & 0.11 & 38 & -3.546 $\pm$ 0.229 & 0.584 $\pm$ 0.019 & 0.11 & 38 \\ 
                            & RRc      & -2.780 $\pm$ 0.515 & 0.723 $\pm$ 0.042 & 0.08 &  7 & -2.779 $\pm$ 0.508 & 0.722 $\pm$ 0.042 & 0.08 &  7 \\ 
                            & RRab+RRc & -3.226 $\pm$ 0.202 & 0.574 $\pm$ 0.019 & 0.12 & 45  & -3.174 $\pm$ 0.206 & 0.572 $\pm$ 0.019 & 0.12 & 45 \\
\hline 
\multirow{3}{*}{$W^{gr}_r$} & RRab     & -3.357 $\pm$ 0.231 & 0.239 $\pm$ 0.019 & 0.11 & 38 & -3.277 $\pm$ 0.237 & 0.237 $\pm$ 0.019 & 0.11 & 38 \\ 
                            & RRc      & -3.086 $\pm$ 0.643 & 0.415 $\pm$ 0.053 & 0.10 &  7 & -3.085 $\pm$ 0.644 & 0.414 $\pm$ 0.053 & 0.10 &  7 \\ 
                            & RRab+RRc & -3.175 $\pm$ 0.195 & 0.232 $\pm$ 0.018 & 0.11 & 45 & -3.123 $\pm$ 0.198 & 0.231 $\pm$ 0.018 & 0.11 & 45 \\ 
\hline
\multirow{3}{*}{$W^{gi}_g$} & RRab     & -3.508 $\pm$ 0.202 & 0.434 $\pm$ 0.016 & 0.10 & 38 & -3.427 $\pm$ 0.211 & 0.432 $\pm$ 0.017 & 0.10 & 38 \\ 
                            & RRc      & -2.914 $\pm$ 0.545 & 0.588 $\pm$ 0.045 & 0.08 &  7 & -2.913 $\pm$ 0.542 & 0.587 $\pm$ 0.044 & 0.08 &  7 \\ 
                            & RRab+RRc & -3.157 $\pm$ 0.182 & 0.424 $\pm$ 0.015 & 0.09 & 45 & -3.151 $\pm$ 0.185 & 0.423 $\pm$ 0.017 & 0.11 & 45 \\ 
\hline
& & \multicolumn{4}{l}{(ABL method):} & \multicolumn{4}{l}{(photo-geometric distances):} \\ 
\hline
\multirow{3}{*}{$g$} & RRab     & -2.318 $\pm$ 0.373 & 0.810 $\pm$ 0.032 & 0.18 & 38 & 
-2.179 $\pm$ 0.382 & 0.803 $\pm$ 0.031 & 0.18 & 38 \\ 
                     & RRc      & -1.862 $\pm$ 0.737 & 0.615 $\pm$ 0.066 & 0.11 &  7 & 
-2.035 $\pm$ 0.743 & 0.607 $\pm$ 0.061 & 0.11 &  7 \\ 
                     & RRab+RRc & -1.506 $\pm$ 0.372 & 0.783 $\pm$ 0.037 & 0.21 & 45 & 
-1.462 $\pm$ 0.367 & 0.771 $\pm$ 0.034 & 0.21 & 45 \\ 
\hline
\multirow{3}{*}{$r$} & RRab     & -2.594 $\pm$ 0.297 & 0.663 $\pm$ 0.026 & 0.14 & 38 & 
-2.465 $\pm$ 0.309 & 0.658 $\pm$ 0.025 & 0.15 & 38 \\ 
                     & RRc      & -2.143 $\pm$ 0.679 & 0.566 $\pm$ 0.061 & 0.10 &  7 & 
-2.317 $\pm$ 0.693 & 0.558 $\pm$ 0.057 & 0.10 &  7 \\ 
                     & RRab+RRc & -1.918 $\pm$ 0.300 & 0.641 $\pm$ 0.030 & 0.17 & 45 & 
-1.891 $\pm$ 0.296 & 0.633 $\pm$ 0.028 & 0.17 & 45 \\ 
\hline
\multirow{3}{*}{$i$} & RRab     & -2.870 $\pm$ 0.269 & 0.644 $\pm$ 0.023 & 0.13 & 38 & 
-2.736 $\pm$ 0.279 & 0.640 $\pm$ 0.023 & 0.13 & 38 \\  
                     & RRc      & -2.280 $\pm$ 0.621 & 0.606 $\pm$ 0.056 & 0.10 &  7 & 
-2.444 $\pm$ 0.634 & 0.599 $\pm$ 0.052 & 0.09 &  7 \\  
                     & RRab+RRc & -2.230 $\pm$ 0.272 & 0.626 $\pm$ 0.027 & 0.15 & 45 & 
-2.211 $\pm$ 0.265 & 0.618 $\pm$ 0.025 & 0.15 & 45 \\  
\hline
\multirow{3}{*}{$W^{ri}_r$} & RRab     & -3.700 $\pm$ 0.227 & 0.587 $\pm$ 0.020 & 0.11 & 38 & -3.563 $\pm$ 0.232 & 0.584 $\pm$ 0.019 & 0.11 & 38 \\ 
                            & RRc      & -2.707 $\pm$ 0.483 & 0.729 $\pm$ 0.045 & 0.08 &  7 & -2.830 $\pm$ 0.500 & 0.725 $\pm$ 0.041 & 0.07 &  7 \\ 
                            & RRab+RRc & -3.171 $\pm$ 0.215 & 0.580 $\pm$ 0.022 & 0.12 & 45 & -3.185 $\pm$ 0.208 & 0.572 $\pm$ 0.019 & 0.12 &  45 \\ 
\hline
\multirow{3}{*}{$W^{gr}_r$} & RRab     & -3.395 $\pm$ 0.237 & 0.241 $\pm$ 0.021 & 0.11 & 38 & -3.294 $\pm$ 0.243 & 0.237 $\pm$ 0.020 & 0.12 & 38 \\ 
                            & RRc      & -2.977 $\pm$ 0.602 & 0.423 $\pm$ 0.056 & 0.10 &  7 & -3.137 $\pm$ 0.650 & 0.416 $\pm$ 0.053 & 0.10 &  7 \\ 
                            & RRab+RRc & -3.146 $\pm$ 0.198 & 0.237 $\pm$ 0.021 & 0.11 & 45 & -3.083 $\pm$ 0.182 & 0.225 $\pm$ 0.017 & 0.10 &  44 \\ 
\hline
\multirow{3}{*}{$W^{gi}_g$} & RRab     & -3.561 $\pm$ 0.205 & 0.435 $\pm$ 0.018 & 0.10 & 38 & -3.445 $\pm$ 0.216 & 0.432 $\pm$ 0.018 & 0.10 & 38 \\ 
                            & RRc      & -2.827 $\pm$ 0.513 & 0.594 $\pm$ 0.048 & 0.08 &  7 & -2.965 $\pm$ 0.540 & 0.590 $\pm$ 0.044 & 0.08 &  7 \\ 
                            & RRab+RRc & -3.127 $\pm$ 0.166 & 0.422 $\pm$ 0.017 & 0.09 & 45 & -3.114 $\pm$ 0.169 & 0.417 $\pm$ 0.016 & 0.10 &  44 \\ 
\hline
\end{tabular}
\tablefoot{
band: the Pan-STARRS $g_{P1}r_{P1}i_{P1}$ bands and Wesenheit indicates constructed 
based on them using reddening vectors from \citet{Green2019}; 
type: variability type of RR~Lyr star; 
$a_{\lambda}$: slope of the fit; 
$b_{\lambda}$: zero point of the fit; 
rms: root mean square of derived relations; 
N: number of stars used for fitting. 
The logarithm of the pivot period used for fitting was $\rm{logP_0} = -0.25$ for 
RRab and RRab+RRc stars, and $\rm{logP_0} = -0.45$ for RRc type stars. 
Fundamentalization according to \citet{Iben1974}: $\rm{logP}_{ab} = \rm{logP}_{c} + 0.127$.
}
\end{table*}

%
\begin{table*}
\caption{Determined PLZ and PWZ relations for Galactic RR~Lyr stars of type RRab and 
        the RRab+RRc population using metallicities from \citet{Crestani2021a,Crestani2021b}.}             
\label{tab:plzr2}      
\centering          
\begin{tabular}{ll|ccccc}     
\hline\hline       
band & type & $a_{\lambda}$ & $b_{\lambda}$ & $c_{\lambda}$ & rms & N  \\
\hline     
\multicolumn{5}{l}{(parallax method):} \\ 
\hline
\multirow{2}{*}{$g$} & RRab     & -0.527 $\pm$ 0.602 & 0.794 $\pm$ 0.028 &  0.264 $\pm$ 0.069 & 0.14 & 31 \\ 
                     & RRab+RRc & -0.284 $\pm$ 0.439 & 0.791 $\pm$ 0.028 &  0.289 $\pm$ 0.053 & 0.14 & 35 \\ 
\hline
\multirow{2}{*}{$r$} & RRab     & -1.230 $\pm$ 0.493 & 0.651 $\pm$ 0.023 &  0.205 $\pm$ 0.056 & 0.12 & 31 \\
                     & RRab+RRc & -1.017 $\pm$ 0.362 & 0.650 $\pm$ 0.023 &  0.228 $\pm$ 0.044 & 0.12 & 35 \\ 
\hline
\multirow{2}{*}{$i$} & RRab     & -1.682 $\pm$ 0.442 & 0.635 $\pm$ 0.021 &  0.174 $\pm$ 0.050 & 0.11 & 31 \\ 
                     & RRab+RRc & -1.469 $\pm$ 0.328 & 0.633 $\pm$ 0.021 &  0.198 $\pm$ 0.040 & 0.11 & 35 \\ 
\hline
\multirow{2}{*}{$W^{ri}_r$} & RRab     & -3.061 $\pm$ 0.370 & 0.584 $\pm$ 0.017 &  0.082 $\pm$ 0.042 & 0.09 & 31 \\ 
                            & RRab+RRc & -2.848 $\pm$ 0.275 & 0.583 $\pm$ 0.018 &  0.105 $\pm$ 0.033 & 0.09 & 35 \\ 
\hline
\multirow{2}{*}{$W^{gr}_r$} & RRab     & -3.273 $\pm$ 0.415 & 0.239 $\pm$ 0.020 & 0.033 $\pm$ 0.047 & 0.10 & 31 \\ 
                            & RRab+RRc & -3.148 $\pm$ 0.305 & 0.238 $\pm$ 0.019 & 0.050 $\pm$ 0.037 & 0.10 & 35 \\ 
\hline 
\multirow{2}{*}{$W^{gi}_g$} & RRab     & -3.154 $\pm$ 0.368 & 0.433 $\pm$ 0.017 & 0.060 $\pm$ 0.042 & 0.09 & 31 \\ 
                            & RRab+RRc & -2.979 $\pm$ 0.271 & 0.432 $\pm$ 0.017 & 0.081 $\pm$ 0.033 & 0.09 & 35 \\ 
\hline                  
\multicolumn{5}{l}{(ABL method):} \\ 
\hline
\multirow{2}{*}{$g$} & RRab     & -0.503 $\pm$ 0.550 & 0.798 $\pm$ 0.028 &  0.266 $\pm$ 0.063 & 0.14 & 31 \\ 
                     & RRab+RRc & -0.290 $\pm$ 0.420 & 0.795 $\pm$ 0.028 &  0.290 $\pm$ 0.049 & 0.14 & 35 \\ 
\hline 
\multirow{2}{*}{$r$} & RRab     & -1.227 $\pm$ 0.457 & 0.655 $\pm$ 0.023 &  0.204 $\pm$ 0.052 & 0.12 & 31 \\ 
                     & RRab+RRc & -1.031 $\pm$ 0.349 & 0.652 $\pm$ 0.023 &  0.227 $\pm$ 0.041 & 0.12 & 35 \\ 
\hline
\multirow{2}{*}{$i$} & RRab     & -1.673 $\pm$ 0.413 & 0.638 $\pm$ 0.021 &  0.175 $\pm$ 0.047 & 0.11 & 31 \\  
                     & RRab+RRc & -1.479 $\pm$ 0.317 & 0.635 $\pm$ 0.022 &  0.197 $\pm$ 0.037 & 0.11 & 35 \\  
\hline
\multirow{2}{*}{$W^{ri}_r$} & RRab     & -3.027 $\pm$ 0.362 & 0.587 $\pm$ 0.019 &  0.084 $\pm$ 0.042 & 0.09 & 31 \\ 
                            & RRab+RRc & -2.837 $\pm$ 0.274 & 0.585 $\pm$ 0.020 &  0.107 $\pm$ 0.032 & 0.09 & 35 \\ 
\hline 
\multirow{2}{*}{$W^{gr}_r$} & RRab     & -3.347 $\pm$ 0.404 & 0.241 $\pm$ 0.021 & 0.019 $\pm$ 0.047 & 0.10 & 31 \\ 
                            & RRab+RRc & -3.178 $\pm$ 0.303 & 0.240 $\pm$ 0.022 & 0.042 $\pm$ 0.036 & 0.10 & 35 \\ 
\hline
\multirow{2}{*}{$W^{gi}_g$} & RRab     & -3.169 $\pm$ 0.362 & 0.435 $\pm$ 0.019 & 0.055 $\pm$ 0.042 & 0.09 & 31 \\ 
                            & RRab+RRc & -2.987 $\pm$ 0.273 & 0.433 $\pm$ 0.019 & 0.078 $\pm$ 0.032 & 0.09 & 35 \\ 
                            \hline
\multicolumn{5}{l}{(geometric distances):} \\ 
\hline
\multirow{2}{*}{$g$} & RRab     & -0.389 $\pm$ 0.617 & 0.792 $\pm$ 0.029 &  0.267 $\pm$ 0.070 & 0.15 & 31 \\ 
                     & RRab+RRc & -0.217 $\pm$ 0.447 & 0.790 $\pm$ 0.029 &  0.285 $\pm$ 0.054 & 0.15 & 35 \\ 
\hline
\multirow{2}{*}{$r$} & RRab     & -1.092 $\pm$ 0.509 & 0.650 $\pm$ 0.024 &  0.208 $\pm$ 0.058 & 0.12 & 31 \\
                     & RRab+RRc & -0.950 $\pm$ 0.371 & 0.648 $\pm$ 0.024 &  0.223 $\pm$ 0.045 & 0.12 & 35 \\ 
\hline
\multirow{2}{*}{$i$} & RRab     & -1.544 $\pm$ 0.461 & 0.633 $\pm$ 0.022 &  0.177 $\pm$ 0.053 & 0.11 & 31 \\ 
                     & RRab+RRc & -1.402 $\pm$ 0.338 & 0.632 $\pm$ 0.022 &  0.193 $\pm$ 0.041 & 0.11 & 35 \\ 
\hline
\multirow{2}{*}{$W^{ri}_r$} & RRab     & -2.923 $\pm$ 0.393 & 0.583 $\pm$ 0.019 &  0.085 $\pm$ 0.045 & 0.09 & 31 \\ 
                            & RRab+RRc & -2.824 $\pm$ 0.300 & 0.580 $\pm$ 0.019 &  0.096 $\pm$ 0.036 & 0.10 & 35 \\ 
\hline 
\multirow{2}{*}{$W^{gr}_r$} & RRab     & -3.135 $\pm$ 0.430 & 0.237 $\pm$ 0.020 & 0.036 $\pm$ 0.049 & 0.10 & 31 \\ 
                            & RRab+RRc & -3.124 $\pm$ 0.327 & 0.236 $\pm$ 0.021 & 0.040 $\pm$ 0.039 & 0.11 & 35 \\ 
\hline
\multirow{2}{*}{$W^{gi}_g$} & RRab     & -3.016 $\pm$ 0.388 & 0.431 $\pm$ 0.018 & 0.063 $\pm$ 0.044 & 0.09 & 31 \\ 
                            & RRab+RRc & -2.955 $\pm$ 0.297 & 0.429 $\pm$ 0.019 & 0.071 $\pm$ 0.036 & 0.10 & 35 \\ 
                            \hline
\multicolumn{5}{l}{(photo-geometric distances):} \\ 
\hline
\multirow{2}{*}{$g$} & RRab     & -0.473 $\pm$ 0.623 & 0.791 $\pm$ 0.030 &  0.257 $\pm$ 0.072 & 0.15 & 31 \\ 
                     & RRab+RRc & -0.260 $\pm$ 0.456 & 0.789 $\pm$ 0.029 &  0.280 $\pm$ 0.055 & 0.15 & 35 \\ 
\hline
\multirow{2}{*}{$r$} & RRab     & -1.177 $\pm$ 0.525 & 0.650 $\pm$ 0.025 &  0.198 $\pm$ 0.060 & 0.13 & 31 \\
                     & RRab+RRc & -0.993 $\pm$ 0.382 & 0.647 $\pm$ 0.024 &  0.218 $\pm$ 0.046 & 0.12 & 35 \\ 
\hline
\multirow{2}{*}{$i$} & RRab     & -1.629 $\pm$ 0.478 & 0.633 $\pm$ 0.023 &  0.168 $\pm$ 0.054 & 0.11 & 31 \\ 
                     & RRab+RRc & -1.445 $\pm$ 0.349 & 0.631 $\pm$ 0.022 &  0.188 $\pm$ 0.042 & 0.11 & 35 \\ 
\hline
\multirow{2}{*}{$W^{ri}_r$} & RRab     & -3.008 $\pm$ 0.411 & 0.582 $\pm$ 0.019 &  0.075 $\pm$ 0.047 & 0.10 & 31 \\ 
                            & RRab+RRc & -2.835 $\pm$ 0.299 & 0.580 $\pm$ 0.019 &  0.095 $\pm$ 0.036 & 0.10 & 35 \\ 
\hline 
\multirow{2}{*}{$W^{gr}_r$} & RRab     & -3.220 $\pm$ 0.450 & 0.236 $\pm$ 0.021 &  0.026 $\pm$ 0.051 & 0.11 & 31 \\ 
                            & RRab+RRc & -3.135 $\pm$ 0.328 & 0.236 $\pm$ 0.021 &  0.039 $\pm$ 0.040 & 0.11 & 35 \\ 
\hline
\multirow{2}{*}{$W^{gi}_g$} & RRab     & -3.101 $\pm$ 0.408 & 0.430 $\pm$ 0.019 &  0.054 $\pm$ 0.047 & 0.10 & 31 \\ 
                            & RRab+RRc & -2.967 $\pm$ 0.296 & 0.430 $\pm$ 0.019 &  0.070 $\pm$ 0.036 & 0.10 & 35 \\ 
\hline
\end{tabular}
\tablefoot{
band: the Pan-STARRS $g_{P1}r_{P1}i_{P1}$ bands and Wesenheit indicates constructed using 
reddening vectors from \citet{Green2019}; 
type: varibility type of RR~Lyr star; 
$a_{\lambda}$: slope of the fit; 
$b_{\lambda}$: zero point of the fit; 
$c_{\lambda}$: metallicity coefficient;
rms: a~root mean square of derived relations; 
N: number of stars used for fitting. 
Pivot logarithm used for fitting was $\rm{logP_0} = -0.25$ for RRab and RRab+RRc stars. 
Pivot metallicity values was chosen to be $\rm{[Fe/H]_0} = -1.5$. 
Fundamentalization according to Iben (1974): $\rm{logP}_{ab} = \rm{logP}_{c} + 0.127$.
}
\end{table*}

%
\begin{table}
\caption{Shifts of $\log P$ in the fundamentalization formula determined for RRc stars.}             
\label{tab:dlogP}      
\centering          
\begin{tabular}{l c | l c }     
\hline\hline       
band & $\Delta \log P$ & band & $\Delta \log P$ \\ 
     & (mag)           &      & (mag)           \\ 
\hline     
\multicolumn{2}{l}{(parallax):} & \multicolumn{2}{l}{(geometric distances):} \\ 
\hline
$g$ & 0.296 $\pm$ 0.034 & $g$ & 0.297 $\pm$ 0.036 \\
\hline
$r$ & 0.247 $\pm$ 0.023 & $r$ & 0.246 $\pm$ 0.025 \\ 
\hline
$i$ & 0.223 $\pm$ 0.019 & $i$ & 0.222 $\pm$ 0.020 \\ 
\hline
$W^{ri}_r$ & 0.174 $\pm$ 0.013 & $W^{ri}_r$ & 0.171 $\pm$ 0.014 \\ 
\hline
$W^{gr}_r$ & 0.151 $\pm$ 0.015 & $W^{gr}_r$ & 0.149 $\pm$ 0.015 \\ 
\hline
$W^{gi}_g$ & 0.164 $\pm$ 0.012 & $W^{gi}_g$ & 0.162 $\pm$ 0.013 \\ 
\hline                  
\multicolumn{2}{l}{(ABL method):} & \multicolumn{2}{l}{(photo-geometric distances):} \\ 
\hline
$g$ & 0.298 $\pm$ 0.035 & $g$ & 0.294 $\pm$ 0.035 \\ 
\hline
$r$ & 0.249 $\pm$ 0.023 & $r$ & 0.244 $\pm$ 0.024 \\ 
\hline
$i$ & 0.226 $\pm$ 0.019 & $i$ & 0.220 $\pm$ 0.020 \\    
\hline 
$W^{ri}_r$ & 0.178 $\pm$ 0.012 & $W^{ri}_r$ & 0.170 $\pm$ 0.014 \\ 
\hline 
$W^{gr}_r$ & 0.153 $\pm$ 0.014 & $W^{gr}_r$ & 0.148 $\pm$ 0.016 \\ 
\hline 
$W^{gi}_g$ & 0.167 $\pm$ 0.012 & $W^{gi}_g$ & 0.161 $\pm$ 0.013 \\ 
\hline
\end{tabular}
\tablefoot{
band: the Sloan--Pan-STARRS $g_{P1}r_{P1}i_{P1}$ bands and Wesenheit indices; 
$\Delta \log P$: a~new shift of $\log P$ for fundamentalization.
}
\end{table}

%
%

%
%

%
%

%
%

%
%

%
%

%
%

\begin{acknowledgements}
      We thank the anonymous referee for valuable comments to the content of the paper.
      We thank prof. Pawel Moskalik for valuable discussion about the RRd stars. 
      The research leading to these results has received funding from the European 
      Research Council (ERC) under the European Union’s Horizon 2020 research and 
      innovation program (grant agreement No. 695099). We also acknowledge support 
      from the National Science Center, Poland grants MAESTRO UMO-2017/26/A/ST9/00446, 
      BEETHOVEN UMO-2018/31/G/ST9/03050 and DIR/WK/2018/09 grants of the Polish Ministry 
      of Science and Higher Education. We also acknowledge financial support from 
      UniverScale grant financed by the European Union’s Horizon 2020 research and 
      innovation programme under the grant agreement number 951549. We gratefully 
      acknowledge financial support for this work from the BASAL Centro de 
      Astrofisica y Tecnologias Afines (CATA) AFB-170002 and the Millenium Institute 
      of Astrophysics (MAS) of the Iniciativa Cientifica Milenio del Ministerio 
      de Economia, Fomento y Turismo de Chile, project IC120009. W.G. also gratefully 
      acknowledges support from the ANID BASAL project ACE210002. 
      P.W. gratefully acknowledges financial support from the Polish National 
      Science Center grant PRELUDIUM 2018/31/N/ST9/02742. 
      This work has made use of data from the European Space Agency (ESA) mission 
      Gaia (\url{https://www.cosmos.esa.int/gaia}), processed by the Gaia Data Processing 
      and Analysis Consortium (DPAC, \url{https://www.cosmos.esa.int/web/gaia/dpac/
      consortium}). Funding for the DPAC has been provided by national institutions, 
      in particular the institutions participating in the Gaia Multilateral Agreement. 
      This publication makes use of VOSA, developed under the Spanish Virtual Observatory 
      (\url{https://svo.cab.inta-csic.es}) project funded by MCIN/AEI/10.13039/501100011033/ 
      through grant PID2020-112949GB-I00. VOSA has been partially updated by using funding 
      from the European Union's Horizon 2020 Research and Innovation Programme, under Grant 
      Agreement n 776403 (EXOPLANETS-A). 
      
      Facilities: LCOGT (0.4m).
      
      Software used in this work: {\tt gaiadr3\_zero-point} \citep{LindegrenBastian2021}, 
          IRAF \citep{Tody1986,Tody1993}, 
          DAOPHOT \citep{Stetson1987}, 
          Astropy \citep{Astropy2013},  
          Sklearn \citep{Pedregosa2011}
          NumPy \citep{Numpy1,Numpy2}, 
          SciPy \citep{Virtanen2020}, 
          Matplotlib \citep{Hunter2007}, 
          VOSA \citep{Bayo2008}.
\end{acknowledgements}

%
%

\bibliographystyle{aa} 
\bibliography{ms_aa} 

\begin{thebibliography}{71}
\expandafter\ifx\csname natexlab\endcsname\relax\def\natexlab#1{#1}\fi

\bibitem[{{Abazajian} {et~al.}(2003){Abazajian}, {Adelman-McCarthy},
  {Ag{\"u}eros}, {Allam}, {Anderson}, {Annis}, {Bahcall}, {Baldry}, {Bastian},
  {Berlind}, {Bernardi}, {Blanton}, {Blythe}, {Bochanski}, {Boroski},
  {Brewington}, {Briggs}, {Brinkmann}, {Brunner}, {Budav{\'a}ri}, {Carey},
  {Carr}, {Castander}, {Chiu}, {Collinge}, {Connolly}, {Covey}, {Csabai},
  {Dalcanton}, {Dodelson}, {Doi}, {Dong}, {Eisenstein}, {Evans}, {Fan},
  {Feldman}, {Finkbeiner}, {Friedman}, {Frieman}, {Fukugita}, {Gal},
  {Gillespie}, {Glazebrook}, {Gonzalez}, {Gray}, {Grebel}, {Grodnicki}, {Gunn},
  {Gurbani}, {Hall}, {Hao}, {Harbeck}, {Harris}, {Harris}, {Harvanek},
  {Hawley}, {Heckman}, {Helmboldt}, {Hendry}, {Hennessy}, {Hindsley}, {Hogg},
  {Holmgren}, {Holtzman}, {Homer}, {Hui}, {Ichikawa}, {Ichikawa}, {Inkmann},
  {Ivezi{\'c}}, {Jester}, {Johnston}, {Jordan}, {Jordan}, {Jorgensen},
  {Juri{\'c}}, {Kauffmann}, {Kent}, {Kleinman}, {Knapp}, {Kniazev}, {Kron},
  {Krzesi{\'n}ski}, {Kunszt}, {Kuropatkin}, {Lamb}, {Lampeitl}, {Laubscher},
  {Lee}, {Leger}, {Li}, {Lidz}, {Lin}, {Loh}, {Long}, {Loveday}, {Lupton},
  {Malik}, {Margon}, {McGehee}, {McKay}, {Meiksin}, {Miknaitis}, {Moorthy},
  {Munn}, {Murphy}, {Nakajima}, {Narayanan}, {Nash}, {Neilsen}, {Newberg},
  {Newman}, {Nichol}, {Nicinski}, {Nieto-Santisteban}, {Nitta}, {Odenkirchen},
  {Okamura}, {Ostriker}, {Owen}, {Padmanabhan}, {Peoples}, {Pier}, {Pindor},
  {Pope}, {Quinn}, {Rafikov}, {Raymond}, {Richards}, {Richmond}, {Rix},
  {Rockosi}, {Schaye}, {Schlegel}, {Schneider}, {Schroeder}, {Scranton},
  {Sekiguchi}, {Seljak}, {Sergey}, {Sesar}, {Sheldon}, {Shimasaku}, {Siegmund},
  {Silvestri}, {Sinisgalli}, {Sirko}, {Smith}, {Smol{\v{c}}i{\'c}}, {Snedden},
  {Stebbins}, {Steinhardt}, {Stinson}, {Stoughton}, {Strateva}, {Strauss},
  {SubbaRao}, {Szalay}, {Szapudi}, {Szkody}, {Tasca}, {Tegmark}, {Thakar},
  {Tremonti}, {Tucker}, {Uomoto}, {Vanden Berk}, {Vandenberg}, {Vogeley},
  {Voges}, {Vogt}, {Walkowicz}, {Weinberg}, {West}, {White}, {Wilhite},
  {Willman}, {Xu}, {Yanny}, {Yarger}, {Yasuda}, {Yip}, {Yocum}, {York},
  {Zakamska}, {Zehavi}, {Zheng}, {Zibetti}, \& {Zucker}}]{Abazajian2003}
{Abazajian}, K., {Adelman-McCarthy}, J.~K., {Ag{\"u}eros}, M.~A., {et~al.}
  2003, \aj, 126, 2081

\bibitem[{{Abbott} {et~al.}(2021){Abbott}, {Adam{\'o}w}, {Aguena}, {Allam},
  {Amon}, {Annis}, {Avila}, {Bacon}, {Banerji}, {Bechtol}, {Becker},
  {Bernstein}, {Bertin}, {Bhargava}, {Bridle}, {Brooks}, {Burke}, {Carnero
  Rosell}, {Carrasco Kind}, {Carretero}, {Castander}, {Cawthon}, {Chang},
  {Choi}, {Conselice}, {Costanzi}, {Crocce}, {da Costa}, {Davis}, {De Vicente},
  {DeRose}, {Desai}, {Diehl}, {Dietrich}, {Drlica-Wagner}, {Eckert},
  {Elvin-Poole}, {Everett}, {Evrard}, {Ferrero}, {Fert{\'e}}, {Flaugher},
  {Fosalba}, {Friedel}, {Frieman}, {Garc{\'\i}a-Bellido}, {Gaztanaga},
  {Gelman}, {Gerdes}, {Giannantonio}, {Gill}, {Gruen}, {Gruendl}, {Gschwend},
  {Gutierrez}, {Hartley}, {Hinton}, {Hollowood}, {Honscheid}, {Huterer},
  {James}, {Jeltema}, {Johnson}, {Kent}, {Kron}, {Kuehn}, {Kuropatkin},
  {Lahav}, {Li}, {Lidman}, {Lin}, {MacCrann}, {Maia}, {Manning}, {Maloney},
  {March}, {Marshall}, {Martini}, {Melchior}, {Menanteau}, {Miquel}, {Morgan},
  {Myles}, {Neilsen}, {Ogando}, {Palmese}, {Paz-Chinch{\'o}n}, {Petravick},
  {Pieres}, {Plazas}, {Pond}, {Rodriguez-Monroy}, {Romer}, {Roodman}, {Rykoff},
  {Sako}, {Sanchez}, {Santiago}, {Scarpine}, {Serrano}, {Sevilla-Noarbe},
  {Smith}, {Smith}, {Soares-Santos}, {Suchyta}, {Swanson}, {Tarle}, {Thomas},
  {To}, {Tremblay}, {Troxel}, {Tucker}, {Turner}, {Varga}, {Walker},
  {Wechsler}, {Weller}, {Wester}, {Wilkinson}, {Yanny}, {Zhang}, {Nikutta},
  {Fitzpatrick}, {Jacques}, {Scott}, {Olsen}, {Huang}, {Herrera}, {Juneau},
  {Nidever}, {Weaver}, {Adean}, {Correia}, {de Freitas}, {Freitas},
  {Singulani}, {Vila-Verde}, \& {Linea Science Server}}]{Abbott2021}
{Abbott}, T.~M.~C., {Adam{\'o}w}, M., {Aguena}, M., {et~al.} 2021, \apjs, 255,
  20

\bibitem[{{Arenou} \& {Luri}(1999)}]{AL1999}
{Arenou}, F. \& {Luri}, X. 1999, in Astronomical Society of the Pacific
  Conference Series, Vol. 167, Harmonizing Cosmic Distance Scales in a
  Post-HIPPARCOS Era, ed. D.~{Egret} \& A.~{Heck}, 13--32

\bibitem[{{Astropy Collaboration} {et~al.}(2013){Astropy Collaboration},
  {Robitaille}, {Tollerud}, {Greenfield}, {Droettboom}, {Bray}, {Aldcroft},
  {Davis}, {Ginsburg}, {Price-Whelan}, {Kerzendorf}, {Conley}, {Crighton},
  {Barbary}, {Muna}, {Ferguson}, {Grollier}, {Parikh}, {Nair}, {Unther},
  {Deil}, {Woillez}, {Conseil}, {Kramer}, {Turner}, {Singer}, {Fox}, {Weaver},
  {Zabalza}, {Edwards}, {Azalee Bostroem}, {Burke}, {Casey}, {Crawford},
  {Dencheva}, {Ely}, {Jenness}, {Labrie}, {Lim}, {Pierfederici}, {Pontzen},
  {Ptak}, {Refsdal}, {Servillat}, \& {Streicher}}]{Astropy2013}
{Astropy Collaboration}, {Robitaille}, T.~P., {Tollerud}, E.~J., {et~al.} 2013,
  \aap, 558, A33

\bibitem[{{Bailer-Jones} {et~al.}(2021){Bailer-Jones}, {Rybizki}, {Fouesneau},
  {Demleitner}, \& {Andrae}}]{BJ2021}
{Bailer-Jones}, C.~A.~L., {Rybizki}, J., {Fouesneau}, M., {Demleitner}, M., \&
  {Andrae}, R. 2021, \aj, 161, 147

\bibitem[{{Bailey}(1902)}]{Bailey1902}
{Bailey}, S.~I. 1902, Annals of Harvard College Observatory, 38, 1

\bibitem[{{Bailey} \& {Pickering}(1913)}]{BP1913}
{Bailey}, S.~I. \& {Pickering}, E.~C. 1913, Annals of Harvard College
  Observatory, 78, 1

\bibitem[{{Baumgardt} \& {Vasiliev}(2021)}]{BaumgardtVasiliev2021}
{Baumgardt}, H. \& {Vasiliev}, E. 2021, \mnras, 505, 5957

\bibitem[{{Bayo} {et~al.}(2008){Bayo}, {Rodrigo}, {Barrado Y Navascu{\'e}s},
  {Solano}, {Guti{\'e}rrez}, {Morales-Calder{\'o}n}, \& {Allard}}]{Bayo2008}
{Bayo}, A., {Rodrigo}, C., {Barrado Y Navascu{\'e}s}, D., {et~al.} 2008, \aap,
  492, 277

\bibitem[{{Bellm} {et~al.}(2019){Bellm}, {Kulkarni}, {Barlow}, {Feindt},
  {Graham}, {Goobar}, {Kupfer}, {Ngeow}, {Nugent}, {Ofek}, {Prince}, {Riddle},
  {Walters}, \& {Ye}}]{Bellm2019}
{Bellm}, E.~C., {Kulkarni}, S.~R., {Barlow}, T., {et~al.} 2019, \pasp, 131,
  068003

\bibitem[{Bellm {et~al.}(2018)Bellm, Kulkarni, Graham, Dekany, Smith, Riddle,
  Masci, Helou, Prince, Adams, Barbarino, Barlow, Bauer, Beck, Belicki, Biswas,
  Blagorodnova, Bodewits, Bolin, Brinnel, Brooke, Bue, Bulla, Burruss, Cenko,
  Chang, Connolly, Coughlin, Cromer, Cunningham, De, Delacroix, Desai, Duev,
  Eadie, Farnham, Feeney, Feindt, Flynn, Franckowiak, Frederick, Fremling,
  Gal-Yam, Gezari, Giomi, Goldstein, Golkhou, Goobar, Groom, Hacopians, Hale,
  Henning, Ho, Hover, Howell, Hung, Huppenkothen, Imel, Ip, Željko Ivezić,
  Jackson, Jones, Juric, Kasliwal, Kaspi, Kaye, Kelley, Kowalski, Kramer,
  Kupfer, Landry, Laher, Lee, Lin, Lin, Lunnan, Giomi, Mahabal, Mao, Miller,
  Monkewitz, Murphy, Ngeow, Nordin, Nugent, Ofek, Patterson, Penprase, Porter,
  Rauch, Rebbapragada, Reiley, Rigault, Rodriguez, van Roestel, Rusholme, van
  Santen, Schulze, Shupe, Singer, Soumagnac, Stein, Surace, Sollerman, Szkody,
  Taddia, Terek, Sistine, van Velzen, Vestrand, Walters, Ward, Ye, Yu, Yan, \&
  Zolkower}]{Bellm2018}
Bellm, E.~C., Kulkarni, S.~R., Graham, M.~J., {et~al.} 2018, Publications of
  the Astronomical Society of the Pacific, 131, 018002

\bibitem[{{Bhardwaj}(2024)}]{Bhardwaj2024}
{Bhardwaj}, A. 2024, IAU Symposium, 376, 250

\bibitem[{{Bhardwaj} {et~al.}(2023){Bhardwaj}, {Marconi}, {Rejkuba}, {de
  Grijs}, {Singh}, {Braga}, {Kanbur}, {Ngeow}, {Ripepi}, {Bono}, {De Somma}, \&
  {Dall'Ora}}]{Bhardwaj2023}
{Bhardwaj}, A., {Marconi}, M., {Rejkuba}, M., {et~al.} 2023, \apjl, 944, L51

\bibitem[{{Bhardwaj} {et~al.}(2021){Bhardwaj}, {Rejkuba}, {Sloan}, {Marconi},
  \& {Yang}}]{Bhardwaj2021}
{Bhardwaj}, A., {Rejkuba}, M., {Sloan}, G.~C., {Marconi}, M., \& {Yang}, S.-C.
  2021, \apj, 922, 20

\bibitem[{{Braga} {et~al.}(2018){Braga}, {Stetson}, {Bono}, {Dall'Ora},
  {Ferraro}, {Fiorentino}, {Iannicola}, {Marconi}, {Marengo}, {Monson},
  {Neeley}, {Persson}, {Beaton}, {Buonanno}, {Calamida}, {Castellani}, {Di
  Carlo}, {Fabrizio}, {Freedman}, {Inno}, {Madore}, {Magurno}, {Marchetti},
  {Marinoni}, {Marrese}, {Matsunaga}, {Minniti}, {Monelli}, {Nonino},
  {Piersimoni}, {Pietrinferni}, {Prada-Moroni}, {Pulone}, {Stellingwerf},
  {Tognelli}, {Walker}, {Valenti}, \& {Zoccali}}]{Braga2018}
{Braga}, V.~F., {Stetson}, P.~B., {Bono}, G., {et~al.} 2018, \aj, 155, 137

\bibitem[{Breuval {et~al.}(2021)Breuval, Kervella, Wielgórski, Gieren,
  Graczyk, Trahin, Pietrzyński, Arenou, Javanmardi, \& Zgirski}]{Breuval2021}
Breuval, L., Kervella, P., Wielgórski, P., {et~al.} 2021, The Astrophysical
  Journal, 913, 38

\bibitem[{{C{\'a}ceres} \& {Catelan}(2008)}]{CaceresCatelan2008}
{C{\'a}ceres}, C. \& {Catelan}, M. 2008, \apjs, 179, 242

\bibitem[{{Catelan}(2009)}]{Catelan2009}
{Catelan}, M. 2009, \apss, 320, 261

\bibitem[{{Crestani} {et~al.}(2021{\natexlab{a}}){Crestani}, {Braga},
  {Fabrizio}, {Bono}, {Sneden}, {Preston}, {Ferraro}, {Iannicola}, {Nonino},
  {Fiorentino}, {Th{\'e}venin}, {Lemasle}, {Prudil}, {Alves-Brito},
  {Altavilla}, {Chaboyer}, {Dall'Ora}, {D'Orazi}, {Gilligan}, {Grebel},
  {Koch-Hansen}, {Lala}, {Marengo}, {Marinoni}, {Marrese},
  {Mart{\'\i}nez-V{\'a}zquez}, {Matsunaga}, {Monelli}, {Mullen}, {Neeley}, {da
  Silva}, {Stetson}, {Salaris}, {Storm}, {Valenti}, \&
  {Zoccali}}]{Crestani2021b}
{Crestani}, J., {Braga}, V.~F., {Fabrizio}, M., {et~al.} 2021{\natexlab{a}},
  \apj, 914, 10

\bibitem[{{Crestani} {et~al.}(2021{\natexlab{b}}){Crestani}, {Fabrizio},
  {Braga}, {Sneden}, {Preston}, {Ferraro}, {Iannicola}, {Bono}, {Alves-Brito},
  {Nonino}, {D'Orazi}, {Inno}, {Monelli}, {Storm}, {Altavilla}, {Chaboyer},
  {Dall'Ora}, {Fiorentino}, {Gilligan}, {Grebel}, {Lala}, {Lemasle}, {Marengo},
  {Marinoni}, {Marrese}, {Mart{\'\i}nez-V{\'a}zquez}, {Matsunaga}, {Mullen},
  {Neeley}, {Prudil}, {da Silva}, {Stetson}, {Th{\'e}venin}, {Valenti},
  {Walker}, \& {Zoccali}}]{Crestani2021a}
{Crestani}, J., {Fabrizio}, M., {Braga}, V.~F., {et~al.} 2021{\natexlab{b}},
  \apj, 908, 20

\bibitem[{Dekany {et~al.}(2020)Dekany, Smith, Riddle, Feeney, Porter, Hale,
  Zolkower, Belicki, Kaye, Henning, Walters, Cromer, Delacroix, Rodriguez,
  Reiley, Mao, Hover, Murphy, Burruss, Baker, Kowalski, Reif, Mueller, Bellm,
  Graham, \& Kulkarni}]{Dekany2020}
Dekany, R., Smith, R.~M., Riddle, R., {et~al.} 2020, Publications of the
  Astronomical Society of the Pacific, 132, 038001

\bibitem[{{Demers} \& {Wehlau}(1977)}]{DW1977}
{Demers}, S. \& {Wehlau}, A. 1977, \aj, 82, 620

\bibitem[{{Drimmel} \& {Spergel}(2001)}]{DrimmelSpergel2001}
{Drimmel}, R. \& {Spergel}, D.~N. 2001, \apj, 556, 181

\bibitem[{{Feast} \& {Catchpole}(1997)}]{FC1997}
{Feast}, M.~W. \& {Catchpole}, R.~M. 1997, \mnras, 286, L1

\bibitem[{{Flaugher} {et~al.}(2015){Flaugher}, {Diehl}, {Honscheid}, {Abbott},
  {Alvarez}, {Angstadt}, {Annis}, {Antonik}, {Ballester}, {Beaufore},
  {Bernstein}, {Bernstein}, {Bigelow}, {Bonati}, {Boprie}, {Brooks},
  {Buckley-Geer}, {Campa}, {Cardiel-Sas}, {Castander}, {Castilla}, {Cease},
  {Cela-Ruiz}, {Chappa}, {Chi}, {Cooper}, {da Costa}, {Dede}, {Derylo},
  {DePoy}, {de Vicente}, {Doel}, {Drlica-Wagner}, {Eiting}, {Elliott}, {Emes},
  {Estrada}, {Fausti Neto}, {Finley}, {Flores}, {Frieman}, {Gerdes},
  {Gladders}, {Gregory}, {Gutierrez}, {Hao}, {Holland}, {Holm}, {Huffman},
  {Jackson}, {James}, {Jonas}, {Karcher}, {Karliner}, {Kent}, {Kessler},
  {Kozlovsky}, {Kron}, {Kubik}, {Kuehn}, {Kuhlmann}, {Kuk}, {Lahav}, {Lathrop},
  {Lee}, {Levi}, {Lewis}, {Li}, {Mandrichenko}, {Marshall}, {Martinez},
  {Merritt}, {Miquel}, {Mu{\~n}oz}, {Neilsen}, {Nichol}, {Nord}, {Ogando},
  {Olsen}, {Palaio}, {Patton}, {Peoples}, {Plazas}, {Rauch}, {Reil}, {Rheault},
  {Roe}, {Rogers}, {Roodman}, {Sanchez}, {Scarpine}, {Schindler}, {Schmidt},
  {Schmitt}, {Schubnell}, {Schultz}, {Schurter}, {Scott}, {Serrano}, {Shaw},
  {Smith}, {Soares-Santos}, {Stefanik}, {Stuermer}, {Suchyta}, {Sypniewski},
  {Tarle}, {Thaler}, {Tighe}, {Tran}, {Tucker}, {Walker}, {Wang}, {Watson},
  {Weaverdyck}, {Wester}, {Woods}, {Yanny}, \& {DES
  Collaboration}}]{Flaugher2015}
{Flaugher}, B., {Diehl}, H.~T., {Honscheid}, K., {et~al.} 2015, \aj, 150, 150

\bibitem[{{Fukugita} {et~al.}(1996){Fukugita}, {Ichikawa}, {Gunn}, {Doi},
  {Shimasaku}, \& {Schneider}}]{Fukugita1996}
{Fukugita}, M., {Ichikawa}, T., {Gunn}, J.~E., {et~al.} 1996, \aj, 111, 1748

\bibitem[{{Gaia Collaboration} {et~al.}(2023){Gaia Collaboration}, {Vallenari},
  {Brown}, {Prusti}, {de Bruijne}, {Arenou}, {Babusiaux}, {Biermann},
  {Creevey}, {Ducourant}, {Evans}, {Eyer}, {Guerra}, {Hutton}, {Jordi},
  {Klioner}, {Lammers}, {Lindegren}, {Luri}, {Mignard}, {Panem}, {Pourbaix},
  {Randich}, {Sartoretti}, {Soubiran}, {Tanga}, {Walton}, {Bailer-Jones},
  {Bastian}, {Drimmel}, {Jansen}, {Katz}, {Lattanzi}, {van Leeuwen}, {Bakker},
  {Cacciari}, {Casta{\~n}eda}, {De Angeli}, {Fabricius}, {Fouesneau},
  {Fr{\'e}mat}, {Galluccio}, {Guerrier}, {Heiter}, {Masana}, {Messineo},
  {Mowlavi}, {Nicolas}, {Nienartowicz}, {Pailler}, {Panuzzo}, {Riclet}, {Roux},
  {Seabroke}, {Sordo}, {Th{\'e}venin}, {Gracia-Abril}, {Portell}, {Teyssier},
  {Altmann}, {Andrae}, {Audard}, {Bellas-Velidis}, {Benson}, {Berthier},
  {Blomme}, {Burgess}, {Busonero}, {Busso}, {C{\'a}novas}, {Carry}, {Cellino},
  {Cheek}, {Clementini}, {Damerdji}, {Davidson}, {de Teodoro}, {Nu{\~n}ez
  Campos}, {Delchambre}, {Dell'Oro}, {Esquej}, {Fern{\'a}ndez-Hern{\'a}ndez},
  {Fraile}, {Garabato}, {Garc{\'\i}a-Lario}, {Gosset}, {Haigron}, {Halbwachs},
  {Hambly}, {Harrison}, {Hern{\'a}ndez}, {Hestroffer}, {Hodgkin}, {Holl},
  {Jan{\ss}en}, {Jevardat de Fombelle}, {Jordan}, {Krone-Martins}, {Lanzafame},
  {L{\"o}ffler}, {Marchal}, {Marrese}, {Moitinho}, {Muinonen}, {Osborne},
  {Pancino}, {Pauwels}, {Recio-Blanco}, {Reyl{\'e}}, {Riello}, {Rimoldini},
  {Roegiers}, {Rybizki}, {Sarro}, {Siopis}, {Smith}, {Sozzetti}, {Utrilla},
  {van Leeuwen}, {Abbas}, {{\'A}brah{\'a}m}, {Abreu Aramburu}, {Aerts},
  {Aguado}, {Ajaj}, {Aldea-Montero}, {Altavilla}, {{\'A}lvarez}, {Alves},
  {Anders}, {Anderson}, {Anglada Varela}, {Antoja}, {Baines}, {Baker},
  {Balaguer-N{\'u}{\~n}ez}, {Balbinot}, {Balog}, {Barache}, {Barbato},
  {Barros}, {Barstow}, {Bartolom{\'e}}, {Bassilana}, {Bauchet}, {Becciani},
  {Bellazzini}, {Berihuete}, {Bernet}, {Bertone}, {Bianchi}, {Binnenfeld},
  {Blanco-Cuaresma}, {Blazere}, {Boch}, {Bombrun}, {Bossini}, {Bouquillon},
  {Bragaglia}, {Bramante}, {Breedt}, {Bressan}, {Brouillet}, {Brugaletta},
  {Bucciarelli}, {Burlacu}, {Butkevich}, {Buzzi}, {Caffau}, {Cancelliere},
  {Cantat-Gaudin}, {Carballo}, {Carlucci}, {Carnerero}, {Carrasco},
  {Casamiquela}, {Castellani}, {Castro-Ginard}, {Chaoul}, {Charlot}, {Chemin},
  {Chiaramida}, {Chiavassa}, {Chornay}, {Comoretto}, {Contursi}, {Cooper},
  {Cornez}, {Cowell}, {Crifo}, {Cropper}, {Crosta}, {Crowley}, {Dafonte},
  {Dapergolas}, {David}, {David}, {de Laverny}, {De Luise}, {De March}, {De
  Ridder}, {de Souza}, {de Torres}, {del Peloso}, {del Pozo}, {Delbo},
  {Delgado}, {Delisle}, {Demouchy}, {Dharmawardena}, {Di Matteo}, {Diakite},
  {Diener}, {Distefano}, {Dolding}, {Edvardsson}, {Enke}, {Fabre}, {Fabrizio},
  {Faigler}, {Fedorets}, {Fernique}, {Fienga}, {Figueras}, {Fournier},
  {Fouron}, {Fragkoudi}, {Gai}, {Garcia-Gutierrez}, {Garcia-Reinaldos},
  {Garc{\'\i}a-Torres}, {Garofalo}, {Gavel}, {Gavras}, {Gerlach}, {Geyer},
  {Giacobbe}, {Gilmore}, {Girona}, {Giuffrida}, {Gomel}, {Gomez},
  {Gonz{\'a}lez-N{\'u}{\~n}ez}, {Gonz{\'a}lez-Santamar{\'\i}a},
  {Gonz{\'a}lez-Vidal}, {Granvik}, {Guillout}, {Guiraud},
  {Guti{\'e}rrez-S{\'a}nchez}, {Guy}, {Hatzidimitriou}, {Hauser}, {Haywood},
  {Helmer}, {Helmi}, {Sarmiento}, {Hidalgo}, {Hilger}, {H{\l}adczuk}, {Hobbs},
  {Holland}, {Huckle}, {Jardine}, {Jasniewicz}, {Jean-Antoine Piccolo},
  {Jim{\'e}nez-Arranz}, {Jorissen}, {Juaristi Campillo}, {Julbe}, {Karbevska},
  {Kervella}, {Khanna}, {Kontizas}, {Kordopatis}, {Korn}, {K{\'o}sp{\'a}l},
  {Kostrzewa-Rutkowska}, {Kruszy{\'n}ska}, {Kun}, {Laizeau}, {Lambert},
  {Lanza}, {Lasne}, {Le Campion}, {Lebreton}, {Lebzelter}, {Leccia}, {Leclerc},
  {Lecoeur-Taibi}, {Liao}, {Licata}, {Lindstr{\o}m}, {Lister}, {Livanou},
  {Lobel}, {Lorca}, {Loup}, {Madrero Pardo}, {Magdaleno Romeo}, {Managau},
  {Mann}, {Manteiga}, {Marchant}, {Marconi}, {Marcos}, {Marcos Santos},
  {Mar{\'\i}n Pina}, {Marinoni}, {Marocco}, {Marshall}, {Martin Polo},
  {Mart{\'\i}n-Fleitas}, {Marton}, {Mary}, {Masip}, {Massari},
  {Mastrobuono-Battisti}, {Mazeh}, {McMillan}, {Messina}, {Michalik}, {Millar},
  {Mints}, {Molina}, {Molinaro}, {Moln{\'a}r}, {Monari}, {Mongui{\'o}},
  {Montegriffo}, {Montero}, {Mor}, {Mora}, {Morbidelli}, {Morel}, {Morris},
  {Muraveva}, {Murphy}, {Musella}, {Nagy}, {Noval}, {Oca{\~n}a}, {Ogden},
  {Ordenovic}, {Osinde}, {Pagani}, {Pagano}, {Palaversa}, {Palicio},
  {Pallas-Quintela}, {Panahi}, {Payne-Wardenaar}, {Pe{\~n}alosa Esteller},
  {Penttil{\"a}}, {Pichon}, {Piersimoni}, {Pineau}, {Plachy}, {Plum}, {Poggio},
  {Pr{\v{s}}a}, {Pulone}, {Racero}, {Ragaini}, {Rainer}, {Raiteri}, {Rambaux},
  {Ramos}, {Ramos-Lerate}, {Re Fiorentin}, {Regibo}, {Richards}, {Rios Diaz},
  {Ripepi}, {Riva}, {Rix}, {Rixon}, {Robichon}, {Robin}, {Robin}, {Roelens},
  {Rogues}, {Rohrbasser}, {Romero-G{\'o}mez}, {Rowell}, {Royer}, {Ruz Mieres},
  {Rybicki}, {Sadowski}, {S{\'a}ez N{\'u}{\~n}ez}, {Sagrist{\`a} Sell{\'e}s},
  {Sahlmann}, {Salguero}, {Samaras}, {Sanchez Gimenez}, {Sanna},
  {Santove{\~n}a}, {Sarasso}, {Schultheis}, {Sciacca}, {Segol}, {Segovia},
  {S{\'e}gransan}, {Semeux}, {Shahaf}, {Siddiqui}, {Siebert}, {Siltala},
  {Silvelo}, {Slezak}, {Slezak}, {Smart}, {Snaith}, {Solano}, {Solitro},
  {Souami}, {Souchay}, {Spagna}, {Spina}, {Spoto}, {Steele},
  {Steidelm{\"u}ller}, {Stephenson}, {S{\"u}veges}, {Surdej}, {Szabados},
  {Szegedi-Elek}, {Taris}, {Taylor}, {Teixeira}, {Tolomei}, {Tonello}, {Torra},
  {Torra}, {Torralba Elipe}, {Trabucchi}, {Tsounis}, {Turon}, {Ulla}, {Unger},
  {Vaillant}, {van Dillen}, {van Reeven}, {Vanel}, {Vecchiato}, {Viala},
  {Vicente}, {Voutsinas}, {Weiler}, {Wevers}, {Wyrzykowski}, {Yoldas}, {Yvard},
  {Zhao}, {Zorec}, {Zucker}, \& {Zwitter}}]{GaiaCol2023}
{Gaia Collaboration}, {Vallenari}, A., {Brown}, A.~G.~A., {et~al.} 2023, \aap,
  674, A1

\bibitem[{{Green} {et~al.}(2019){Green}, {Schlafly}, {Zucker}, {Speagle}, \&
  {Finkbeiner}}]{Green2019}
{Green}, G.~M., {Schlafly}, E., {Zucker}, C., {Speagle}, J.~S., \&
  {Finkbeiner}, D. 2019, \apj, 887, 93

\bibitem[{{Groenewegen}(2021)}]{Groenewegen2021}
{Groenewegen}, M.~A.~T. 2021, \aap, 654, A20

\bibitem[{{Harris} {et~al.}(2020){Harris}, {Millman}, {van der Walt},
  {Gommers}, {Virtanen}, {Cournapeau}, {Wieser}, {Taylor}, {Berg}, {Smith},
  {Kern}, {Picus}, {Hoyer}, {van Kerkwijk}, {Brett}, {Haldane}, {del R{\'\i}o},
  {Wiebe}, {Peterson}, {G{\'e}rard-Marchant}, {Sheppard}, {Reddy}, {Weckesser},
  {Abbasi}, {Gohlke}, \& {Oliphant}}]{Numpy2}
{Harris}, C.~R., {Millman}, K.~J., {van der Walt}, S.~J., {et~al.} 2020, \nat,
  585, 357

\bibitem[{{Hunter}(2007)}]{Hunter2007}
{Hunter}, J.~D. 2007, Computing in Science and Engineering, 9, 90

\bibitem[{{Iben}(1974)}]{Iben1974}
{Iben}, I., J. 1974, \araa, 12, 215

\bibitem[{{Iben} \& {Huchra}(1971)}]{IbenHuchra1971}
{Iben}, I., J. \& {Huchra}, J. 1971, \aap, 14, 293

\bibitem[{{Ivezi{\'c}} {et~al.}(2019){Ivezi{\'c}}, {Kahn}, {Tyson}, {Abel},
  {Acosta}, {Allsman}, {Alonso}, {AlSayyad}, {Anderson}, {Andrew}, {Angel},
  {Angeli}, {Ansari}, {Antilogus}, {Araujo}, {Armstrong}, {Arndt}, {Astier},
  {Aubourg}, {Auza}, {Axelrod}, {Bard}, {Barr}, {Barrau}, {Bartlett}, {Bauer},
  {Bauman}, {Baumont}, {Bechtol}, {Bechtol}, {Becker}, {Becla}, {Beldica},
  {Bellavia}, {Bianco}, {Biswas}, {Blanc}, {Blazek}, {Blandford}, {Bloom},
  {Bogart}, {Bond}, {Booth}, {Borgland}, {Borne}, {Bosch}, {Boutigny},
  {Brackett}, {Bradshaw}, {Brandt}, {Brown}, {Bullock}, {Burchat}, {Burke},
  {Cagnoli}, {Calabrese}, {Callahan}, {Callen}, {Carlin}, {Carlson},
  {Chandrasekharan}, {Charles-Emerson}, {Chesley}, {Cheu}, {Chiang}, {Chiang},
  {Chirino}, {Chow}, {Ciardi}, {Claver}, {Cohen-Tanugi}, {Cockrum}, {Coles},
  {Connolly}, {Cook}, {Cooray}, {Covey}, {Cribbs}, {Cui}, {Cutri}, {Daly},
  {Daniel}, {Daruich}, {Daubard}, {Daues}, {Dawson}, {Delgado}, {Dellapenna},
  {de Peyster}, {de Val-Borro}, {Digel}, {Doherty}, {Dubois},
  {Dubois-Felsmann}, {Durech}, {Economou}, {Eifler}, {Eracleous}, {Emmons},
  {Fausti Neto}, {Ferguson}, {Figueroa}, {Fisher-Levine}, {Focke}, {Foss},
  {Frank}, {Freemon}, {Gangler}, {Gawiser}, {Geary}, {Gee}, {Geha}, {Gessner},
  {Gibson}, {Gilmore}, {Glanzman}, {Glick}, {Goldina}, {Goldstein}, {Goodenow},
  {Graham}, {Gressler}, {Gris}, {Guy}, {Guyonnet}, {Haller}, {Harris},
  {Hascall}, {Haupt}, {Hernandez}, {Herrmann}, {Hileman}, {Hoblitt}, {Hodgson},
  {Hogan}, {Howard}, {Huang}, {Huffer}, {Ingraham}, {Innes}, {Jacoby}, {Jain},
  {Jammes}, {Jee}, {Jenness}, {Jernigan}, {Jevremovi{\'c}}, {Johns}, {Johnson},
  {Johnson}, {Jones}, {Juramy-Gilles}, {Juri{\'c}}, {Kalirai}, {Kallivayalil},
  {Kalmbach}, {Kantor}, {Karst}, {Kasliwal}, {Kelly}, {Kessler}, {Kinnison},
  {Kirkby}, {Knox}, {Kotov}, {Krabbendam}, {Krughoff}, {Kub{\'a}nek},
  {Kuczewski}, {Kulkarni}, {Ku}, {Kurita}, {Lage}, {Lambert}, {Lange},
  {Langton}, {Le Guillou}, {Levine}, {Liang}, {Lim}, {Lintott}, {Long},
  {Lopez}, {Lotz}, {Lupton}, {Lust}, {MacArthur}, {Mahabal}, {Mandelbaum},
  {Markiewicz}, {Marsh}, {Marshall}, {Marshall}, {May}, {McKercher}, {McQueen},
  {Meyers}, {Migliore}, {Miller}, {Mills}, {Miraval}, {Moeyens}, {Moolekamp},
  {Monet}, {Moniez}, {Monkewitz}, {Montgomery}, {Morrison}, {Mueller},
  {Muller}, {Mu{\~n}oz Arancibia}, {Neill}, {Newbry}, {Nief}, {Nomerotski},
  {Nordby}, {O'Connor}, {Oliver}, {Olivier}, {Olsen}, {O'Mullane}, {Ortiz},
  {Osier}, {Owen}, {Pain}, {Palecek}, {Parejko}, {Parsons}, {Pease},
  {Peterson}, {Peterson}, {Petravick}, {Libby Petrick}, {Petry},
  {Pierfederici}, {Pietrowicz}, {Pike}, {Pinto}, {Plante}, {Plate}, {Plutchak},
  {Price}, {Prouza}, {Radeka}, {Rajagopal}, {Rasmussen}, {Regnault}, {Reil},
  {Reiss}, {Reuter}, {Ridgway}, {Riot}, {Ritz}, {Robinson}, {Roby}, {Roodman},
  {Rosing}, {Roucelle}, {Rumore}, {Russo}, {Saha}, {Sassolas}, {Schalk},
  {Schellart}, {Schindler}, {Schmidt}, {Schneider}, {Schneider}, {Schoening},
  {Schumacher}, {Schwamb}, {Sebag}, {Selvy}, {Sembroski}, {Seppala}, {Serio},
  {Serrano}, {Shaw}, {Shipsey}, {Sick}, {Silvestri}, {Slater}, {Smith},
  {Smith}, {Sobhani}, {Soldahl}, {Storrie-Lombardi}, {Stover}, {Strauss},
  {Street}, {Stubbs}, {Sullivan}, {Sweeney}, {Swinbank}, {Szalay}, {Takacs},
  {Tether}, {Thaler}, {Thayer}, {Thomas}, {Thornton}, {Thukral}, {Tice},
  {Trilling}, {Turri}, {Van Berg}, {Vanden Berk}, {Vetter}, {Virieux},
  {Vucina}, {Wahl}, {Walkowicz}, {Walsh}, {Walter}, {Wang}, {Wang}, {Warner},
  {Wiecha}, {Willman}, {Winters}, {Wittman}, {Wolff}, {Wood-Vasey}, {Wu},
  {Xin}, {Yoachim}, \& {Zhan}}]{Ivezic2019}
{Ivezi{\'c}}, {\v{Z}}., {Kahn}, S.~M., {Tyson}, J.~A., {et~al.} 2019, \apj,
  873, 111

\bibitem[{{Jerzykiewicz} \& {Wenzel}(1977)}]{JW1977}
{Jerzykiewicz}, M. \& {Wenzel}, W. 1977, \actaa, 27, 35

\bibitem[{{Kaiser} {et~al.}(2010){Kaiser}, {Burgett}, {Chambers}, {Denneau},
  {Heasley}, {Jedicke}, {Magnier}, {Morgan}, {Onaka}, \& {Tonry}}]{Kaiser2010}
{Kaiser}, N., {Burgett}, W., {Chambers}, K., {et~al.} 2010, in Society of
  Photo-Optical Instrumentation Engineers (SPIE) Conference Series, Vol. 7733,
  Ground-based and Airborne Telescopes III, ed. L.~M. {Stepp}, R.~{Gilmozzi},
  \& H.~J. {Hall}, 77330E

\bibitem[{{Kapteyn}(1890)}]{Kapteyn1890}
{Kapteyn}, J.~C. 1890, Astronomische Nachrichten, 125, 165

\bibitem[{{Karczmarek} {et~al.}(2015){Karczmarek}, {Pietrzy{\'n}ski}, {Gieren},
  {Suchomska}, {Konorski}, {G{\'o}rski}, {Pilecki}, {Graczyk}, \&
  {Wielg{\'o}rski}}]{Karczmarek2015}
{Karczmarek}, P., {Pietrzy{\'n}ski}, G., {Gieren}, W., {et~al.} 2015, \aj, 150,
  90

\bibitem[{{Karczmarek} {et~al.}(2017){Karczmarek}, {Pietrzy{\'n}ski},
  {G{\'o}rski}, {Gieren}, \& {Bersier}}]{Karczmarek2017}
{Karczmarek}, P., {Pietrzy{\'n}ski}, G., {G{\'o}rski}, M., {Gieren}, W., \&
  {Bersier}, D. 2017, \aj, 154, 263

\bibitem[{{Koll{\'a}th} {et~al.}(2002){Koll{\'a}th}, {Buchler}, {Szab{\'o}}, \&
  {Csubry}}]{Kollth2002}
{Koll{\'a}th}, Z., {Buchler}, J.~R., {Szab{\'o}}, R., \& {Csubry}, Z. 2002,
  \aap, 385, 932

\bibitem[{{Lindegren} {et~al.}(2021){Lindegren}, {Bastian}, {Biermann},
  {Bombrun}, {de Torres}, {Gerlach}, {Geyer}, {Hern{\'a}ndez}, {Hilger},
  {Hobbs}, {Klioner}, {Lammers}, {McMillan}, {Ramos-Lerate},
  {Steidelm{\"u}ller}, {Stephenson}, \& {van Leeuwen}}]{LindegrenBastian2021}
{Lindegren}, L., {Bastian}, U., {Biermann}, M., {et~al.} 2021, \aap, 649, A4

\bibitem[{{Madore}(1982)}]{Madore1982}
{Madore}, B.~F. 1982, \apj, 253, 575

\bibitem[{{Majaess} {et~al.}(2012){Majaess}, {Turner}, {Gieren}, \&
  {Lane}}]{Majaess2012}
{Majaess}, D., {Turner}, D., {Gieren}, W., \& {Lane}, D. 2012, \apjl, 752, L10

\bibitem[{{Marconi} {et~al.}(2006){Marconi}, {Cignoni}, {Di Criscienzo},
  {Ripepi}, {Castelli}, {Musella}, \& {Ruoppo}}]{Marconi2006}
{Marconi}, M., {Cignoni}, M., {Di Criscienzo}, M., {et~al.} 2006, \mnras, 371,
  1503

\bibitem[{{Marconi} {et~al.}(2022){Marconi}, {Molinaro}, {Dall'Ora}, {Ripepi},
  {Musella}, {Bono}, {Braga}, {Di Criscienzo}, {Fiorentino}, {Leccia}, \&
  {Monelli}}]{Marconi2022}
{Marconi}, M., {Molinaro}, R., {Dall'Ora}, M., {et~al.} 2022, \apj, 934, 29

\bibitem[{{Monson} {et~al.}(2017){Monson}, {Beaton}, {Scowcroft}, {Freedman},
  {Madore}, {Rich}, {Seibert}, {Kollmeier}, \& {Clementini}}]{Monson2017}
{Monson}, A.~J., {Beaton}, R.~L., {Scowcroft}, V., {et~al.} 2017, \aj, 153, 96

\bibitem[{{Narloch} {et~al.}(2023){Narloch}, {Hajdu}, {Pietrzy{\'n}ski},
  {Gieren}, {Wielg{\'o}rski}, {Zgirski}, {Karczmarek}, {G{\'o}rski}, \&
  {Graczyk}}]{Narloch2023}
{Narloch}, W., {Hajdu}, G., {Pietrzy{\'n}ski}, G., {et~al.} 2023, \apj, 953, 14

\bibitem[{{Navarrete} {et~al.}(2017){Navarrete}, {Catelan}, {Contreras Ramos},
  {Alonso-Garc{\'\i}a}, {Gran}, {D{\'e}k{\'a}ny}, \& {Minniti}}]{Navarrete2017}
{Navarrete}, C., {Catelan}, M., {Contreras Ramos}, R., {et~al.} 2017, \aap,
  604, A120

\bibitem[{{Nemec}(1985)}]{Nemec1985}
{Nemec}, J.~M. 1985, \aj, 90, 204

\bibitem[{{Nemec} {et~al.}(2024){Nemec}, {Linnell Nemec}, {Moskalik},
  {Moln{\'a}r}, {Plachy}, {Szab{\'o}}, \& {Kolenberg}}]{Nemec2024}
{Nemec}, J.~M., {Linnell Nemec}, A.~F., {Moskalik}, P., {et~al.} 2024, \mnras,
  529, 296

\bibitem[{{Ngeow} {et~al.}(2022){Ngeow}, {Bhardwaj}, {Dekany}, {Duev},
  {Graham}, {Groom}, {Mahabal}, {Masci}, {Medford}, \& {Riddle}}]{Ngeow2022RRL}
{Ngeow}, C.-C., {Bhardwaj}, A., {Dekany}, R., {et~al.} 2022, \aj, 163, 239

\bibitem[{{Pedregosa} {et~al.}(2011){Pedregosa}, {Varoquaux}, {Gramfort},
  {Michel}, {Thirion}, {Grisel}, {Blondel}, {M{\"u}ller}, {Nothman}, {Louppe},
  {Prettenhofer}, {Weiss}, {Dubourg}, {Vanderplas}, {Passos}, {Cournapeau},
  {Brucher}, {Perrot}, \& {Duchesnay}}]{Pedregosa2011}
{Pedregosa}, F., {Varoquaux}, G., {Gramfort}, A., {et~al.} 2011, Journal of
  Machine Learning Research, 12, 2825

\bibitem[{{Pickering}(1901)}]{Pickering1901}
{Pickering}, E.~C. 1901, Astronomische Nachrichten, 154, 423

\bibitem[{{Pietrzy{\'n}ski} {et~al.}(2008){Pietrzy{\'n}ski}, {Gieren},
  {Szewczyk}, {Walker}, {Rizzi}, {Bresolin}, {Kudritzki}, {Nalewajko}, {Storm},
  {Dall'Ora}, \& {Ivanov}}]{Pietrzynski2008}
{Pietrzy{\'n}ski}, G., {Gieren}, W., {Szewczyk}, O., {et~al.} 2008, \aj, 135,
  1993

\bibitem[{{Pritzl} {et~al.}(2005){Pritzl}, {Venn}, \& {Irwin}}]{Pritzl2005}
{Pritzl}, B.~J., {Venn}, K.~A., \& {Irwin}, M. 2005, \aj, 130, 2140

\bibitem[{{Riess} {et~al.}(2021){Riess}, {Casertano}, {Yuan}, {Bowers},
  {Macri}, {Zinn}, \& {Scolnic}}]{Riess2021}
{Riess}, A.~G., {Casertano}, S., {Yuan}, W., {et~al.} 2021, \apjl, 908, L6

\bibitem[{{Sandage} {et~al.}(1981){Sandage}, {Katem}, \& {Sandage}}]{SKS1981}
{Sandage}, A., {Katem}, B., \& {Sandage}, M. 1981, \apjs, 46, 41

\bibitem[{{Schlafly} \& {Finkbeiner}(2011)}]{SF2011}
{Schlafly}, E.~F. \& {Finkbeiner}, D.~P. 2011, \apj, 737, 103

\bibitem[{{Sesar} {et~al.}(2017){Sesar}, {Hernitschek}, {Mitrovi{\'c}},
  {Ivezi{\'c}}, {Rix}, {Cohen}, {Bernard}, {Grebel}, {Martin}, {Schlafly},
  {Burgett}, {Draper}, {Flewelling}, {Kaiser}, {Kudritzki}, {Magnier},
  {Metcalfe}, {Tonry}, \& {Waters}}]{Sesar2017}
{Sesar}, B., {Hernitschek}, N., {Mitrovi{\'c}}, S., {et~al.} 2017, \aj, 153,
  204

\bibitem[{{Shapley}(1918)}]{Shapley1918}
{Shapley}, H. 1918, \pasp, 30, 42

\bibitem[{{Stetson}(1987)}]{Stetson1987}
{Stetson}, P.~B. 1987, \pasp, 99, 191

\bibitem[{{Suchomska} {et~al.}(2015){Suchomska}, {Graczyk}, {Smolec},
  {Pietrzy{\'n}ski}, {Gieren}, {St{\c{e}}pie{\'n}}, {Konorski}, {Pilecki},
  {Villanova}, {Thompson}, {G{\'o}rski}, {Karczmarek}, {Wielg{\'o}rski}, \&
  {Anderson}}]{Suchomska2015}
{Suchomska}, K., {Graczyk}, D., {Smolec}, R., {et~al.} 2015, \mnras, 451, 651

\bibitem[{{Tody}(1986)}]{Tody1986}
{Tody}, D. 1986, in Society of Photo-Optical Instrumentation Engineers (SPIE)
  Conference Series, Vol. 627, Instrumentation in astronomy VI, ed. D.~L.
  {Crawford}, 733

\bibitem[{{Tody}(1993)}]{Tody1993}
{Tody}, D. 1993, in Astronomical Society of the Pacific Conference Series,
  Vol.~52, Astronomical Data Analysis Software and Systems II, ed. R.~J.
  {Hanisch}, R.~J.~V. {Brissenden}, \& J.~{Barnes}, 173

\bibitem[{{Tonry} {et~al.}(2018){Tonry}, {Denneau}, {Flewelling}, {Heinze},
  {Onken}, {Smartt}, {Stalder}, {Weiland}, \& {Wolf}}]{Tonry2018}
{Tonry}, J.~L., {Denneau}, L., {Flewelling}, H., {et~al.} 2018, \apj, 867, 105

\bibitem[{{Tonry} {et~al.}(2012){Tonry}, {Stubbs}, {Lykke}, {Doherty},
  {Shivvers}, {Burgett}, {Chambers}, {Hodapp}, {Kaiser}, {Kudritzki},
  {Magnier}, {Morgan}, {Price}, \& {Wainscoat}}]{Tonry2012}
{Tonry}, J.~L., {Stubbs}, C.~W., {Lykke}, K.~R., {et~al.} 2012, \apj, 750, 99

\bibitem[{{van der Walt} {et~al.}(2011){van der Walt}, {Colbert}, \&
  {Varoquaux}}]{Numpy1}
{van der Walt}, S., {Colbert}, S.~C., \& {Varoquaux}, G. 2011, Computing in
  Science and Engineering, 13, 22

\bibitem[{{Virtanen} {et~al.}(2020){Virtanen}, {Gommers}, {Oliphant},
  {Haberland}, {Reddy}, {Cournapeau}, {Burovski}, {Peterson}, {Weckesser},
  {Bright}, {van der Walt}, {Brett}, {Wilson}, {Millman}, {Mayorov}, {Nelson},
  {Jones}, {Kern}, {Larson}, {Carey}, {Polat}, {Feng}, {Moore}, {VanderPlas},
  {Laxalde}, {Perktold}, {Cimrman}, {Henriksen}, {Quintero}, {Harris},
  {Archibald}, {Ribeiro}, {Pedregosa}, {van Mulbregt}, \& {SciPy 1. 0
  Contributors}}]{Virtanen2020}
{Virtanen}, P., {Gommers}, R., {Oliphant}, T.~E., {et~al.} 2020, Nature
  Methods, 17, 261

\bibitem[{{Vivas} {et~al.}(2017){Vivas}, {Saha}, {Olsen}, {Blum}, {Olszewski},
  {Claver}, {Valdes}, {Axelrod}, {Kaleida}, {Kunder}, {Narayan}, {Matheson}, \&
  {Walker}}]{Vivas2017}
{Vivas}, A.~K., {Saha}, A., {Olsen}, K., {et~al.} 2017, \aj, 154, 85

\bibitem[{Wielgórski {et~al.}(2022)Wielgórski, Pietrzyński, Pilecki, Gieren,
  Zgirski, Górski, Hajdu, Narloch, Karczmarek, Smolec, Kervella, Storm,
  Gallenne, Breuval, Lewis, Kałuszyński, Graczyk, Pych, Suchomska, Taormina,
  Garcia, Kotek, Chini, Nũnez, Noroozi, Figaredo, Haas, Hodapp, Mikołajczyk,
  Kotysz, Moździerski, \& Kołaczek-Szymański}]{Wielgorski2022}
Wielgórski, P., Pietrzyński, G., Pilecki, B., {et~al.} 2022, The
  Astrophysical Journal, 927, 89

\bibitem[{{Zgirski} {et~al.}(2023){Zgirski}, {Pietrzy{\'n}ski}, {G{\'o}rski},
  {Gieren}, {Wielg{\'o}rski}, {Karczmarek}, {Hajdu}, {Lewis}, {Chini},
  {Graczyk}, {Ka{\l}uszy{\'n}ski}, {Narloch}, {Pilecki}, {Garc{\'\i}a},
  {Suchomska}, \& {Taormina}}]{Zgirski2023}
{Zgirski}, B., {Pietrzy{\'n}ski}, G., {G{\'o}rski}, M., {et~al.} 2023, \apj,
  951, 114

\end{thebibliography}

%
%
%
%
%
%
%
%
%

%

\begin{appendix} 

\section{The Sloan band light curves of Galactic RR~Lyr stars analazyed in this work} 
\label{app:lc}

%
Figure~\ref{fig:fig1} presents the Sloan--Pan-STARSS $g_{P1}r_{P1}i_{P1}$ bands light 
curves of $44$ RRab stars used in this study. Figure~\ref{fig:fig2} shows the analogous 
light curves for nine RRc stars. 
The figures are also available at Zenodo data repository: 
\url{https://doi.org/10.5281/zenodo.11565605}. 
The presented light curves are available at the webpage of the Araucaria Project: 
\url{https://araucaria.camk.edu.pl/} and the CDS.

\begin{figure*}[h]
    \centering
    \includegraphics[width=\textwidth]{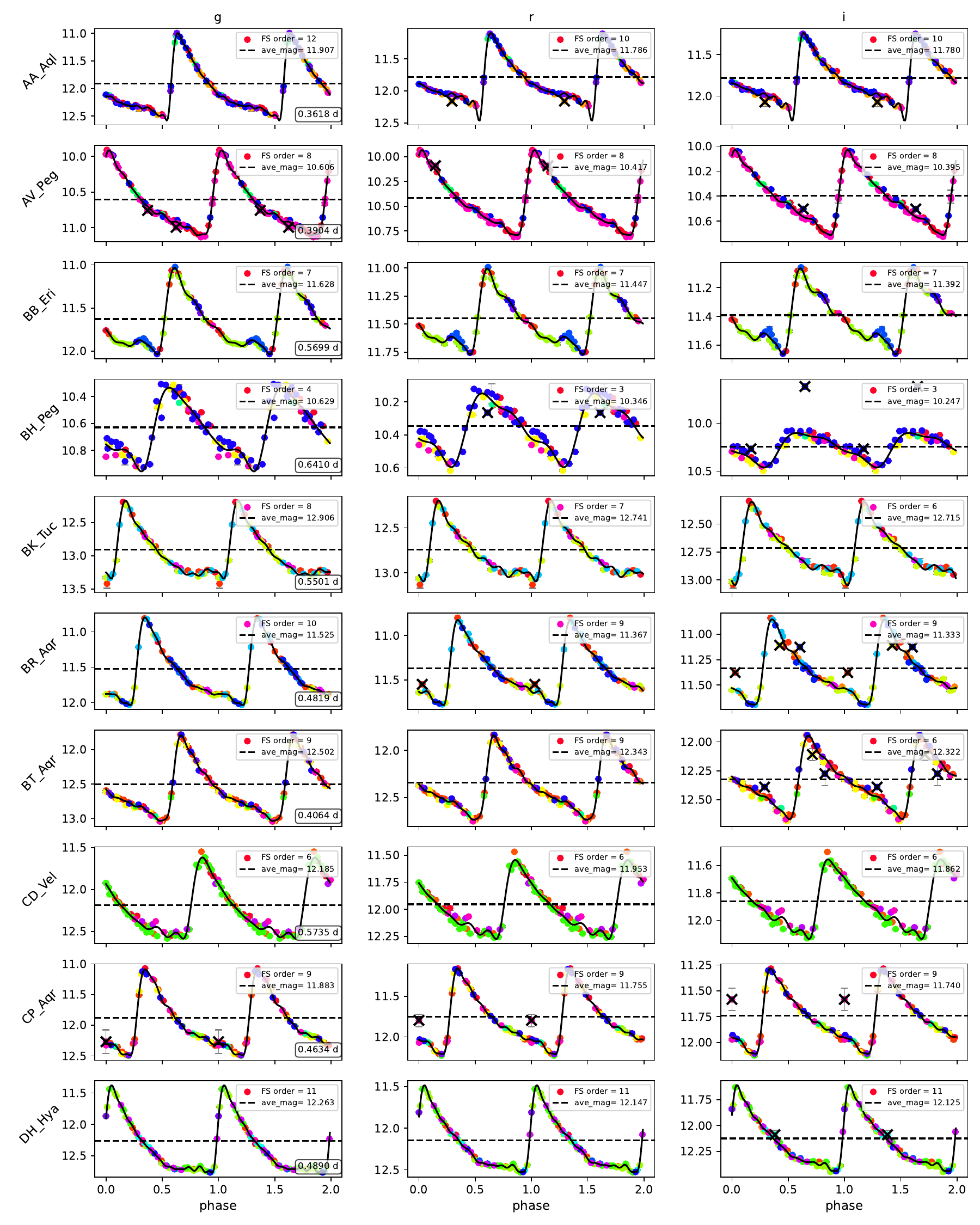}
    \caption{\label{fig:fig1} 
    Sloan--Pan-STARRS $g_{P1}r_{P1}i_{P1}$ band light curves of RRab stars 
    analyzed in this work. 
    Horizontal dashed, black lines correspond to the determined mean magnitudes. 
    Different colors of points mark different telescopes used during the data 
    collection, while black crosses mark points rejected during the fitting. 
    Black lines show the best fit Fourier series.}
\end{figure*}

\begin{figure*}
    \ContinuedFloat
    \centering
    \includegraphics[width=\textwidth]{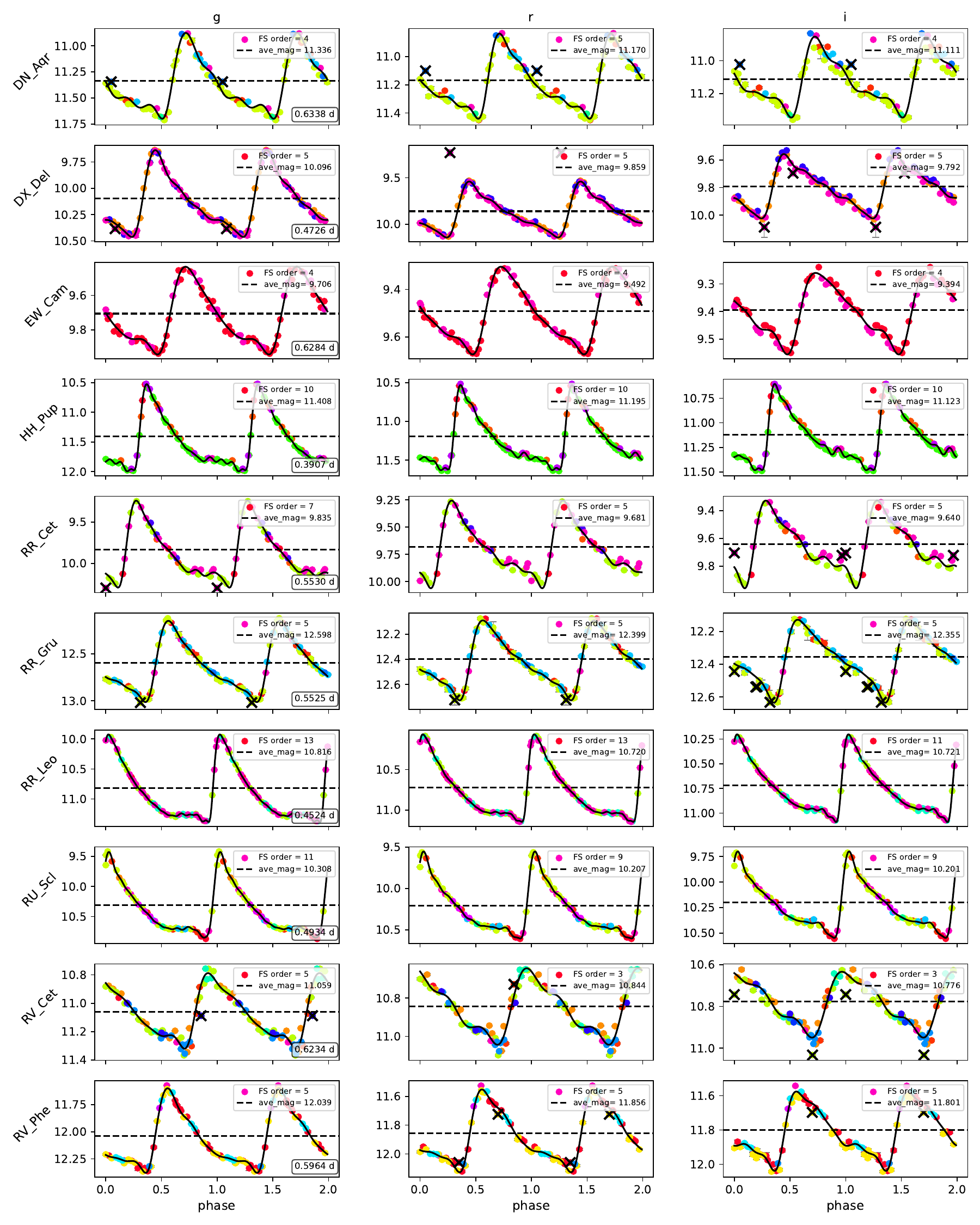}
    \caption{Continued from the previous page.}
\end{figure*}

\begin{figure*}
    \ContinuedFloat
    \centering
    \includegraphics[width=\textwidth]{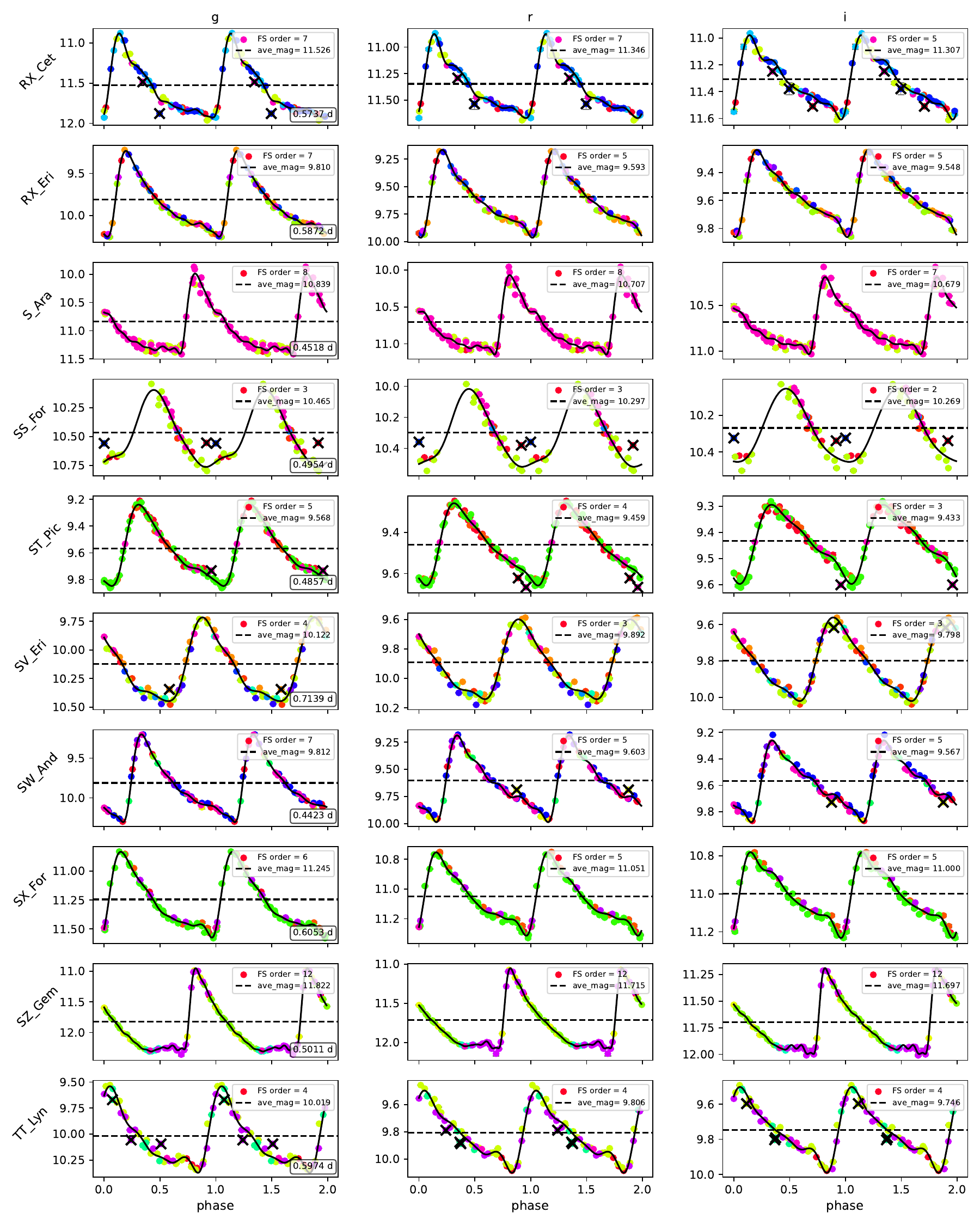}
    \caption{Continued from the previous page.}
\end{figure*}

\begin{figure*}
    \ContinuedFloat
    \centering
    \includegraphics[width=\textwidth]{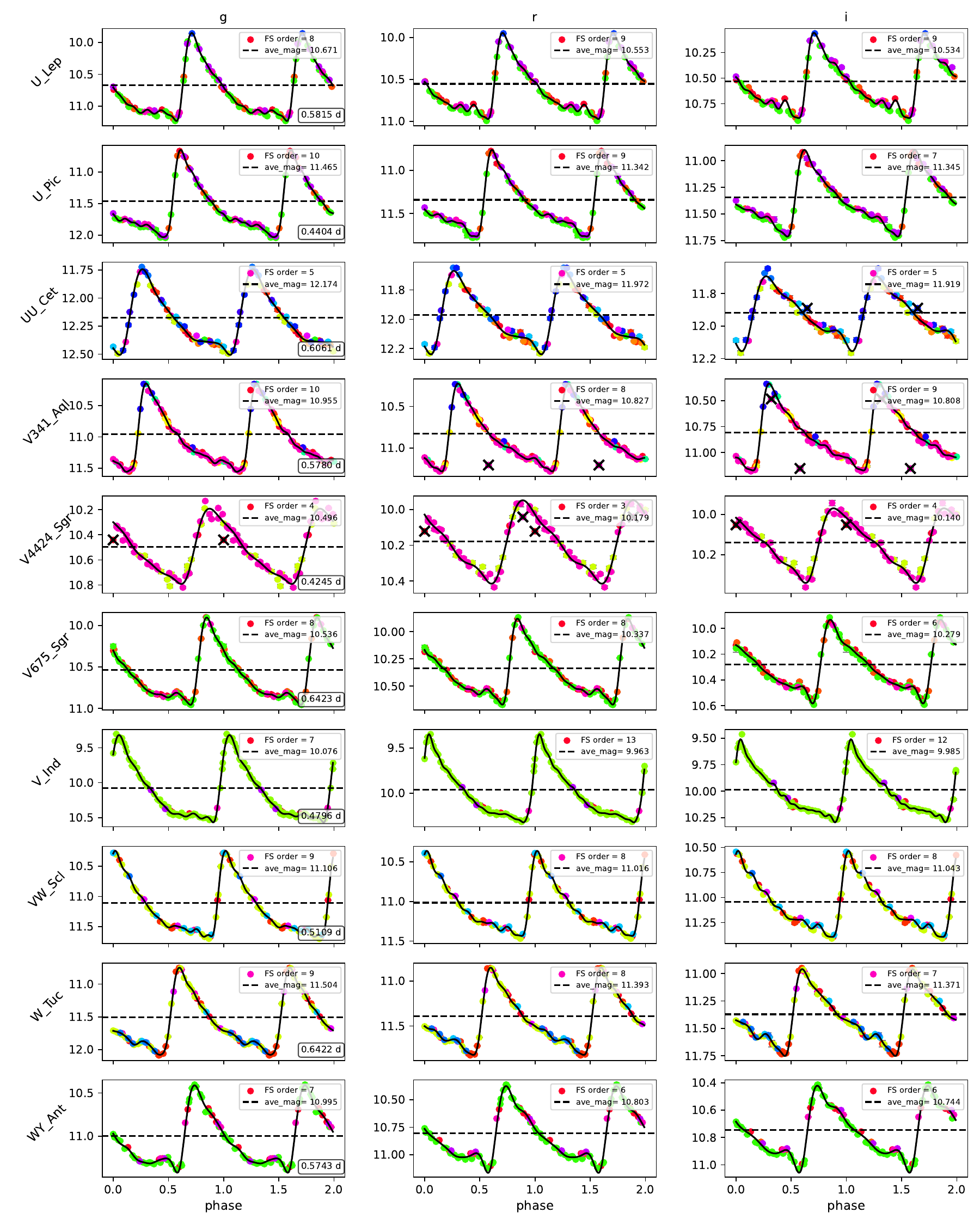}
    \caption{Continued from the previous page.}
\end{figure*}

\begin{figure*}
    \ContinuedFloat
    \centering
    \includegraphics[width=\textwidth]{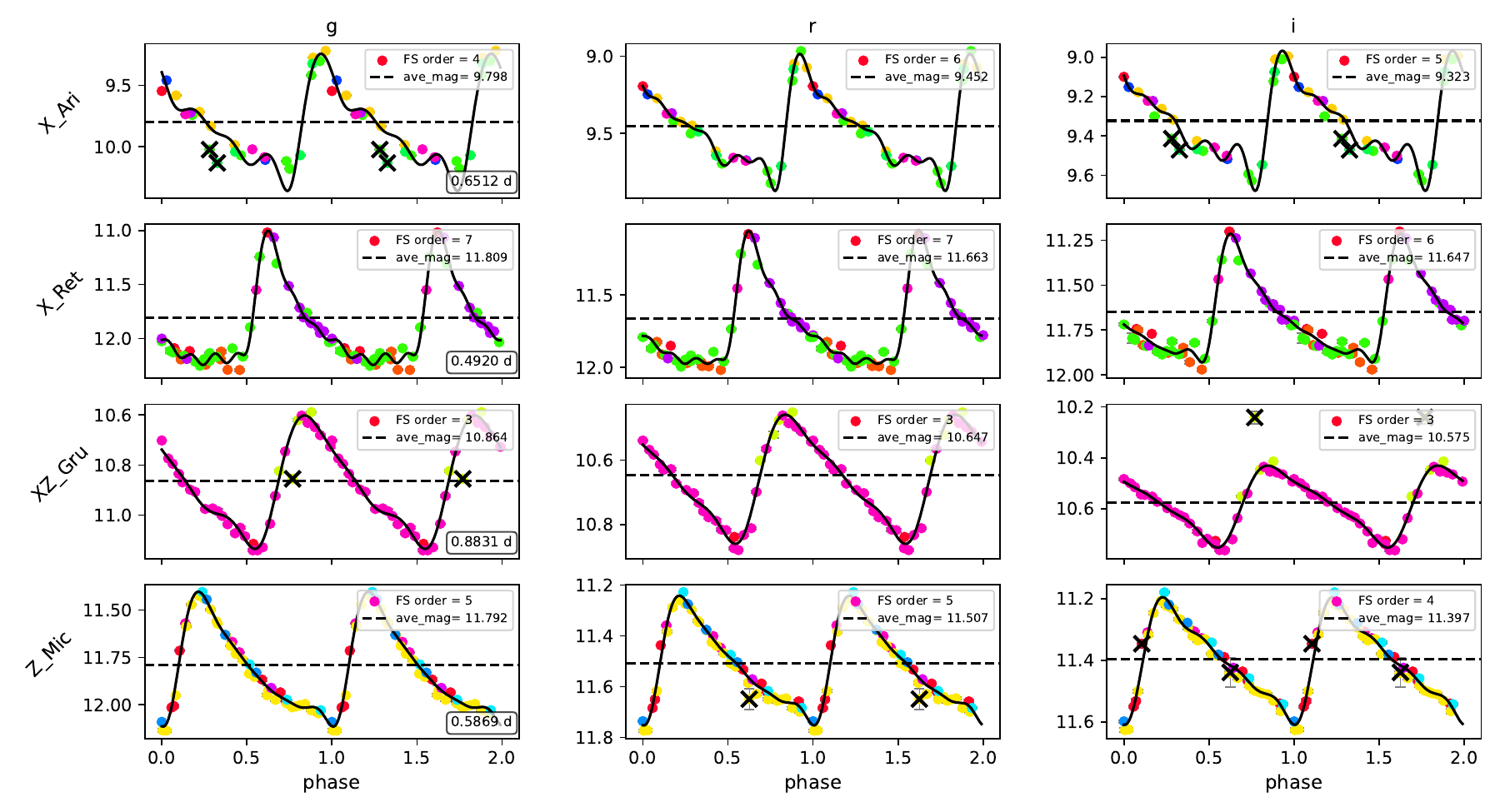}
    \caption{Continued from the previous page.}
\end{figure*}

\begin{figure*}[h]
    \centering
    \includegraphics[width=\textwidth]{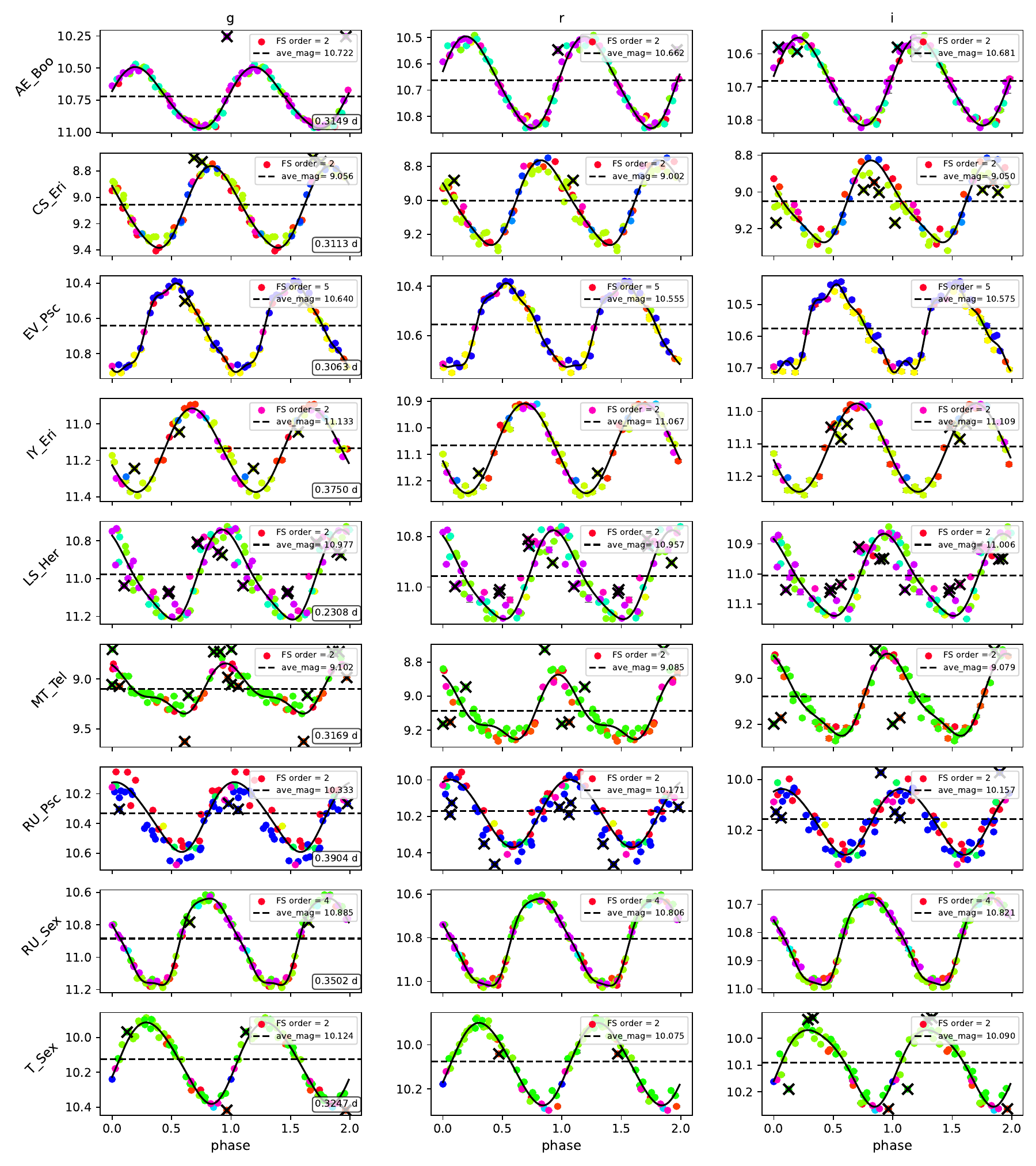}
    \caption{\label{fig:fig2} 
    Sloan--Pan-STARRS $g_{P1}r_{P1}i_{P1}$ band light curves of RRc stars analyzed 
    in this work. 
    Horizontal dashed, black lines correspond to the determined mean magnitudes. 
    Different colors of points mark different telescopes used during the data collection, 
    while black crosses mark points rejected during the fitting. 
    Black lines show the best fit Fourier series.}
\end{figure*}

\section{The ZP shift of PL relations}
\label{app:ZPshift}

Tables~\ref{tab:ZPshift_teoretical} and \ref{tab:ZPshift_empirical} present the 
ZP shift between PL relations obtained in this work with four methods of deriving absolute 
magnitudes described in Section~\ref{ssec:dist}, and theoretical and empirical PL relations 
from the literature. 

%
\begin{table*}
\scriptsize
\caption{ZP shift from the comparison with the theoretical PLZ relation from \citet{CaceresCatelan2008}.}             
\label{tab:ZPshift_teoretical}      
\centering          
\begin{tabular}{c|cccc}     
\hline\hline       
\hline     
 $\rm{[\alpha/Fe]}$ & parallax & ABL method & geometric distances & photo-geometric distances \\
  (dex) &  &  &  &  \\
\hline 
\multicolumn{5}{l}{$\rm{[Fe/H]}_a=-0.5$~dex; $\log P_a = -0.25$~days} \\ 
\hline
$-0.1$ & 0.176 & 0.177 & 0.170 & 0.164 \\ 
$+0.2$ & 0.132 & 0.133 & 0.126 & 0.120 \\ 
$+0.5$ & 0.080 & 0.081 & 0.074 & 0.068 \\ 
\hline
\multicolumn{5}{l}{$\rm{[Fe/H]}_a=-2.0$~dex; $\log P_a = -0.25$~days} \\ 
\hline
$-0.1$ & 0.207 & 0.210 & 0.209 & 0.210 \\ 
$+0.2$ & 0.163 & 0.166 & 0.165 & 0.166 \\ 
$+0.5$ & 0.111 & 0.114 & 0.113 & 0.114 \\ 
\hline
\end{tabular}
\tablefoot{ 
The ZP shifts were calculated for PLZ relations of RRab+RRc stars, for adopted iron abundance 
$\rm{[Fe/H]}_a$, constant period $\log P_a$ for filter~SSDS~$i$.
}
\end{table*}

%
\begin{table*}
\scriptsize
\caption{ZP shift from the comparison with empirical PLZ relations.}             
\label{tab:ZPshift_empirical}      
\centering          
\begin{tabular}{c|cccc}     
\hline\hline       
\hline     
\multicolumn{5}{l}{\citet{Sesar2017}} \\ 
\hline
 filter & parallax & ABL method & geometric distances & photo-geometric distances \\
\hline 
\multicolumn{5}{l}{$\rm{[Fe/H]}_a=-0.5$~dex; $\log P_a = -0.25$~days} \\ 
\hline
$M^{P1}_{g}$ (RRab) & 0.241 & 0.247 & 0.242 & 0.231 \\ 
$M^{P1}_{r}$ (RRab) & 0.211 & 0.214 & 0.213 & 0.203 \\ 
$M^{P1}_{i}$ (RRab) & 0.219 & 0.223 & 0.220 & 0.211 \\ 
\hline
\multicolumn{5}{l}{$\rm{[Fe/H]}_a=-2.0$~dex; $\log P_a = -0.25$~days; $M_{i}$} \\ 
\hline
 $\rm{[\alpha/Fe]}$ (dex) & parallax & ABL method & geometric distances & photo-geometric distances \\
\hline 
$M^{P1}_{g}$ (RRab) & -0.036 & -0.033 & -0.039 & -0.035 \\ 
$M^{P1}_{r}$ (RRab) & 0.039 & 0.043 & 0.036 & 0.041 \\ 
$M^{P1}_{i}$ (RRab) & 0.078 & 0.081 & 0.075 & 0.079 \\ 
\hline\hline 
\multicolumn{5}{l}{\citet{Vivas2017}} \\ 
\hline
\multicolumn{5}{l}{$\rm{[Fe/H]}_a=-1.25$~dex (M5); $\log P_a = -0.25$~days; $\rm{DM}_a = 14.44$~mag} \\ 
\hline
$M^{P1}_{g}$ (RRab) & 0.335 & 0.339 & 0.334 & 0.330 \\ 
$M^{P1}_{r}$ (RRab) & 0.252 & 0.256 & 0.252 & 0.249 \\ 
$M^{P1}_{i}$ (RRab) & 0.226 & 0.229 & 0.225 & 0.223 \\ 
\hline
\multicolumn{5}{l}{$\rm{[Fe/H]}_a=-1.25$~dex (M5); $\log P_a = -0.25$~days; $\rm{DM}_a = 14.37$~mag} \\ 
\hline
$M^{P1}_{g}$ (RRab) & 0.265 & 0.269 & 0.264 & 0.260 \\ 
$M^{P1}_{r}$ (RRab) & 0.182 & 0.186 & 0.182 & 0.179 \\ 
$M^{P1}_{i}$ (RRab) & 0.156 & 0.159 & 0.155 & 0.153 \\ 
\hline\hline 
\multicolumn{5}{l}{\citet{Bhardwaj2021}} \\ 
\hline
\multicolumn{5}{l}{$\rm{[Fe/H]}_a=-2.33$~dex (M15); $\log P_a = -0.25$~days; $\rm{DM}_a = 15.15$~mag} \\ 
\hline
$M^{SDSS}_{g}$ (RRab) & 0.174 & 0.176 & 0.170 & 0.177 \\ 
$M^{SDSS}_{i}$ (RRab) & 0.210 & 0.213 & 0.206 & 0.213 \\ 
$M^{SDSS}_{g}$ (RRab+RRc) & 0.179 & 0.182 & 0.181 & 0.185 \\ 
$M^{SDSS}_{i}$ (RRab+RRc) & 0.197 & 0.200 & 0.200 & 0.203 \\ 
\hline\hline
\multicolumn{5}{l}{\citet{Ngeow2022RRL}} \\ 
\hline
\multicolumn{5}{l}{$\rm{[Fe/H]}_a=-0.5$~dex; $\log P_a = -0.25$~days} \\ 
\hline
$M^{P1}_{g}$ (RRab) & 0.413 & 0.419 & 0.414 & 0.403 \\ 
$M^{P1}_{r}$ (RRab) & 0.316 & 0.319 & 0.318 & 0.308 \\ 
$M^{P1}_{i}$ (RRab) & 0.280 & 0.284 & 0.281 & 0.272 \\ 
$W^{P1}_{ri}$ (RRab) & 0.019 & 0.024 & 0.021 & 0.010 \\ 
$W^{P1}_{gr}$ (RRab) & 0.111 & 0.099 & 0.112 & 0.101 \\ 
$W^{P1}_{gi}$ (RRab) & 0.035 & 0.032 & 0.036 & 0.026 \\ 
$M^{P1}_{g}$ (RRab+RRc) & 0.366 & 0.371 & 0.361 & 0.355 \\ 
$M^{P1}_{r}$ (RRab+RRc) & 0.304 & 0.305 & 0.297 & 0.291 \\ 
$M^{P1}_{i}$ (RRab+RRc) & 0.299 & 0.300 & 0.293 & 0.287 \\ 
$W^{P1}_{ri}$ (RRab+RRc) & 0.060 & 0.068 & 0.052 & 0.051 \\ 
$W^{P1}_{gr}$ (RRab+RRc) & 0.198 & 0.192 & 0.186 & 0.185 \\ 
$W^{P1}_{gi}$ (RRab+RRc) & 0.085 & 0.083 & 0.072 & 0.072 \\ 
\hline
\multicolumn{5}{l}{$\rm{[Fe/H]}_a=-2.0$~dex; $\log P_a = -0.25$~days} \\ 
\hline
$M^{P1}_{g}$ (RRab) & 0.255 & 0.258 & 0.252 & 0.256 \\ 
$M^{P1}_{r}$ (RRab) & 0.217 & 0.221 & 0.214 & 0.219 \\ 
$M^{P1}_{i}$ (RRab) & 0.235 & 0.238 & 0.232 & 0.236 \\ 
$W^{P1}_{ri}$ (RRab) & 0.186 & 0.188 & 0.184 & 0.188 \\ 
$W^{P1}_{gr}$ (RRab) & 0.150 & 0.148 & 0.136 & 0.140 \\ 
$W^{P1}_{gi}$ (RRab) & 0.158 & 0.163 & 0.154 & 0.158 \\ 
$M^{P1}_{g}$ (RRab+RRc) & 0.218 & 0.221 & 0.219 & 0.220 \\ 
$M^{P1}_{r}$ (RRab+RRc) & 0.194 & 0.197 & 0.195 & 0.196 \\ 
$M^{P1}_{i}$ (RRab+RRc) & 0.176 & 0.177 & 0.176 & 0.177 \\ 
$W^{P1}_{ri}$ (RRab+RRc) & 0.130 & 0.131 & 0.131 & 0.132 \\ 
$W^{P1}_{gr}$ (RRab+RRc) & 0.138 & 0.144 & 0.141 & 0.142 \\ 
$W^{P1}_{gi}$ (RRab+RRc) & 0.116 & 0.118 & 0.118 & 0.119 \\ 
\hline
\end{tabular}
\tablefoot{ 
The shifts were calculated for adopted iron abundance $\rm{[Fe/H]}_a$, constant period $\log P_a$.
}
\end{table*}

\end{appendix}

\end{document}